\documentclass[twocolumn,aps,pra,superscriptaddress,amsmath,amssym,floatfix,longbibliography,nofootinbib]{revtex4-1}

\usepackage{graphicx}
\usepackage{bm}
\usepackage{color}
\usepackage[usenames,dvipsnames]{xcolor}
\usepackage{tikz}
\usetikzlibrary{%
    decorations.pathreplacing,%
    decorations.pathmorphing,%
    arrows
}
\usepackage{amssymb}
\usepackage{enumerate}
\usepackage{comment}
\usepackage{listings}

\usepackage{accents}
\newcommand{\ubar}[1]{\underaccent{\bar}{#1}}

\usepackage[colorlinks=true,linkcolor=MidnightBlue,citecolor=blue,filecolor=green,urlcolor=PineGreen]{hyperref}

\def\be{\begin{equation}}
\def\ee{\end{equation}}
\newcommand{\partl}[3]{ \frac{\partial^{#3}#1}{ \partial #2^{#3}} }	
\newcommand{\limit}[2]{\underset{#1 \rightarrow #2}{\text{lim}} \;}
\newcommand{\ket}[1]{\left\vert #1 \right\rangle}
\newcommand{\bra}[1]{\left\langle #1 \right\vert}
\newcommand{\ip}[2]{\langle #1 \vert #2 \rangle}
\newcommand{\dxval}[1]{\left\langle #1 \right\rangle}
\newcommand{\exval}[3]{\langle #1 \vert #2 \vert #3 \rangle}

\definecolor{codegreen}{rgb}{0,0.6,0}
\definecolor{codegray}{rgb}{0.5,0.5,0.5}
\definecolor{codepurple}{rgb}{0.58,0,0.82}
\definecolor{backcolour}{rgb}{0.95,0.95,0.95}
\definecolor{internationalorange}{rgb}{1.0, 0.31, 0.0}
\definecolor{cadetgrey}{rgb}{0.57, 0.64, 0.69}
\definecolor{ceruleanblue}{rgb}{0.16, 0.32, 0.75}
\definecolor{darkspringgreen}{rgb}{0.09, 0.45, 0.27}
\definecolor{deeplilac}{rgb}{0.6, 0.33, 0.73}
\definecolor{goldenyellow}{rgb}{1.0, 0.87, 0.0}
\definecolor{cobalt}{rgb}{0.0, 0.28, 0.67}
\definecolor{selectiveyellow}{rgb}{1.0, 0.73, 0.0}
\definecolor{turquoiseblue}{rgb}{0.0, 1.0, 0.94}
\definecolor{vividviolet}{rgb}{0.62, 0.0, 1.0}
\definecolor{neongreen}{rgb}{0.22, 0.88, 0.08}
\definecolor{huntergreen}{rgb}{0.21, 0.37, 0.23}
\definecolor{lavenderpurple}{rgb}{0.59, 0.48, 0.71}
\definecolor{coquelicot}{rgb}{1.0, 0.22, 0.0}
\definecolor{crimsonglory}{rgb}{0.75, 0.0, 0.2}
\definecolor{deeppink}{rgb}{1.0, 0.08, 0.58}
\definecolor{electricviolet}{rgb}{0.56, 0.0, 1.0}
\definecolor{electricgreen}{rgb}{0.0, 1.0, 0.0}
\definecolor{mint}{rgb}{0.24, 0.71, 0.54}
\definecolor{dodgerblue}{rgb}{0.12, 0.56, 1.0}
\definecolor{lincolngreen}{rgb}{0.11, 0.35, 0.02}
\definecolor{persianblue}{rgb}{0.11, 0.22, 0.73}
\definecolor{patriarch}{rgb}{0.5, 0.0, 0.5}
\definecolor{amaranth}{rgb}{0.9, 0.17, 0.31}

\lstdefinestyle{mystyle}{
    backgroundcolor=\color{backcolour},   
    commentstyle=\tt\color{Cerulean},
    keywordstyle=\tt\color{Fuchsia},
    numberstyle=\it\tiny\color{MidnightBlue},
    stringstyle=\tt\color{red},
    basicstyle=\tt\footnotesize,
    breakatwhitespace=false,         
    breaklines=true,                 
    captionpos=b,                    
    keepspaces=true,                 
    numbers=left,                    
    numbersep=5pt,                  
    showspaces=false,                
    showstringspaces=false,
    showtabs=false,                  
    tabsize=2
}
 
\newcommand{\blue}[1]{{\color{blue}#1}}

\begin{document}

\title{
Measuring Fluorescence to Track a Quantum Emitter's State: A Theory Review
}
\author{Philippe Lewalle} 
\email{plewalle@ur.rochester.edu}
\affiliation{Department of Physics and Astronomy, University of Rochester, Rochester, NY 14627, USA}
\affiliation{Center for Coherence and Quantum Optics, University of Rochester, Rochester, NY 14627, USA}
\author{Sreenath K. Manikandan} 
\affiliation{Department of Physics and Astronomy, University of Rochester, Rochester, NY 14627, USA}
\affiliation{Center for Coherence and Quantum Optics, University of Rochester, Rochester, NY 14627, USA}
\author{Cyril Elouard} 
\affiliation{Department of Physics and Astronomy, University of Rochester, Rochester, NY 14627, USA}
\affiliation{Center for Coherence and Quantum Optics, University of Rochester, Rochester, NY 14627, USA}
\author{Andrew N. Jordan} 
\affiliation{Department of Physics and Astronomy, University of Rochester, Rochester, NY 14627, USA}
\affiliation{Center for Coherence and Quantum Optics, University of Rochester, Rochester, NY 14627, USA}
\affiliation{Institute for Quantum Studies, Chapman University, Orange, CA 92866, USA}
\date{\today}

\begin{abstract}
    We review the continuous monitoring of a qubit through its spontaneous emission, at an introductory level. Contemporary experiments have been able to collect the fluorescence of an artificial atom in a cavity and transmission line, and then make measurements of that emission to obtain diffusive quantum trajectories in the qubit's state. We give a straightforward theoretical overview of such scenarios, using a framework based on Kraus operators derived from a Bayesian update concept; we apply this flexible framework across common types of measurements including photodetection, homodyne, and heterodyne monitoring, and illustrate its equivalence to the stochastic master equation formalism throughout. Special emphasis is given to homodyne (phase--sensitive) monitoring of fluorescence. The examples we develop are used to illustrate basic methods in quantum trajectories, but also to introduce some more advanced topics of contemporary interest, including the arrow of time in quantum measurement, and trajectories following optimal measurement records derived from a variational principle. The derivations we perform lead directly from the development of a simple model to an understanding of recent experimental results.
\end{abstract}

\maketitle

\section{Introduction}

The literature on quantum theory and quantum optics is replete with works concerning the spontaneous emission of atoms, across virtually all of its century--long history \cite{EinsteinAB,DiracFlor, Weisskopf1930, Mollow1969, AckerhaltEberly, Kimble1976, MilonniFlor, HarocheReview}. 
The generic case, in which the excited--state population of an emitter decays exponentially on average due to the spontaneous emission of a photon, is a paradigmatic phenomenon in quantum optics.
More recently, both the theory \cite{BookCarmichael, BookGardiner, BookPercival, BookWiseman, BookBarchielli, BookJacobs, Gisin1984, Barchielli1986, Caves1987, Diosi1988, Gisin1992-1, Gisin1992-2, Wiseman1993prl, Wiseman1993pra1, Wiseman1993pra2, Wiseman1996Review, Brun2001Teach, Gough2004, Jacobs2006, Clerk2010Review, Santos2011, Gough2012, Jordan2013rev, Chantasri2013, Bolund2014, Chantasri2015, Korotkov2016, Gross_2018, Gough2018, Gough2018_JMP, Gough2019-1, Gough2019-2, Crowder2019} and experiments \cite{Gisin1993, Gambetta2008, Murch2013, Weber2014, Leigh2016} about continuous quantum measurement have received considerable attention, and seen rapid progress, revealing new phenomena and insights into the quantum measurement process \cite{Leigh2016, Lewalle2016, DresselArrow, SreenathReversal, Lewalle2018, SreenathFluctuation}, and applications to quantum control \cite{Taylor2017, Gourgy2018, Minev2018}. 
In any such generalized measurement(s) \cite{BookWiseman, BookNielsen} of some primary system of interest, there must necessarily be some series of interactions between that system and its environment, which allows for information to flow from the primary system to some meter(s) which record the measurement outcome(s) \cite{VonNeumann}. 
The interaction between the system and environment will necessarily disturb the system of interest in a random way, but inferences about that evolution of the system of interest can be drawn as long as our measurement brings us the information that the environment ``learned'' by interacting with the system \cite{HarocheBook}.
Generalized measurements can be weak (a small amount of information is acquired about the system state, with correspondingly little disruption to its prior behavior) \cite{AAV1988, Tamir2013, Fuchs1996, Fuchs2001, Oreshkov2005}, or strong (e.g.~the system is ``collapsed'' to an eigenstate of the measurement operator by a projective measurement, such that we have acquired a lot of information at once and disturbed the state by corresponding ramifications in the process)  \cite{VonNeumann, BookWiseman}. 
Thus, in contrast with the closed--system quantum mechanics described by the Schr\"{o}dinger equation alone, a physical description of the measurement process necessarily requires that we consider our primary system as being open. 
A ``quantum trajectory'' arises when a sequence of measurements are made in time, such that we have a time--series of measurement outcomes, and a corresponding time--series of inferred quantum states of our primary system, based on that information. The process is necessarily stochastic, as there is randomness present in each successive measurement outcome. 

\par
\begin{figure}
    \centering
    \begin{tikzpicture}[
    wave/.style={decorate,decoration={snake,post length=0.1cm,amplitude=0.2cm,segment length=0.3cm}},
    wave2/.style={decorate,decoration={snake,post length=0.6cm,amplitude=0.3cm,segment length=0.3cm}}, wave3/.style={decorate,decoration={snake,post length=0.2cm,amplitude=0.15cm,segment length=0.2cm}},
    wave4/.style={decorate,decoration={snake,post length=0.0cm,amplitude=0.5cm,segment length=1.5cm}}]
    \draw[fill = black!60, color = black!60, rounded corners = 0.3 cm] (0,0) -- (0,3) -- (8.5,3) -- (8.5,0) -- cycle;
    \draw[fill = black!60, color = black!60, rounded corners = 0.3 cm] (0,3.2) -- (0,7.2) -- (8.5,7.2) -- (8.5,3.2) -- cycle;
    \draw[fill = black!60, color = black!60, rounded corners = 0.3 cm] (0,7.4) -- (0,9.4) -- (4,9.4) -- (4,7.4) -- cycle;
    \draw[fill = black!60, color = black!60, rounded corners = 0.3 cm] (4.2,7.4) -- (4.2,9.4) -- (8.5,9.4) -- (8.5,7.4) -- cycle;
    \filldraw[even odd rule,inner color=red ,outer color=black!60,color = black!60] (2,8.4) circle (1.0);
    \draw[red,wave3,line width = 0.05cm,->, opacity = 0.4] (2,8.4) -- (2,9.3);
    \draw[red,wave3,line width = 0.05cm,->, opacity = 0.4] (2,8.4) -- (3.2,9.1);
    \draw[red,wave3,line width = 0.05cm,->, opacity = 0.4] (2,8.4) -- (3.2,7.7);
    \draw[red,wave3,line width = 0.05cm,->, opacity = 0.4] (2,8.4) -- (2,7.5);
    \draw[red,wave3,line width = 0.05cm,->, opacity = 0.4] (2,8.4) -- (0.8,9.1);
    \draw[red,wave3,line width = 0.05cm,->, opacity = 0.4] (2,8.4) -- (0.8,7.7);
    \draw[color = white, line width = 0.03cm] (1.8,8.5) -- (2.2,8.5);
    \draw[color = white, line width = 0.03cm] (1.8,8.3) -- (2.2,8.3);
    \draw[fill = black,color= black,rounded corners = 0.08cm] (7.5,8.9) arc[radius = 0.5, start angle = 90, end angle = -90] -- cycle;
    \draw[fill = black!50, color = black!50] (5.9,7.9) -- (6.3,8.4) -- (5.9,8.9) -- cycle;
    \filldraw[even odd rule,inner color=red ,outer color=black!50,color = black!50] (5.3,8.4) circle (0.8);
    \draw[red,wave,line width = 0.05cm,->] (5.3,8.4) -- (6.3,8.4);
    \draw[color = white, line width = 0.05cm] (5.0,8.6) -- (5.6,8.6);
    \draw[color = white, line width = 0.05cm] (5.0,8.2) -- (5.6,8.2);
    \draw[color = red, line width = 0.05cm, ->] (6.3,8.4) -- (7.5,8.4);
    \draw[fill =  black!50, color =  black!50] (1.6,5.7) -- (2,6.2) -- (1.6,6.7) -- cycle;
    \draw[fill =  black!50, color =  black!50] (0.5,5.6) -- (1,5.2) -- (1.5,5.6) -- cycle;
    \filldraw[even odd rule,inner color=red ,outer color= black!50,color =  black!50] (1,6.2) circle (0.8);
    \draw[red,wave,line width = 0.05cm,->] (1,6.2) -- (2,6.2);
    \draw[Dandelion,wave2,line width = 0.05cm,->] (1,3.5) -- (1,5.2);
    \draw[color = white, line width = 0.05cm] (0.7,6.4) -- (1.3,6.4);
    \draw[color = white, line width = 0.05cm] (0.7,6.0) -- (1.3,6.0);
    \draw[color = Dandelion, line width = 0.04cm, ->] (1.2,6.05) arc[radius = 0.35, start angle = -25, end angle = 25];
    \draw[color = Dandelion, line width = 0.04cm, ->] (0.8,6.35) arc[radius = 0.35, start angle = 155, end angle = 205];
    \draw[fill = Orchid!60!black, color = Orchid!60!black, rounded corners = 0.3 cm] (2.4,4.2) rectangle (4.0,7);
    \draw[color = red, dotted, line width = 0.05cm, ->] (2,6.2) -- (2.8,6.2);
    \draw[top color = MidnightBlue,bottom color = Cerulean,color = Cerulean,rounded corners = 0.08cm] (2.8,5.95) -- (2.8,6.5) -- (3.4,6.5) -- cycle;
    \draw[top color = Cerulean,bottom color = RoyalBlue,color = Cerulean,rounded corners = 0.08cm] (2.85,5.9) -- (3.45,5.9) -- (3.45,6.45) -- cycle;
    \draw[red,wave2,line width = 0.05cm,->] (3.15,4.6) -- (3.15,5.9);
    \draw[red,line width = 0.05cm,->] (3.15,6.5) -- (3.15,6.7);
    \draw[red,line width = 0.05cm,->] (3.45,6.2) -- (3.65,6.2);
    \draw[fill = black,color= black,rounded corners = 0.03cm] (3.65,6.45) arc[radius = 0.25, start angle = 90, end angle = -90] -- cycle;
    \draw[fill = black,color= black,rounded corners = 0.03cm] (2.9,6.7) arc[radius = 0.25, start angle = 180, end angle = 0] -- cycle;
    \draw[line width = 0.05cm,->] (7.0,6.0) -- (7.7,6.0);
    \draw[color = Orchid!05!black, fill = Orchid!60!black, rounded corners = 0.2 cm, line width = 0.05cm] (5.7,5.3) -- (5.7,6.7) -- (7.2,6.0) -- cycle;
    \draw[color=red, dotted, line width = 0.05cm, ->] (4.3,6.3) -- (5.7,6.3);
    \draw[color=red, line width = 0.05cm, ->] (5.0,5.7) -- (5.7,5.7);
    \draw[line width = 0.05cm, ->, rounded corners = 0.2cm] (6.5,4.0) -- (6.7,4.4) -- (7.7,4.4);
    \draw[line width = 0.05cm, ->, rounded corners = 0.2cm] (6.5,4.0) -- (6.7,3.6) -- (7.7,3.6);
    \draw[color = Orchid!05!black, fill = Orchid!60!black, rounded corners = 0.2 cm, line width = 0.05cm] (5.7,3.3) -- (5.7,4.7) -- (7.2,4.0) -- cycle;
    \draw[color=red, dotted, line width = 0.05cm, ->] (4.3,4.3) -- (5.7,4.3);
    \draw[color=internationalorange, line width = 0.05cm, ->] (5.0,3.7) -- (5.7,3.7);
    \draw[fill =  black!50, color =  black!50] (1.6,1) -- (2,1.5) -- (1.6,2) -- cycle;
    \filldraw[even odd rule,inner color=red ,outer color=black!50,color = black!50] (1,1.5) circle (0.8);
    \draw[red,wave,line width = 0.05cm,->] (1,1.5) -- (2,1.5);
    \draw[color = white, line width = 0.05cm] (0.7,1.7) -- (1.3,1.7);
    \draw[color = white, line width = 0.05cm] (0.7,1.3) -- (1.3,1.3);
    \draw[color = red, line width = 0.05cm, ->] (2,1.5) -- (3,1.5);
    \draw[top color = MidnightBlue,bottom color = Cerulean,color = Cerulean,rounded corners = 0.08cm] (3,1.05) -- (3,2.05) -- (4,2.05) -- cycle;
    \draw[top color = Cerulean,bottom color = RoyalBlue,color = Cerulean,rounded corners = 0.08cm] (3.1,0.95) -- (4.1,0.95) -- (4.1,1.95) -- cycle;
    \draw[color = red, line width = 0.05cm, opacity = 0.6, ->] (4.1,1.5) -- (6,1.5);
    \draw[color = red, line width = 0.05cm, opacity = 0.6, ->] (3.55,2.05) -- (3.55,2.8);
    \draw[color = red, line width = 0.05cm, dotted] (3,1.5) -- (3.55,1.5);
    \draw[color = red, line width = 0.05cm, dotted, opacity = 0.6] (3.55,1.5) -- (4.1,1.5);
    \draw[color = red, line width = 0.05cm, dotted, opacity = 0.6] (3.55,1.5) -- (3.55,2.05);
    \draw[color = black, line width = 0.05cm, ->] (3.55,0.2) -- (3.55, 0.95);
    \draw[fill = Orchid!60!black, color = Orchid!60!black, rounded corners = 0.3 cm] (6,0.8) rectangle (8,2.2);
    \end{tikzpicture} \\
    \begin{picture}(0,0)
    \put(-117,268){\color{white} (a)}
    \put(106,268){\color{white} (b)}
    \put(106,205){\color{white} (c)}
    \put(106,85){\color{white} (d)}
    \put(-110,184.5){\color{Dandelion} $\Omega$}
    \put(-40,132){\color{red} \bf LO,$\boldsymbol{\theta}$}
    \put(-45.5,120){\color{Orchid!30} balanced}
    \put(-38,108){\color{Orchid!30} dyne}
    \put(45,178){QLA}
    \put(0,205){\color{Orchid!30} \bf hom.$\sim$phase--sensitive}
    \put(-3,170){\color{red} \bf LO,$\boldsymbol{\theta}$}
    \put(104,179){$r$}
    \put(45,121){QLA}
    \put(0,148){\color{Orchid!30} \bf het.$\sim$phase--preserving}
    \put(-3,113){\color{internationalorange} \bf LO,$\boldsymbol{\theta}$}
    \put(104,134){$r_I$}
    \put(104,111){$r_Q$}
    \put(-17,76){ {\color{Cerulean!50} $\boldsymbol{\sqrt{1-\eta}}$}~{\color{red} $\hat{\boldsymbol{a}}^{\boldsymbol{\dag}}_{\boldsymbol{\ell}} $} }
    \put(0,42){ {\color{Cerulean!50} $\boldsymbol{\sqrt{\eta}}$}~{\color{red} $\hat{\boldsymbol{a}}^{\boldsymbol{\dag}}_{\mathbf{s} }$} }
    \put(-55,57){\color{red} $\hat{\boldsymbol{a}}^{\boldsymbol{\dag}}$}
    \put(-45,20){vac.}
    \put(61,57){any ideal}
    \put(63,45){detector}
    \end{picture} \vspace{-10pt}
    \caption{We show schematics representing the possible experimental situations we consider. Historically, treatments of spontaneous emission have often focused on the average dynamics of emitters, without any monitoring of individual emitters and their individual emission events, represented in (a). Our emphasis here is on devices in which the emission is captured in a cavity / transmission line, and routed to a detector, such as (b) a photodetector, or (c) a homodyne or heterodyne setup which measures one or both of the signal field's quadratures. Optically, the single--photon signal is mixed with a strong local oscillator (LO) on a 50/50 beamsplitter for quadrature detection, leading to a readout of one (homodyne) or both (heterodyne) quadratures of the field. Contemporary circuit--QED experiments using microwave photons typically perform these measurements using a quantum--limited amplifier (QLA) built from Joesphson junctions (see e.g.~\cite{Clerk2010Review, Bergeal2010, Flurin_Thesis, DevoretAmp}). The measurement axis in the quadrature phase space is determined by the relative phase $\theta$ between the signal and LO / amplifier pump tone. 
    Relevant experiments often include a drive characterized by the Rabi frequency $\Omega$. In (d) we illustrate a simple model for measurement inefficiency, in which an unbalanced beamsplitter splits the ideal signal into a measured portion with probability $\eta \in [0,1]$ and a lost portion (where $\hat{a}^\dag$ denotes a photon creation operator). See Sec.~\ref{sec-inefficient} for further details on this last point. }
    \label{fig-exp-diagrams}
\end{figure}
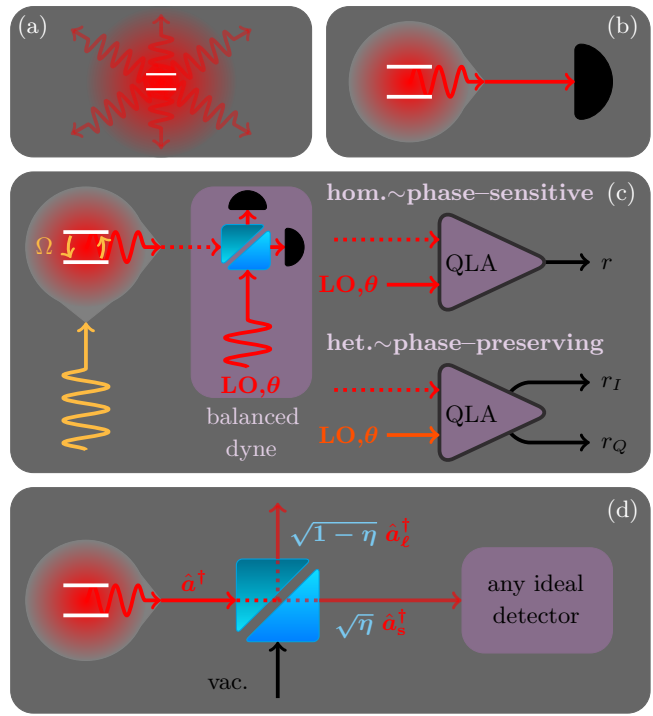

 Our emphasis will be on tracking the state of a qubit (a two--level quantum system), through its spontaneous emission; this means that the qubit is coupled to a field mode, which is its ``environment'' in this scheme, and by interrogating this mode in a variety of ways, we will be able to infer a corresponding evolution of the qubit's state. 
A quantized electromagnetic field mode is represented by a quantum harmonic oscillator; we will discuss the cases where we interrogate the output field by photodetection (effectively an energy measurement), or by quadrature\footnote{The quadrature space of the field is effectively the phase space of the quantum harmonic oscillator describing the field mode in question. In other words, a quadrature is analoguous to the ``position'' or the ``momentum'' of a quantum harmonic oscillator, and the product of the noise in orthogonal directions in quadrature space is bounded by the Heisenberg uncertainty principle. A reader unfamiliar with a quadrature phase space representation of a field mode may benefit from perusing e.g.~Refs.~\cite{Leuchs_1988_Squeezing, Silberhorn2007}.} measurements (homodyne or heterodyne detection are analoguous to making ``position'' and/or ``momentum'' measurements of the oscillator). 
We are motivated in large part by recent experimental work, in which a superconducting transmon qubit is continuously monitored by homodyne or heterodyne detection of its spontaneous emission, leading to diffusive quantum trajectories \cite{Wiseman1996Review, Campagne-Ibarcq2013, Bolund2014, Jordan2015flor, Campagne-Ibarcq2016, PCI-2016-2, Naghiloo2016flor, Mahdi2016, Tan2017, Mahdi2017Qtherm, Ficheux2018}. 
Following the relevant circuit--QED experiments, the physical setup we have in mind throughout this work involves a single qubit placed inside a cavity, such that microwave photons emitted by the qubit via spontaneous emission are coupled into a transmission line leading to a measurement device. 
The setup is designed such that photons emitted by the qubit via spontaneous emission are transmitted to the detector, but photons in other modes (e.g.~to implement some unitary Rabi rotations on the qubit) are not routed towards it. 
Such devices allow for high collection efficiency of emitted photons, in contrast with situations in which an atom emits into free space. See Fig.~\ref{fig-exp-diagrams} for an illustration, and guides to the experimental details can be found e.g.~in Refs.~\cite{Campagne-Ibarcq2016, PCI-2016-2, Naghiloo2016flor, Mahdi2016, Tan2017, Mahdi2017Qtherm, Ficheux2018, FicheuxThesis, MahdiTeachThesis}. 

We will proceed by splitting our manuscript into two main parts. In the first part, we describe a qubit open to a decay channel and subsequent measurements from several different perspectives. We carry out the formal treatment of an unmonitored decay channel, using first the typical quantum--mechanical analysis in Sec.~\ref{sec-QMdecay}, and then by introducing the corresponding master equation in Sec.~\ref{sec-OpenSysME}. 
We transition towards diffusive trajectories in Sec.~\ref{sec-transition}, introducing the Kraus operators that will serve as our primary tool, along with the Stochastic Master Equation (SME). We examine the cases of heterodyne or homodyne detection in detail, in sections \ref{sec-1QHet} and \ref{sec-1QHom}, respectively. We first discuss ideal measurements in which all information is collected, and then describe inefficient measurements in which some information is collected and some is lost in sec.~\ref{sec-inefficient}. 
Each of these steps is represented graphically in Fig.~\ref{fig-exp-diagrams}.
In the second part, we use the examples developed in the first part as a springboard to introduce certain concepts and methods of interest in the current research literature. For example, we are able to use these examples to introduce ideas related to the arrow of time in quantum trajectories in Sec.~\ref{sec-timereverse}, and discuss ``optimal paths'' (OPs) \cite{Chantasri2013, Chantasri2015, Jordan2015flor, Areeya_Thesis, Lewalle2016, Lewalle2018} in Secs.~\ref{sec-OPmain} and \ref{sec-OPineff}, which are quantum trajectories which connect given states according to an extremal--probability readout, derived according to a variational principle. 
Summary, outlook, and further discussion are included in Sec.~\ref{sec-conclude}. 

\section{Un--monitored Decay \label{sec-unmon}}

We review the case of a single qubit whose fluorescence goes unmonitored in two parts. First we review the standard treatment of Weisskopf and Wigner \cite{Weisskopf1930}; next we introduce an equivalent master equation description of the system \cite{HarocheBook, BookNielsen, Breuer1997}. 

\begin{figure*}
\centering
\includegraphics[width = .44\textwidth]{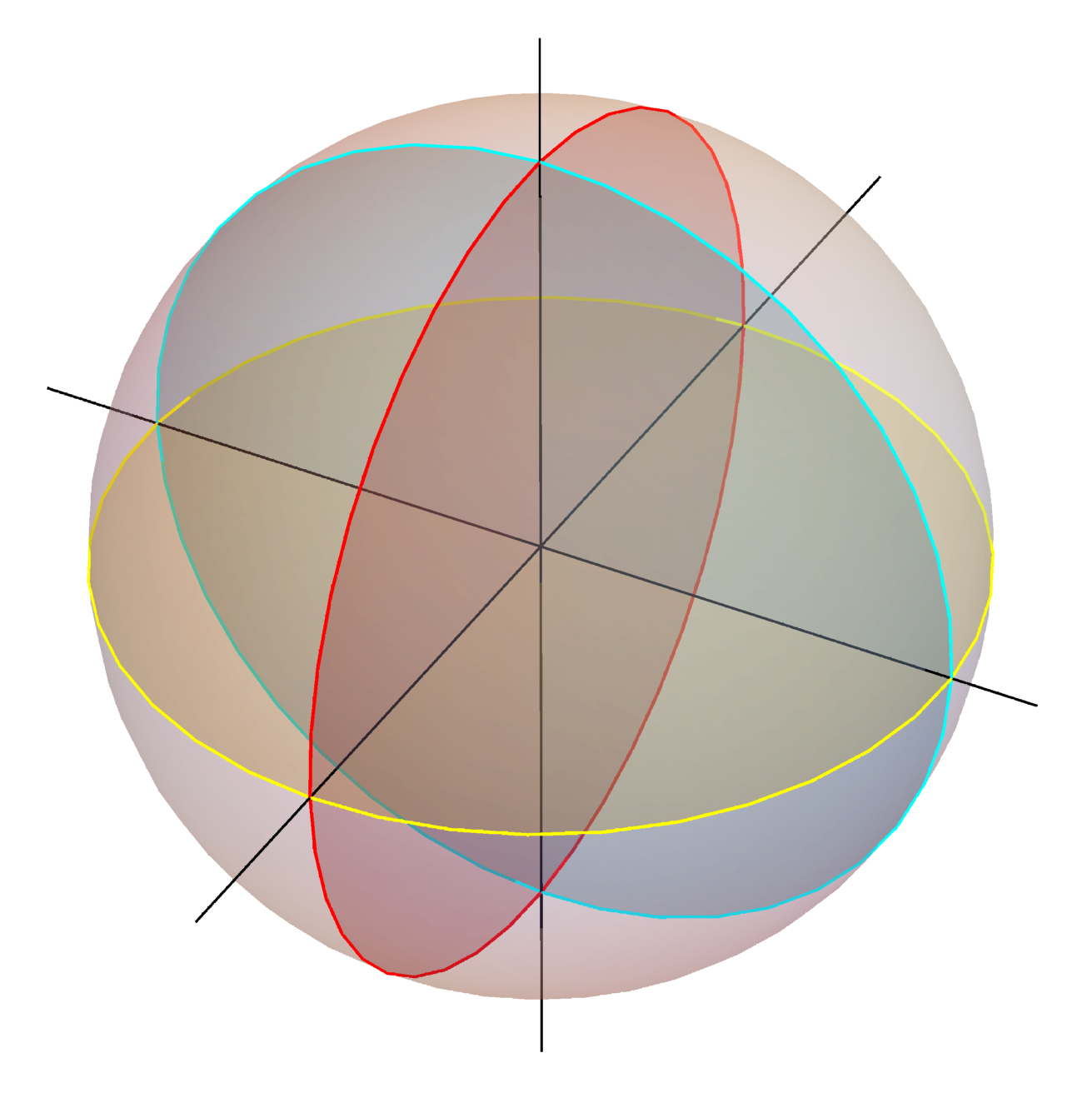} \hspace{1cm}
\includegraphics[width = .44\textwidth]{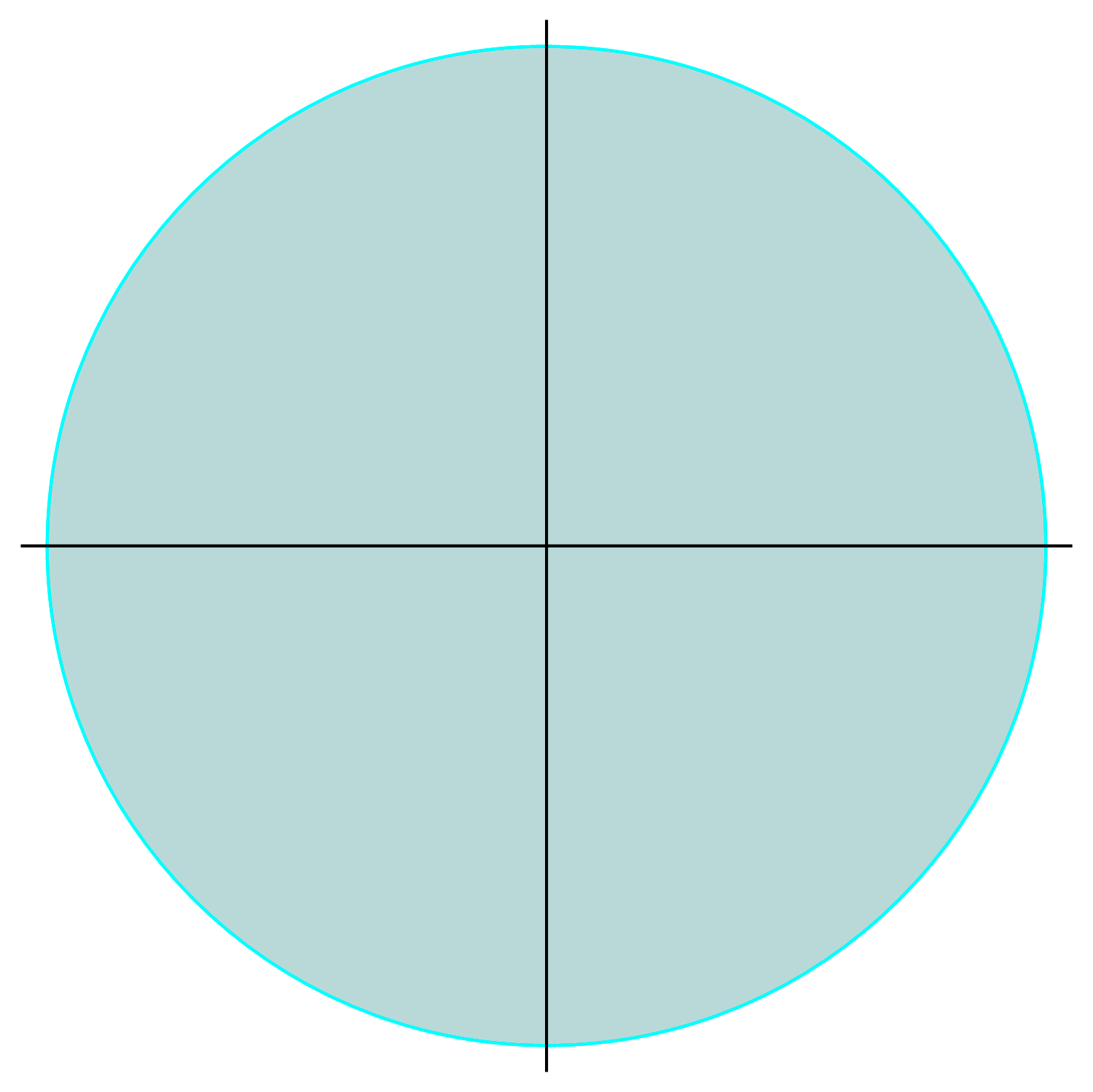} \\
\begin{picture}(0,0)
\put(-220,220){(a)}
\put(40,220){(b)}
\put(-140,224){$z$}
\put(-128,224){$\ket{e}$}
\put(-128,23){$\ket{g}$}
\put(-59,207){$y$}
\put(-23,89){$x$}
\put(222,126){$x$}
\put(229,120.5){{\color{deeppink} $\blacktriangleright$}}
\put(235,115){$1$}
\put(134,218){$z$}
\put(126,223){$\color{coquelicot} \blacktriangle$}
\put(23,120.5){\color{crimsonglory} $\blacktriangleleft$}
\put(13,115){$-1$}
\put(120,227){$1$}
\put(126,17){{\color{electricviolet} $\blacktriangledown$}}
\put(116,10){$-1$}
\put(39,173){\color{blue} $\bullet$}
\put(159,23){\color{red} $\bullet$}
\put(111,83){\color{patriarch} $\bullet$}
\put(49,172){\color{blue} $\ket{\psi}\bra{\psi}$}
\put(179,22){\color{red} $\ket{\Upsilon}\bra{\Upsilon}$}
\put(80,133){\color{red} $\wp_\Upsilon$}
\put(136,35){\color{blue} $\wp_\psi$}
\put(132,80){{\color{patriarch} $\rho_m = $}{\color{blue} $~\wp_\psi\ket{\psi}\bra{\psi}$}}
\put(152,68){$+${\color{red}$\wp_\Upsilon \ket{\Upsilon}\bra{\Upsilon}$}}
\put(136,145){$\vartheta$}
\end{picture} \\ \vspace{-10pt} 
\begin{tikzpicture}[overlay]
\draw[line width = 1pt, red, ->] (1.54,6.10) -- (3.92,3.08);
\draw[line width = 1pt, blue, ->] (5.59,1.0) -- (4.05,2.94);
\draw[line width = 0.5pt] (4.55,4.33) -- (6.67,7.25);
\draw[line width = 0.5pt,->] (4.55,5) arc(90:55:0.67);
\end{tikzpicture}
\caption{
We show the Bloch sphere (a), with $\ket{e}$ at the ``north pole'' ($z=1$), and $\ket{g}$ at the ``south pole'' ($z = -1$). Any pure qubit state can be represented by a point on the sphere's surface, while mixed states (i.e.~weighted averages of two or more pure--state density operators) live inside the sphere's surface. In (b), we highlight some of these features in more detail, focusing on the $xz$--plane of the Bloch sphere. Some special states $\ket{e} = {\color{coquelicot} \blacktriangle}$, $\ket{g} = {\color{electricviolet} \blacktriangledown}$, $\tfrac{1}{\sqrt{2}}(\ket{e}-\ket{g}) = {\color{crimsonglory} \blacktriangleleft}$, and $\tfrac{1}{\sqrt{2}}(\ket{e}+\ket{g}) = {\color{deeppink} \blacktriangleright}$ are marked. A mixed state {\color{patriarch}$\rho_m$} is shown as the weighted sum of two pure state density operators {\color{blue}$\rho_\psi$} and {\color{red}$\rho_\Upsilon$}(where the vectors drawn are proportional in length to the probability of drawing their respective pure state from an ensemble). We also illustrate the coordinate $\vartheta$ used later in the text.
}\label{fig-bloch}
\end{figure*}
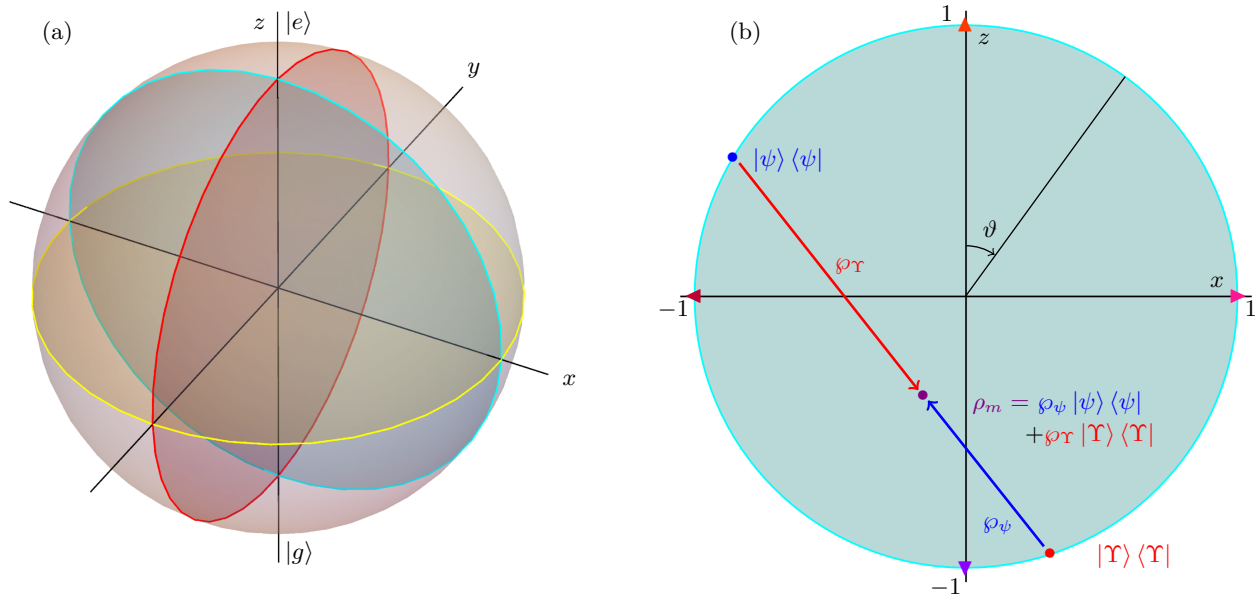

\par A qubit is any two--level quantum system; mathematically--speaking this means that it is described like a spin--$\tfrac{1}{2}$. Physically speaking, a qubit might be any of e.g.~a particular transition in an atom or ion, a spin in a quantum dot or diamond nitrogen--vacancy center, or the lowest--two levels of the superconducting ``artificial atoms'' now used in many experiments. 
The state of a qubit can be represented as living in the Bloch sphere; we will generically parameterize our single--qubit density matrix with Bloch coordinates according to 
\be \label{qubit-rho}
\rho = \frac{1}{2}\left(\begin{array}{cc} 1+z & x-iy \\ x+iy & 1-z \end{array} \right) = \tfrac{1}{2}\left(\openone + x\hat{\sigma}_x + y\hat{\sigma}_y + z\hat{\sigma}_z \right)
\ee
throughout the forthcoming derivations, where $(1+z)/2 = \rho_{ee}$ denotes the excited state population. See Fig.~\ref{fig-bloch}.
It is also necessary that we introduce a distinction between pure states, which can be represented by a state vector $\ket{\psi}$, and mixed states which require the use of $\rho$.
Pure qubit states $\rho = \ket{\psi}\bra{\psi}$ live on the outer surface of the unit sphere, while more general states $\rho = \sum_i \wp_i \ket{\psi_i}\bra{\psi_i}$ may live inside the sphere; a ``mixed'' state with more than one non--zero $\wp_i$ may be used to describe a system which is imperfectly isolated from its surrounding environment, where the $\wp_i$ are effectively probabilities. 
This description implies that we have a 'classical' statistical mixture, in which we have a probability $\wp_i$ of finding the pure state $\ket{\psi_i}$ in an ensemble; in contrast with a coherent pure state superposition, elements in such a mixture do not interfere with each other.
For example, we may consider states which lead to a probability $\tfrac{1}{2}$ for a $\sigma_z$--measurement to return $\ket{e}$ or $\ket{g}$, such as $\ket{x+} = \tfrac{1}{\sqrt{2}}(\ket{e}+\ket{g})$.  The density matrix for this pure state (which contains the possibility for quantum interference) is \emph{not} the same as the classical statistical mixture of $\ket{e}$ and $\ket{g}$ (where the off--diagonal ``coherences'' are suppressed), i.e.
\be \begin{split}
\ket{x+}\bra{x+} &= \frac{1}{2}\left(\begin{array}{cc} 1 & 1 \\ 1 & 1 \end{array} \right)\\ &\neq \frac{1}{2}\left( \begin{array}{cc} 1 & 0 \\ 0 & 1 \end{array}\right) = \tfrac{1}{2} \ket{e}\bra{e} + \tfrac{1}{2}\ket{g}\bra{g}.
\end{split} \ee
By using the density matrix to describe our state, we may account for all of these options, which is necessary when we have an open system and the possibility of information loss.

\par Note that $\rho$ is Hermitian ($\rho = \rho^\dag$), and normalization requires that $\text{tr}\left( \rho \right) = 1$. We will often represent a qubit's state and dynamics in terms of the Bloch vector $\mathbf{q} = \lbrace x,y,z \rbrace$ below; such coordinates should be understood in the context of \eqref{qubit-rho}. We suppress the ``hat'' notation on our density operators $\rho$.

\subsection{Standard quantum--mechanical treatment \label{sec-QMdecay}}
We summarize the typical approach, originally by Weisskopf and Wigner, as a point of departure in describing spontaneous emission. A more complete derivation of the results we summarize can be found in Ref.~\cite{Jacobs2006}, and we will stick closely to their conventions for clarity. The Hamiltonian describing the joint system including the qubit and a single mode of the electromagnetic field is of the form
\be \begin{split}
\frac{\hat{h}}{\hbar} =& \underbrace{\omega_\text{qb} \hat{\sigma}^+ \hat{\sigma}_-}_{\text{qubit}} + \underbrace{\omega_{f} \left(\hat{a}^\dag \hat{a} + \tfrac{1}{2} \right)}_\text{single field mode} +  \underbrace{\left( g \hat{\sigma}^+ \hat{a} + g^\ast \hat{\sigma}_- \hat{a}^\dag \right)}_{\text{interaction } \hat{h}_{int}},
\end{split} \ee
where $\hbar\omega_\text{qb}$ is the energy separation between the two qubit levels of interest, $\omega_{f}$ denotes the frequency of the field mode, and $g$ is a coupling constant between the field and qubit. The first two terms represent the qubit and field, respectively, while the third term describes their interaction. One can regard this model as corresponding to a two level system and quantum harmonic oscillator which are able to exchange excitations. 
We have raising and lowering operators on the qubit $\hat{\sigma}^+ = \ket{e}\bra{g}$ and $\hat{\sigma}_- = \ket{g} \bra{e}$, and on a field mode $\hat{a}\ket{n} = \sqrt{n}\ket{n-1}$ and $\hat{a}^\dag \ket{n} = \sqrt{n+1}\ket{n+1}$ for Fock states $\ket{n}$. The interaction term $\hbar g \hat{\sigma}^+ \hat{a}$ describes the possibility for the atom to become excited by absorbing a photon (which is removed from the field); the adjoint of this term denotes the reverse process, in which the qubit loses energy, emitting a photon which is added to the field mode. 
We may use the Schr\"{o}dinger equation to compute the evolution of the joint system and subsequent qubit state amplitudes under the influence of this Hamiltonian.
The electro--magnetic environment of the qubit contains many modes, and the apparently incoherent evolution of the qubit associated with spontaneous emission emerges when summing up the action of the couplings to all these modes, each described like $\hat{h}_{int}$ above. 
Such analysis requires that we average the evolution predicted by the Schr\"odinger equation over the available density of free--space field modes and sum over polarizations \cite{Jacobs2006}. 
To good approximation (i.e.~the approximations first made by Weisskopf and Wigner), we may simplify the dynamics at timescales much longer than the field periods, eliminating environmental modes from the description. For a generic qubit state $\zeta \ket{e} + \phi \ket{g}$, we obtain the evolution of the excited state amplitude
\be \label{zetadot}
\dot{\zeta} = \left(-i \omega_\text{qb} - \tfrac{\gamma}{2} \right) \zeta.
\ee
We have introduced the spontaneous emission rate $\gamma = T_1^{-1}$ which is equal to the density of modes coupled resonantly to the qubit via $\gamma = \sum_j \vert g_j\vert^2 \delta(\omega_\text{qb}-\omega_j)$. Here, $g_j$ stands for the strength of the coupling to the $j^\text{th}$ mode of the electromagnetic reservoir. The contributions of all the non--resonant modes oscillate and quickly average to zero over a typical time $\tau_\text{corr} \ll \gamma^{-1}$; this condition is key in allowing us to obtain a Markovian description of the qubit evolution \eqref{zetadot}, from which the reservoir dynamics have been completely eliminated. A precise discussion of the approximations leading to Eq.~\eqref{zetadot} and the order of the associated errors can be found in Ref.~\cite{Cohen-Tannoudji98}, Chapter 4.

\par 
The excited state population is described by the density matrix element 
\be \label{rotframe}
\rho_{ee} = \zeta^\ast \zeta = (z+1)/2. 
\ee
It is straightforward to compute that \eqref{zetadot} and \eqref{rotframe} imply
\be 
\dot{\rho}_{ee} = - \gamma \rho_{ee} \quad\leftrightarrow\quad \dot{z} = - \gamma \left(z+1\right).\label{rhoeedot}
\ee
The result that spontaneous emission leads to exponential decay of the excited state population at rate $\gamma$ (or with characteristic time $T_1 = \gamma^{-1}$), absent other dynamics, is among the most fundamental phenomena in the quantum optics literature. 

\begin{figure}
    \centering
    \includegraphics[width=\columnwidth]{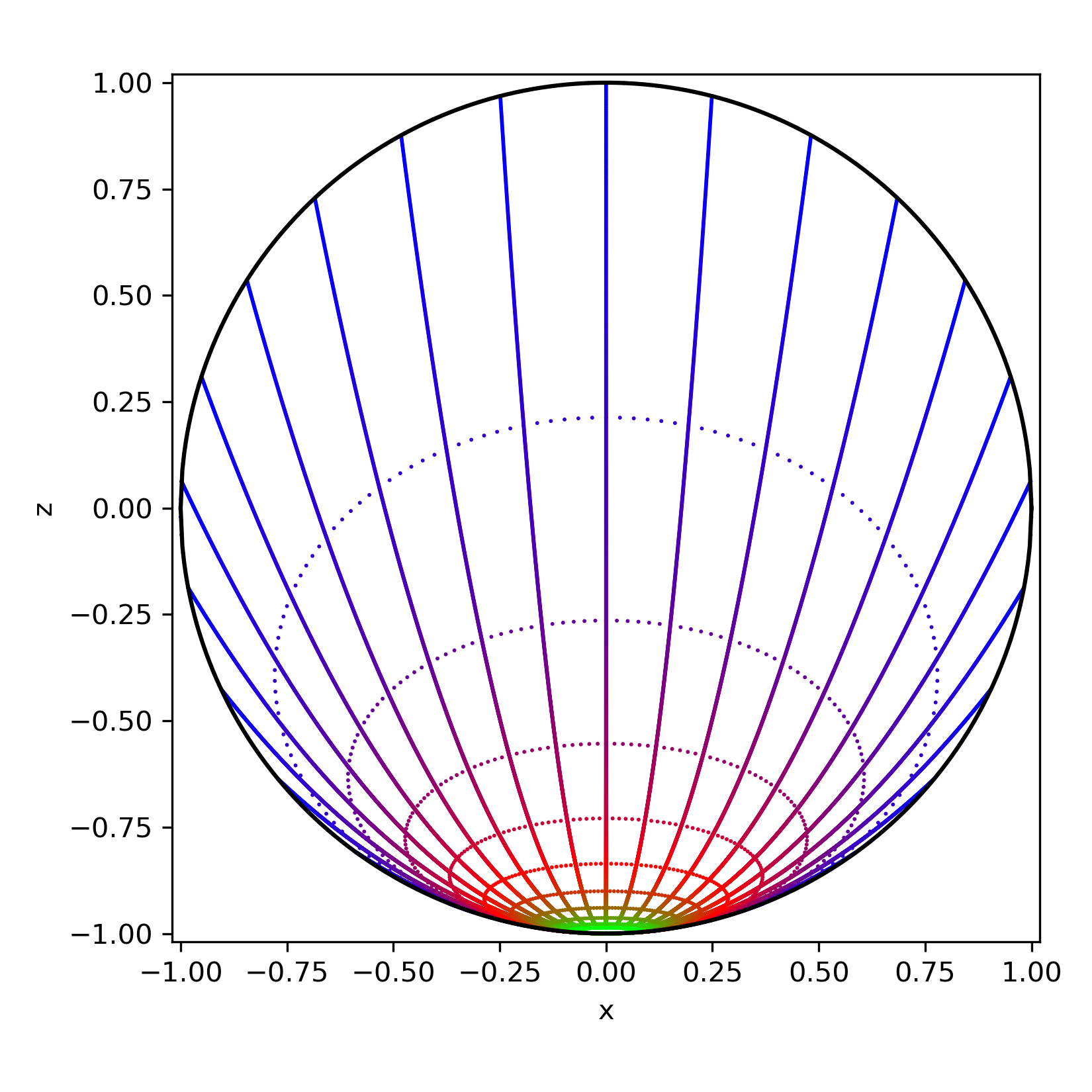}\\
    \includegraphics[width = .8\columnwidth]{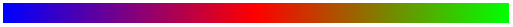} \\
    \begin{picture}(0,0)
    \put(-108,0){$t=0$}
    \put(88,0){$t=5T_1$}
    \end{picture}
    \caption{We plot the evolution of the qubit state under the unmonitored fluorescence dynamics \eqref{unmonitored-Bloch-eqs} in the $xz$--plane of the Bloch sphere, originating from a variety of initial pure states. The excited state is at the top of the sphere, and all paths converge towards the ground state at the bottom. Color denotes the time evolution along each path. The trajectories in the qubit state tend to become impure / mixed, because no information is collected about fluorescence output; we have an open system with lost information in this case. These dynamics are often represented with an equivalent picture in which the Bloch ball contracts into an ellipsoid near the ground state under the influence of a decay channel \cite{BookNielsen}; we sample a greater number of initial states at intervals of $T_1/2$ to illustrate this.}
    \label{fig-unmon}
\end{figure}
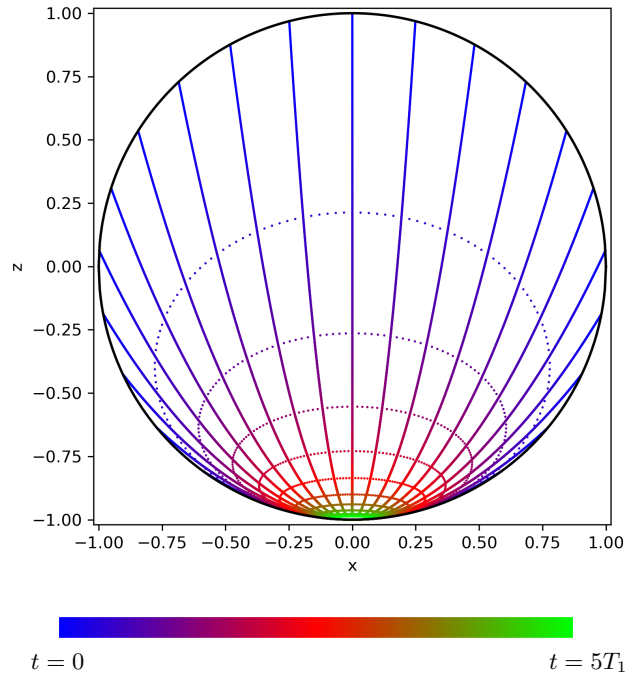

\subsection{Master equation treatment \label{sec-OpenSysME}}


In order to express the complete information about the qubit state at any time in a compact way, and straightforwardly generalize our system (e.g.~we might consider coherently driving qubit), it is convenient to formulate the dynamics of spontaneous emission as a master equation for the density operator $\rho$ introduced above.
The evolution of an open quantum system in contact with a Markovian environment (i.e.~with an environment of very short correlation time $\tau_\text{corr}$ with respect to the other time scales in the problem) can, in general, be written as a Lindblad equation; such a master equation is of the form \cite{HarocheBook}
\be \label{masterequation}
\dot{\rho} = \underbrace{\frac{i}{\hbar} [\rho,\hat{H}]}_\text{Unitary Evol.} + \underbrace{\sum_c \left( \hat{L}_c \rho \hat{L}_c^\dag - \tfrac{1}{2} \hat{L}_c^\dag \hat{L}_c \rho - \tfrac{1}{2} \rho \hat{L}_c^\dag \hat{L}_c \right)}_\text{Lindblad Dissipation},
\ee
where $\rho$ is the density matrix of the system of primary interest (in this case, the qubit), and each operator $\hat{L}_c$ describes a coupling between the system and its environment (in this case, the decay channel).
We see a term describing the unitary evolution $\dot{\rho} = \tfrac{i}{\hbar}[\rho,\hat{H}]$, plus the Lindblad term which accounts for information leaking into the environment through any channels to which the system is open.

The arguments of the previous section can be used to show that the case of spontaneous emission corresponds to a single channel characterized by operator $\hat{L} = \sqrt{\gamma} \hat{\sigma}_-$ \cite{Cohen-Tannoudji98}, which indicates that the qubit may lose its excitation with an effective coupling rate $\gamma$. The master equation capturing spontaneous emission of a qubit is then
\be \label{unmonitored-meq-fixed}
\dot{\rho} = \frac{i}{\hbar}[\rho,\hat H]+\gamma \hat{\sigma}_- \rho \hat{\sigma}^+  - \tfrac{\gamma}{2}\left( \hat{\sigma}^+\hat{\sigma}_- \rho + \rho \hat{\sigma}^+\hat{\sigma}_- \right),
\ee
where the qubit Hamiltonian is $\hat H = \hbar\omega_\text{qb}\hat{\sigma}_z/2$. The unitary part solely induces a rotation at frequency $\omega_\text{qb}$ of the qubit state around the $z$-axis of the Bloch sphere, and it is convenient to work in a frame in which this rotation is suppressed. This rotating frame is formally the interaction picture with respect to $\hat H$, associated with the transformation $\rho \to e^{i\hat H t/\hbar}\rho e^{-i\hat H t/\hbar}$. In the following, we will always work in such frame, where the master equation reads
\be \label{unmonitored-meq}
\dot{\rho} = \gamma \hat{\sigma}_- \rho \hat{\sigma}^+  - \tfrac{\gamma}{2}\left( \hat{\sigma}^+\hat{\sigma}_- \rho + \rho \hat{\sigma}^+\hat{\sigma}_- \right).
\ee
We can get equations of motion in the Bloch coordinates (in the rotating frame) by computing $\dot{\mathbf{q}} = \text{tr}(\hat{\sigma}_\mathbf{q} \dot{\rho})$, yielding
\be \label{unmonitored-Bloch-eqs}
\dot{x} = - \frac{\gamma}{2} x, \quad \dot{y} = - \frac{\gamma}{2} y, \quad \dot{z} = -\gamma(1+z),
\ee
in perfect agreement with the treatment above \eqref{rhoeedot}. The decaying solutions of these equations, initialized from different pure states on the edge of the Bloch sphere, are illustrated in Fig.~\ref{fig-unmon}.

For many application, the qubit needs to be driven; to describe such a situation, we must modify the qubit Hamiltonian $\hat H$, adding time--dependent terms. In general, the derivation of the master equation describing the dynamics of the qubit's density operator needs to be carefully redone in presence of this new Hamiltonian, and one may find that the Lindblad term is modified due to the presence of the drive \cite{BookBreuer}. 
Such a modification typically occurs, for example, when the drive causes the qubit to become sensitive to modes of the environment at different frequencies, which are sufficiently separated from each other so as to have a different density of states. 
For instance, in the case of a quasi--resonant monochromatic drive inducing Rabi oscillations, the emission spectrum of the qubit contains multiple peaks (the famous Mollow triplet \cite{Mollow1969}) separated by the Rabi frequency (which is related to the intensity of the driving). 
If the environmental density of state varies around the qubit frequency $\omega_\text{qb}$ on the Rabi frequency scale, the form of the damping is drastically modified. In Ref.~\cite{Murch12Cavity} this effect was exploited to stabilize an arbitrary state of the Bloch sphere. 
However, provided the drives are weak enough (Rabi frequencies much smaller than the qubit frequency and the inverse correlation time of the reservoir $\tau_\text{corr}$), and there is no cavity or other resonance close to the driven qubit's emission spectrum peaks to cause especially fast variations in the environment spectrum \cite{BookCarmichaelStat1}, this effect is negligible. 
Within these conditions, the action of a drive can be simply captured by adding a unitary term $(i/\hbar)[\rho,\hat H_\text{dr}(t)]$ in Eq.~\eqref{unmonitored-meq-fixed}; the treatment we develop below assumes this simplest case. While some modification to this simplest scheme may be necessary in adapting it to situations beyond the stated constraints, experiments not explicitly focused on engineering more exotic effects will typically obey these simplifying constraints by default; this simplest scheme we lay out below is thus widely applicable.

\section{Quantum Trajectories \label{sec-transition}}

\par The treatment of spontaneous emission in the previous section, and in particular the master equation Eq.~\eqref{unmonitored-meq}, captures the dynamics of the qubit under the assumption that any information emitted by the qubit (leaking into the environment) during the qubit-field mode interaction is lost forever. We are, however, primarily interested in the case where we, the observer(s), recover some (or ideally all) of this information through measurement(s) on the field mode. 
In this section, we present the formalism of Kraus operators, which describes the update of the qubit's state conditioned on acquiring such information. 

\subsection{Kraus Operator Formalism \label{sec-KrausIntro}}

The basic idea is that there exist a set of Kraus operators $\hat{M}_r$, which describe how the state of our system should be updated, each of them conditioned on acquiring one of the possible measurement outcomes $r$ in the environment during a measurement of duration $dt$, according to \cite{BookNielsen, BookWiseman, HarocheBook}
\be \label{state-update}
\rho(t+dt) = \frac{\hat{M}_r \rho(t) \hat{M}_r^\dag}{\text{tr}\left( \hat{M}_r \rho(t) \hat{M}_r^\dag \right)}.
\ee
We require that either $\sum_r \hat{M}_r^\dag \hat{M}_r = \openone$, or $\int dr \hat{M}_r^\dag \hat{M}_r = \openone$, depending on whether the possible measurement outcomes $r$ are discrete or continuous; such a condition tells us that we have a valid (completely positive) transformation on $\rho$, and insures that we have considered a complete, self--consistent set of measurement outcomes. 
As dictated by the axioms of quantum mechanics, the outcome $r$ is obtained randomly among its possible values based on Born's rule, which here yields probabilities (or a probability density) $\wp(r|\rho) = \text{tr}\left( \hat{M}_r \rho \hat{M}_r^\dag \right)$. Note that the denominator of \eqref{state-update}, which serves to ensure the updated density matrix is properly normalized, exactly matches this probability. If measurement(s) on the environment is(are) repeated (in our case every $dt$), the successive outcomes and subsequent state updates define a stochastic sequence of states called a quantum trajectory.


Following Ref.~\cite{Jordan2015flor}, we construct the particular $\hat{M}$ of interest for the case of a spontaneously--emitting qubit, using a Bayesian probability argument; it is useful to consider a pure state of the qubit and an effective field mode it emits into (initially in vacuum state $\ket{0}$)
\be \label{pure-initstate}
\ket{\psi_0} = \left(\zeta \ket{e} + \phi \ket{g} \right) \otimes \ket{0},
\ee
where $z = 2|\zeta|^2-1$, and $\ket{e}$ and $\ket{g}$ are the excited and ground states of the qubit, respectively. There is a probability $\wp(e) = |\zeta|^2$ to find the qubit in $\ket{e}$ and probability $\wp(g) = |\phi|^2$ to find the qubit in $\ket{g}$, with $\wp(e) + \wp(g) = 1$. On phenomenological grounds, we suppose that the probability for an emission event in a time interval $dt$ is given by $\wp(1|e) = \epsilon = \gamma \: dt$, where $\gamma$ is some characteristic rate at which the qubit fluoresces (i.e.~$\gamma = 1/T_1$ is a measurable quantity for a qubit--cavity system). Then $\wp(1|e)\wp(e) = \epsilon |\zeta|^2 = \wp(e|1)\wp(1)$ and/or $\wp(0|e)\wp(e) = (1-\epsilon)|\zeta|^2 = \wp(e|0)\wp(0)$ according to Bayes' theorem. A quantum--coherent state assignment after the short interval $dt$ which reflects these probabilistic considerations is
\be \label{pure-firststepstate}
\ket{\psi_1} = \sqrt{1-\epsilon} \zeta \ket{e,0} + \phi \ket{g,0} + \sqrt{\epsilon} \zeta \ket{g,1}. 
\ee
In other words, there is some probability for an emission event which involves a photon being created in the output mode ($0\rightarrow 1$), and which shifts qubit population from $\ket{e}\rightarrow\ket{g}$, reflecting a common sense understanding of spontaneous emission. 
Below we will always assume that the measurement time (in practice, a detector integration time) is much faster than the characteristic decay time of the qubit, i.e.~we have $dt \ll T_1$, or $\epsilon \ll 1$. 
This is a key condition which will ensure that the quadrature measurements we will eventually consider are weak measurements, and that the subsequent quantum trajectories are diffusive. 
We also assume that the information we acquire applies to the qubit in real time, which implies that that photon travel time between the qubit and measurement apparatus should be negligible.
We may rewrite the change of state from above as
\be \label{pure-stateup-bayes}
\ket{\psi_1} = \left( \begin{array}{cc}
\sqrt{1-\epsilon} & 0 \\ \sqrt{\epsilon} \hat{a}^\dag & 1
\end{array} \right) \ket{\psi_0} \text{ for } \ket{\psi_0} = \left( \begin{array}{c} \zeta \\ \phi \end{array} \right) \otimes \ket{0},
\ee
where $a^\dag$ creates a photon in the relevant cavity/field output ($a^\dag \ket{0} = \ket{1}$). The Kraus operators $\hat{M}_\text{r}$ in \eqref{state-update} act only on the qubit state, and are obtained by projecting out the field mode in a final state corresponding to some outcome from measuring the field, i.e.
\be \label{gen-kraus}
\hat{M}_r = \bra{\psi_r}  \left( \begin{array}{cc}
\sqrt{1-\epsilon} & 0 \\ \sqrt{\epsilon} \hat{a}^\dag & 1
\end{array} \right) \ket{0},
\ee
where $\{\ket{\psi_r}\}$ could be any basis of states of the field mode, which should be chosen based on the kind of measurement being performed and result $r$. All of the examples we consider below rely on a Kraus operator of the form \eqref{gen-kraus}. Much of what we do below will revolve around relating different measurements to the appropriate choice of $\ket{\psi_r}$, and then exploring the ramifications that choice has on the measurement backaction and quantum trajectories.

\subsection{Photodetection and quantum jump trajectories}

As a first example, suppose that we choose our $\ket{\psi_r}$ in the Fock basis, i.e.~we consider outcomes of the type $\ket{1}$ (a photon exits in the field mode in the given timestep), or $\ket{0}$ (no photon exits), which correspond to making a photodetection measurement. In other words, we imagine counting the photons emitted by the qubit into the field mode, in a time--resolved manner, with a detector integration time $dt\ll \gamma^{-1}$ (equivalently, $\epsilon \ll 1$).
 
We may define Kraus operators $\hat{M}_1$ ($\hat{M}_0$) for the single--qubit state update conditioned on a click (no--click) in the detector, according to 
\be 
\hat{M}_0 = \bra{0} \left( \begin{array}{cc} 
\sqrt{1 - \epsilon} & 0 \\ \sqrt{\epsilon} \hat{a}^\dag & 1
\end{array} \right) \ket{0} = \left(\begin{array}{cc} \sqrt{1-\epsilon} & 0 \\ 0 & 1 \end{array} \right),
\ee \be 
\hat{M}_1 = \bra{1} \left( \begin{array}{cc} 
\sqrt{1 - \epsilon} & 0 \\ \sqrt{\epsilon} \hat{a}^\dag & 1
\end{array} \right) \ket{0} = \left( \begin{array}{cc} 0 & 0 \\ \sqrt{\epsilon} & 0 \end{array} \right). 
\ee
It is easy to verify that $\hat{M}_0^\dag \hat{M}_0 + \hat{M}_1^\dag \hat{M}_1 = \openone$, such that these measurement operators form a positive operator valued measure (POVM) \cite{BookNielsen}. 
We can say that under continuous photodetection, the qubit state is updated every $dt$ by
\be 
\rho(t+dt) = \frac{\hat{M}_1 \rho(t) \hat{M}_1^\dag}{\text{tr}\left( \hat{M}_1 \rho(t) \hat{M}_1^\dag \right)},
\ee 
if the detector registers that a photon emerged between $t$ and $t+dt$, or according to
\be 
\rho(t+dt) = \frac{\hat{M}_0 \rho(t) \hat{M}_0^\dag}{\text{tr}\left( \hat{M}_0 \rho(t) \hat{M}_0^\dag \right)}
\ee
if no photon reaches the detector. 
The probability of a click in any given timestep is $\wp_1 = \text{tr}( \hat{M}_1 \rho(t) \hat{M}_1^\dag) = \gamma \: dt \: (1+z)/2$, and the probability of no--click is $\wp_0 = \text{tr}( \hat{M}_0 \rho(t) \hat{M}_0^\dag)$, with $\wp_0 + \wp_1 = 1$. These expressions reflect the common--sense result that $\wp_1$ must vanish when the qubit is in the ground state, i.e.~$\wp_1 = 0$ for $z = -1$. 
Thus, a single quantum trajectory for this photodetection scenario is characterized by a time series of outcomes $r\in\{0,1\}$. 
Simulation of such a trajectory can be performed by drawing a click/no--click readout from a binomial distribution at each short timestep of duration $dt \ll T_1$, and subsequently updating the qubit state $\rho$ according to the appropriate rule above. 
Results of such a simulation are shown in Fig.~\ref{fig-simul3meas}(a). 
The trajectories generated by photodetection are an example of ``quantum jump'' trajectories, for which the qubit state immediately jumps to $\ket{g}$ when a click event occurs (this is related to the discrete nature of the possible outcomes $r$).

Before moving on to different types of measurements on the output mode, we bridge the gap between our Kraus operator description and the un-monitored decay channel we discussed in the previous section. The situation in which the outcome of the measurement performed on the field mode between $t$ and $t+dt$ is actually unavailable can be captured by averaging the state update over both outcomes, i.e.
\be \label{no-measurement}
\rho(t+dt) = \frac{\hat{M}_0 \rho(t) \hat{M}_0^\dag + \hat{M}_1 \rho(t) \hat{M}_1^\dag}{\text{tr}\left( \hat{M}_0 \rho(t) \hat{M}_0^\dag + \hat{M}_1 \rho(t) \hat{M}_1^\dag\right)}.
\ee
An equation of motion can be obtained by taking
\be 
\dot{\rho} \approx \frac{\rho(t+dt) - \rho(t)}{dt},
\ee
where the numerator on the RHS is expanded to $O(dt)$. It is then straightforward to verify that the equations \eqref{unmonitored-Bloch-eqs} reappear exactly, i.e.~the procedure just described to obtain $\dot{\rho}$ leads to exactly the same expression as the master equation as described above, and as shown in Fig.~\ref{fig-unmon}. 
A similar procedure allows to show that for any measurement basis $\ket{\psi_r}$ chosen for the field, the master equation is recovered when averaging over all of the outcomes we could have obtained from measurement; we will soon be able to elaborate further on this point.

\subsection{Diffusive trajectories and stochastic master equation}

In the remainder of this article, we are concerned about measurements on the environment leading to a continuous--valued outcome $r$, e.g.~a voltage or current from a detector, leading to ``diffusive'' trajectories (in contrast with the ``jump'' trajectories we have just discussed). 
The specifics of the two most common examples, heterodyne and homodyne measurements, are presented in detail in the following section. 
Because the evolution during $dt$ is infinitesimal, it is common to write the change in the density operator of the qubit, conditionned on the outcome $r$ obtained at time $t$, under the form of a stochastic master equation (SME); the SME can be seen as an extension of Eq.~\eqref{masterequation}, in which we add a term which accounts for the measurement outcome\footnote{Photodetection, as considered above, constitutes a particular ``unraveling'' of the master equation into stochastic trajectories; the heterodyne and homodyne measurements we subsequently consider are additional possible ``unravelings''.}.
The SME may generically be obtained by expanding an expression of the form \eqref{state-update} to $O(dt)$ \cite{Brun2001Teach, Jacobs2006} (detailed examples of this process follow below). 
The addition of a stochastic element into a differential equation is not trivial, because a genuinely stochastic element is not really differentiable, the way a smooth and well--behaved function is. 

\par Generically, what we will momentarily consider is a type of Langevin equation, or first--order stochastic differential equation of the form
\be 
\dot{q} = \mathsf{a}(q) + \mathsf{b}(q)\xi(t);
\ee 
the term $\mathsf{a}$ is often called the drift term, whereas $\mathsf{b}$ functions as a diffusion constant, and together with the randomly--varying $\xi(t)$, gives stochastic evolution.
Equations of this type were first written down to model Brownian motion of small particles \cite{FeynmanBrownian}, where complex mechanical forces lead to effectively random kicks in a particle's position. In our present case, we care about the evolution of a quantum state, and the stochasticity denoted by $\xi(t)$ is a result of the randomness inherent in the quantum measurement process. 
The particular type of random evolution we consider is delta--correlated Gaussian white noise, obeying $\xi(t) = dW(t)/dt$, where $W(t)$ is called a Wiener process. The Wiener increment $dW(t) = W(t+dt)-W(t)$ is a Gaussian random variable, independent on any past values $dW(s)$ for $s<t$ and characterized by a mean of zero and variance equal to $dt$. These properties lead to a noise term $\xi(t)$ of zero expectation value $\langle\langle \xi(t)\rangle\rangle= 0$ and co-variance obeying $\langle\langle\xi(t)\xi(t')\rangle\rangle = \delta(t-t')$, where the double bracket indicates the ensemble average over realizations of the process.
This is suitable for describing the quantum noise arising from measurement in a variety of physical situations, including those we consider below\footnote{Strictly speaking, writing $\xi(t) = dW(t)/dt$ is an odd mathematical statement, because $W(t)$ is pure noise and non--differentiable. In practice such substitutions does not cause us a problem in writing down sensible stochastic calculus however. For details, refer e.g.~to the books by Gardiner \cite{BookGardiner, BookGardiner2}, or other references on stochastic differential equations, such as \cite{BookKloedenPlaten}.}. 
Some physical justification for the appropriateness of the use of a Gaussian $\xi$ for the examples below is provided in the following sections, and in Appendix~\ref{app-Noise}.
For $\mathsf{a} = 0$ and constant $\mathsf{b}$ (simple diffusion without drift), the variance of an ensemble of diffusing trajectories scales like time; this is summarized by the It\^{o} stochastic calculus rule $dW(t)^2=dt$ (or equivalently $\xi(t)^2 = 1/dt$).

The general form of the SME that we use for diffusive quantum trajectories, in units $\hbar\rightarrow 1$, reads \cite{Jacobs2006, BookWiseman}
\be \label{SME}
d\rho = i[\rho,\hat{H}] dt + \sum_c \left( \hat{\mathcal{L}}[\rho,\hat{L}_c] dt + \sqrt{\eta_c} \hat{\mathcal{M}}[\rho,\hat{L}_c] dW_c \right).
\ee
The super--operators are the Lindblad dissipation term, from \eqref{masterequation},
\be 
\hat{\mathcal{L}}[\rho,\hat{L}_c] \equiv \hat{L}_c \rho \hat{L}_c^\dag - \tfrac{1}{2}\left(\hat{L}_c^\dag \hat{L}_c \rho + \rho \hat{L}_c^\dag \hat{L}_c \right),
\ee
and the newly--added measurement backaction term
\be 
\hat{\mathcal{M}}[\rho,\hat{L}_c] \equiv \hat{L}_c \rho + \rho \hat{L}_c^\dag - \rho \: \text{tr}\left(\hat{L}_c \rho + \rho \hat{L}_c^\dag \right).
\ee
As before, a Hamiltonian $\hat{H}$ may describe any unitary processes applied to the system (e.g.~Rabi drive on a qubit). Each of the operators $\hat{L}_c$ describes a particular measurement channel, which is monitored with efficiency $\eta_c \in [0,1]$ (where $1$ denotes perfect measurement efficiency, and $\eta_c$ is dimensionless). 
The measurement record associated with any monitored channel, contains one outcome every $dt$, going like $r_c \propto \sqrt{\eta_c} \langle \hat{L}_c + \hat{L}_c^\dag \rangle + \xi_c$, where the brackets denote the expectation value in state $\rho$. 
Such an expression for the readout is easy to interpret as a signal $\langle \hat{L}_c + \hat{L}_c^\dag \rangle$, attenuated due to inefficiency by a factor $\sqrt{\eta_c}$, plus quantum noise $\xi_c$ intrinsic to the measurement process. A more detailed introductory guide to SME can be found in Ref.~\cite{Jacobs2006}.

\par Channels which are open to the environment, but un-monitored (e.g.~typical dephasing mechanisms, or the decay channel in the un-monitored case), can be modeled by placing an operator in the sum over $c$ which is monitored with efficiency $\eta_c = 0$. 
The master equation \eqref{masterequation} can be recovered from the SME by taking an ensemble average over stochastic trajectories 
Applying these concepts to the example of a single decay channel introduced above, we see that 1) opening the qubit to an unmonitored decay channel $\hat{L} = \sqrt{\gamma} \hat{\sigma}_-$, 2) measuring the qubit fluorescence/decay according to $\hat{L} = \sqrt{\gamma}\hat{\sigma}_-$ with efficiency zero, or 3) the average dynamics over an ensemble of stochastic trajectories obtained by continuously monitoring the qubit fluorescence as per $\hat{L} = \sqrt{\gamma}\hat{\sigma}_-$, are all equivalent situations. 
This view from the master equation is also entirely equivalent to that for the Kraus operators, as presented in and around \eqref{no-measurement}. 
In Fig.~\ref{fig-simul3meas} we observe this in simulations, observing that the average over many quantum trajectories reproduces the dynamics of the unmonitored case \eqref{unmonitored-Bloch-eqs}, regardless of the character of the individual measurements. 

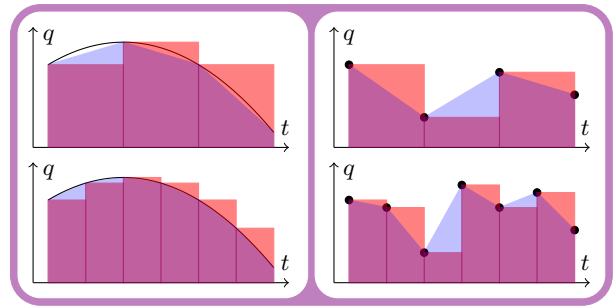
\begin{figure}
    \centering
    \begin{tikzpicture}
    \draw[fill = patriarch!50, draw = patriarch!50, rounded corners = 0.4cm] (0,0) rectangle (8,4);
    \draw[fill = white, draw = white,rounded corners = 0.3cm] (0.1,0.1) rectangle (3.95,3.9);
    \draw[fill = white, draw = white,rounded corners = 0.3cm] (4.05,0.1) rectangle (7.9,3.9);
    \draw[->] (0.3,0.3) -- (3.7,0.3);
    \draw[->] (0.3,2.1) -- (3.7,2.1);
    \draw[->] (0.3,0.3) -- (0.3,1.9);
    \draw[->] (0.3,2.1) -- (0.3,3.7);
    \draw[->] (4.3,0.3) -- (7.7,0.3);
    \draw[->] (4.3,2.1) -- (7.7,2.1);
    \draw[->] (4.3,0.3) -- (4.3,1.9);
    \draw[->] (4.3,2.1) -- (4.3,3.7);
    \draw[smooth,samples = 100,domain = 0.5:3.5,variable=\x] plot({\x},{3.5-0.3*(\x-1.5)^2});
    \draw[smooth,samples = 100,domain = 0.5:3.5,variable=\x] plot({\x},{1.7-0.3*(\x-1.5)^2});
    \draw[draw = red, fill = red, opacity = 0.5] (0.5,2.1) rectangle (1.5,3.2);
    \draw[draw = blue, fill = blue, opacity = 0.25] (0.5,2.1) -- (0.5,3.2) -- (1.5,3.5) -- (1.5,2.1) -- cycle;
    \draw[draw = red, fill = red, opacity = 0.5] (1.5,2.1) rectangle (2.5,3.5);
    \draw[draw = blue, fill = blue, opacity = 0.25] (1.5,2.1) -- (1.5,3.5) -- (2.5,3.2) -- (2.5,2.1) -- cycle;
    \draw[draw = red, fill = red, opacity = 0.5] (2.5,2.1) rectangle (3.5,3.2);
    \draw[draw = blue, fill = blue, opacity = 0.25] (2.5,2.1) -- (2.5,3.2) -- (3.5,2.3) -- (3.5,2.1) -- cycle;
    \draw[draw = red, fill = red, opacity = 0.5] (0.5,0.3) rectangle (1.0,1.4);
    \draw[draw = blue, fill = blue, opacity = 0.25] (0.5,0.3) -- (0.5,1.4) -- (1.0,1.625) -- (1.0,0.3) -- cycle;
    \draw[draw = red, fill = red, opacity = 0.5] (1.0,0.3) rectangle (1.5,1.625);
    \draw[draw = blue, fill = blue, opacity = 0.25] (1.0,0.3) -- (1.0,1.625) -- (1.5,1.7) -- (1.5,0.3) -- cycle;
    \draw[draw = red, fill = red, opacity = 0.5] (1.5,0.3) rectangle (2,1.7);
    \draw[draw = blue, fill = blue, opacity = 0.25] (1.5,0.3) -- (1.5,1.7) -- (2.0,1.625) -- (2.0,0.3) -- cycle;
    \draw[draw = red, fill = red, opacity = 0.5] (2,0.3) rectangle (2.5,1.625);
    \draw[draw = blue, fill = blue, opacity = 0.25] (2.0,0.3) -- (2.0,1.625) -- (2.5,1.4) -- (2.5,0.3) -- cycle;
    \draw[draw = red, fill = red, opacity = 0.5] (2.5,0.3) rectangle (3,1.4);
    \draw[draw = blue, fill = blue, opacity = 0.25] (2.5,0.3) -- (2.5,1.4) -- (3.0,1.025) -- (3.0,0.3) -- cycle;
    \draw[draw = red, fill = red, opacity = 0.5] (3,0.3) rectangle (3.5,1.025);
    \draw[draw = blue, fill = blue, opacity = 0.25] (3,0.3) -- (3,1.025) -- (3.5,0.5) -- (3.5,0.3) -- cycle;
    \draw[fill = black] (4.5,3.2) circle [radius = 0.05];
    \draw[fill = black] (5.5,2.5) circle [radius = 0.05];
    \draw[fill = black] (6.5,3.1) circle [radius = 0.05];
    \draw[fill = black] (7.5,2.8) circle [radius = 0.05];
    \draw[fill = black] (4.5,1.4) circle [radius = 0.05];
    \draw[fill = black] (5,1.3) circle [radius = 0.05];
    \draw[fill = black] (5.5,0.7) circle [radius = 0.05];
    \draw[fill = black] (6,1.6) circle [radius = 0.05];
    \draw[fill = black] (6.5,1.3) circle [radius = 0.05];
    \draw[fill = black] (7,1.5) circle [radius = 0.05];
    \draw[fill = black] (7.5,1.0) circle [radius = 0.05];
    \draw[draw = red, fill = red, opacity = 0.5] (4.5,2.1) rectangle (5.5,3.2);
    \draw[draw = blue, fill = blue, opacity = 0.25] (4.5,2.1) -- (4.5,3.2) -- (5.5,2.5) -- (5.5,2.1) -- cycle;
    \draw[draw = red, fill = red, opacity = 0.5] (5.5,2.1) rectangle (6.5,2.5);
    \draw[draw = blue, fill = blue, opacity = 0.25] (5.5,2.1) -- (5.5,2.5) -- (6.5,3.1) -- (6.5,2.1) -- cycle;
    \draw[draw = red, fill = red, opacity = 0.5] (6.5,2.1) rectangle (7.5,3.1);
    \draw[draw = blue, fill = blue, opacity = 0.25] (6.5,2.1) -- (6.5,3.1) -- (7.5,2.8) -- (7.5,2.1) -- cycle;
    \draw[draw = red, fill = red, opacity = 0.5] (4.5,0.3) rectangle (5,1.4);
    \draw[draw = blue, fill = blue, opacity = 0.25] (4.5,0.3) -- (4.5,1.4) -- (5,1.3) -- (5,0.3) -- cycle;
    \draw[draw = red, fill = red, opacity = 0.5] (5,0.3) rectangle (5.5,1.3);
    \draw[draw = blue, fill = blue, opacity = 0.25] (5,0.3) -- (5,1.3) -- (5.5,0.7) -- (5.5,0.3) -- cycle;
    \draw[draw = red, fill = red, opacity = 0.5] (5.5,0.3) rectangle (6,0.7);
    \draw[draw = blue, fill = blue, opacity = 0.25] (5.5,0.3) -- (5.5,0.7) -- (6,1.6) -- (6,0.3) -- cycle;
    \draw[draw = red, fill = red, opacity = 0.5] (6,0.3) rectangle (6.5,1.6);
    \draw[draw = blue, fill = blue, opacity = 0.25] (6,0.3) -- (6,1.6) -- (6.5,1.3) -- (6.5,0.3) -- cycle;
    \draw[draw = red, fill = red, opacity = 0.5] (6.5,0.3) rectangle (7,1.3);
    \draw[draw = blue, fill = blue, opacity = 0.25] (6.5,0.3) -- (6.5,1.3) -- (7,1.5) -- (7,0.3) -- cycle;
    \draw[draw = red, fill = red, opacity = 0.5] (7,0.3) rectangle (7.5,1.5);
    \draw[draw = blue, fill = blue, opacity = 0.25] (7,0.3) -- (7,1.5) -- (7.5,1.0) -- (7.5,0.3) -- cycle;
    \end{tikzpicture} \\ 
    \begin{picture}(0,0)
    \put(-10,24){$t$}
    \put(-10,75){$t$}
    \put(-100,60){$q$}
    \put(-100,111){$q$}
    \put(103.5,24){$t$}
    \put(103.5,75){$t$}
    \put(13.5,60){$q$}
    \put(13.5,111){$q$}
    \end{picture} \vspace{-10pt}
    \caption{We illustrate some of the concepts implicit in \eqref{discrete_stochastic}. The It\^{o}--like choice $\beta = 0$ is illustrated with red boxes in all subfigures, while the Stratonovich--like choice $\beta = \tfrac{1}{2}$ is illustrated with the blue trapezoids in all subfigures. We see these applied to an ordinary differential equation ($\mathsf{b} = 0$) on the left, and to a stochastic differential equation ($\mathsf{b} \neq 0$) on the right, with timesteps decreasing as we go from the top down. All choices of $\beta$ will converge to the same area under the curve in the time--continuum limit for smooth function on the left. For the stochastic process depicted on the right, however, (which remains stochastic at any timescale, such that the kind of picture on the left never emerges from it), different choices of $\beta$ will not necessarily lead to the same solution. Some ramifications of this are discussed in the main text, leading to equations \eqref{ito-generalform} through \eqref{ito-strato-convert}.}
    \label{fig-ItoStrato}
\end{figure}

\par In the sections below, we will formallly compare equations of motion derived from our Kraus operator methods to those from the SME; in order to do this, it is necessary that we briefly comment on a technical issue pertaining to stochastic calculus and the integration of stochastic differential equations. 
The calculus used to derive and/or manipulate a Langevin of the type above is closely tied to the type of Riemann sum used as the basis of any subsequent integration. 
If we were integrating an ordinary differential equation, any valid choice of Riemann sum would lead to the same result in the time--continuum limit. 
This is not so in the stochastic case however; if we suppose that $dW$ is stochastic at \emph{every} time--scale, different Riemann sums will \emph{not} converge to the same solutions in the limit anymore! 
To get the idea, we may consider a discrete update step
\be \label{discrete_stochastic}
q_{k+1}-q_k = \mathsf{a}(\tilde{q}_k) \Delta t + \mathsf{b}(\tilde{q}_k) \Delta W_k,
\ee
where we have $\tilde{q}_k = \beta q_{k+1} + (1-\beta) q_k$, and the indices $k$, $k+1$ correspond to times $\Delta t$ apart\footnote{Formally, $q(t) - q(0) = \int_0^t dt' \mathsf{a}[q(t'),t'] + \int_0^t dW(t') \mathsf{b}[q(t'),t']$ is more appropriate; the way the integration of the diffusion term, over $dW$, is carried out is both significant and potentially ambiguous. See chapter 4 of \cite{BookGardiner2} for rigorous derivations and more detailed comments.}. 
We highlight two very common conventions: The It\^{o} convention uses $\beta = 0$, such that we evaluate drift and diffusion coefficients at the beginning of a timestep, whereas the Stratonovich convention corresponds to $\beta = \tfrac{1}{2}$, such that functions are evaluated according to a trapezoidal rule (see Fig.~\ref{fig-ItoStrato}). 
The form of the SME \eqref{SME} assumes a derivation based on It\^{o} calculus, in which expansions are made to $O(dt)$ using the rule $dW^2 = dt$ (i.e.~expansions to $O(dt)$ must include explicit expansions to $O(dW^2)$ in that formalism \cite{Jacobs2006}). For an accessible and intuitive explanation of this rule, we encourage the interested reader to look at section 4 of Ref.~\cite{Gough2018}. Expansions made with regular calculus will lead a Stratonovich equation instead however.
In other words, we have to consider two different stochastic calculus conventions, each leading to different differential equations; they give consistent results, however, when paired with the correct integration rules.
Specifically: Integrating the equation
\be \label{ito-generalform}
d\mathbf{q} = \mathbf{a}(\mathbf{q})\:dt + \mathbf{b}(\mathbf{q}) \: d\boldsymbol{W} 
\ee
according to the It\^{o} sense ($\beta = 0$), is equivalent to performing a Stratonovich integration ($\beta = \tfrac{1}{2}$) on
\be \label{strato-generalform}
\dot{\mathbf{q}} = \mathbf{A}(\mathbf{q}) + \mathbf{b}(\mathbf{q}) \: \boldsymbol{\xi},
\ee
where the two drift terms $\mathbf{a}$ and $\mathbf{A}$ are related by the transformation
\be \label{ito-strato-convert}
A_q = a_q - \tfrac{1}{2} \sum_{j,n} b_{jn} \partial_n b_{jq};
\ee
$n$ indexes the coordinates (components of $\mathbf{q}$), and $j$ indexes the independent noise(s) on each measurement channel, which are summed.
For justification and details see e.g.~\cite{BookGardiner2, BookKloedenPlaten}. We will use this conversion rule to connect different descriptions of the quantum measurement scenarios we consider below.

We make a final remark about numerical simulations before moving on. 
The appeal of the SME as a theoretical tool is that it expresses quantum trajectory dynamics as a differential equation, similar to how physicists are accustomed to describing classical dynamics; furthermore, the SME readily splits those dynamics into three terms, which make qualitatively distinct contributions to the dynamics. 
It is worth noting, however, that compared with the case of ordinary differential equations \cite{NumRecC}, methods for the numerical integration of stochastic differential equations \cite{BookKloedenPlaten} are more complex, and are accurate only to substantially lower order in $dt$. 
Additionally, direct numerical integration of the SME does not necessarily preserve the properties of a valid density matrix beyond $O(dt)$, leading to problematic numerical errors unless $dt$ is extremely small; it is consequently numerically preferable to execute simulations of stochastic quantum trajectories by direct application of a positive mapping, as in \eqref{state-update} or similar, when possible. 
The interested reader may find further comments in this vein e.g.~in Ref.~\cite{Rouchon2015}.


\section{Single--Qubit Heterodyne Trajectories \label{sec-1QHet}}

We now begin looking at diffusive quantum trajectories due to heterodyne detection. What follows is essentially a review of the simplest non--trivial case described more extensively in Ref.~\cite{Jordan2015flor}, and corresponding to the experimental implementation e.g.~of Ref.~\cite{Campagne-Ibarcq2016}. 
In the language of quantum--limited amplifiers (QLAs), which are essential to realizing experiments involving individual quantum trajectories, our meaning of ``heterodyne'' corresponds to ``phase--preserving'' amplification (e.g.~see \cite{Clerk2010Review, Bergeal2010, Flurin_Thesis, DevoretAmp} or similar, regarding implementations in circuit QED scenarios). See Fig.~\ref{fig-exp-diagrams}.
Owing to the mixing of the fluorescence signal with a coherent state of the field (the ``local oscillator'', or LO), the heterodyne measurement gives access to both quadratures of the field, with a symmetric uncertainty. A reader unfamiliar with a quadrature phase space representation of a field mode may benefit from perusing e.g.~Ref.~\cite{Silberhorn2007}. When performed with an ideal QLA, this scheme is formally equivalent to projecting the field mode into the basis of the coherent states \cite{Jordan2015flor}.

\subsection{Stochastic Master Equation Treatment}

The SME is given in Eq.~\eqref{SME}, and provides one of the most--used approaches to modeling diffusive quantum trajectories arising from continuous weak measurement \cite{Brun2001Teach, Jacobs2006}. We will consider an idealized measurement in the rotating frame, characterized by $\hat{H} = 0$ (no unitary dynamics), $\hat{L}_X = \hat{\sigma}_- \sqrt{\gamma/2}$, and $\hat{L}_P = i\hat{\sigma}_-\sqrt{\gamma/2}$, where there is no dephasing channel and the measurement efficiency $\eta = 1$
is perfect. We can make qualitative sense of the two operators $\hat{L}_X$ and $\hat{L}_P$ by understanding that $\hat{\sigma}_-$ indicates that our measurement is being made through a decay channel, and that $\hat{L}_X$ and $\hat{L}_P$ are associated, respectively, with the information encoded in the two quadratures $\hat{X}$ and $\hat{P}$ of the field read out by the heterodyne measurement; the factor $i$ between $\hat{L}_X$ and $\hat{L}_P$ is the $90^\circ$ phase between these two orthogonal directions in the $XP$--plane (often also conventially labeled as the $IQ$--plane). 

The resulting SME is then
\be 
\dot{\rho}= \hat{\mathcal{L}}[\rho,\hat{L}_X] + \hat{\mathcal{L}}[\rho,\hat{L}_P] + \hat{\mathcal{M}}[\rho,\hat{L}_X] \xi_X + \hat{\mathcal{M}}[\rho,\hat{L}_P] \xi_P,
\ee
where $\hat{\mathcal{L}}$ and $\hat{\mathcal{M}}$ are still the Lindblad dissipation, and measurement backaction terms, respectively. The Gaussian white noise for the measurement channels is characterized by each $\xi(t) \sim dW/dt$. We may obtain equations of motion in terms of Bloch sphere coordinates using $\dot{\mathbf{q}} = \text{tr}(\hat{\sigma}_\mathbf{q} \: \dot{\rho})$, yielding
\begin{subequations} \label{ito-Het}
\be 
\dot{x} = - \tfrac{\gamma}{2} x + \sqrt{\tfrac{\gamma}{2}} \left[ \left( 1 + z - x^2 \right) \xi_X - x \: y \: \xi_P \right],
\ee \be 
\dot{y} = - \tfrac{\gamma}{2} y + \sqrt{\tfrac{\gamma}{2}} \left[ \left( 1 + z - y^2 \right) \xi_X - x \: y \: \xi_P \right],
\ee \be
\dot{z} = - \gamma (1+z) - \sqrt{\tfrac{\gamma}{2}} (1+z) \left[ x \: \xi_X + y \: \xi_P \right],
\ee
\end{subequations}
in agreement with the result in eq.~(25) of \cite{Jordan2015flor} (for $u = 1+z$, $\eta = 1$, and $\gamma_\phi = 0$, in their notation). The stochastic readouts (signals arising from the measurement process) are given by
\begin{subequations}  \label{hetd-ro-sme} \be
r_X = \langle \hat{L}_X + \hat{L}_X^\dag \rangle + \xi_X = \sqrt{\tfrac{\gamma}{2}} x + \xi_X,
\ee \be 
r_P = \langle \hat{L}_P + \hat{L}_P^\dag \rangle + \xi_P = \sqrt{\tfrac{\gamma}{2}} y + \xi_P.
\ee \end{subequations}
Notice that the average path given by these equations (where averages over an ensemble lead to $\xi \rightarrow 0$, since these are zero--mean stochastic variables) obeys the same basic fluorescence relations \eqref{unmonitored-Bloch-eqs}. This is a typical example of the relationship between an un-monitored and continuously--monitored system, as we have discussed in general above. 

\par We will interpret the equations \eqref{ito-Het} as being equations suitable for It\^{o} integration and stochastic calculus (consistent with the assumptions used to derive \eqref{SME} in the first place \cite{Jacobs2006}). It will also be useful to have the corresponding Stratonovich version of this system of equations, which can be manipulated using regular calculus. In this case, the conversion \eqref{ito-strato-convert} can be written as 
\be \label{Het-strato-drift}
\mathbf{A} = \mathbf{a} - \tfrac{1}{2}(\mathbf{b}_X\cdot\nabla)\mathbf{b}_X - \tfrac{1}{2}(\mathbf{b}_P \cdot \nabla)\mathbf{b}_P;
\ee
a trio of Stratonovich equations corresponding to the It\^{o} equations \eqref{ito-Het} 
are obtained by substituting this new drift vector \eqref{Het-strato-drift} into \eqref{strato-generalform}.

\subsection{Kraus Operator Treatment}
We now consider the corresponding Kraus operator treatment of this situation. As discussed previously, a heterodyne measurement effectively projects the fluorescence signal onto a coherent state \eqref{coherent-state} at each measurement timestep, such that we write down an operator
\be \label{kraus-het-gen}
\hat{M}_\alpha = \bra{\alpha} \left( \begin{array}{cc} 
\sqrt{1 - \epsilon} & 0 \\ \sqrt{\epsilon} \hat{a}^\dag & 1
\end{array} \right) \ket{0} = e^{-|\alpha|^2/2} \left(\begin{array}{cc}  \sqrt{1-\epsilon} & 0 \\ \sqrt{\epsilon} \alpha^\ast & 1 \end{array} \right).
\ee
We will use a substitution for the readouts given by
\be \label{het-rodef}
\alpha = \sqrt{\frac{dt}{2}}(r_X - i r_P);
\ee
the prefactor $\sqrt{dt/2}$ is chosen because it generates statistics consistent with the shot--noise of the coherent state LO; for clarification see \cite{Jordan2015flor} and/or appendix~\ref{app-Noise}.
With this substitution, we have a Kraus operator
\be \begin{split}
\hat{M}_\alpha &=  \exp\left[ - \tfrac{dt}{4} (r_X^2+r_P^2) \right]\left( \begin{array}{cc} 
\sqrt{1 - \gamma \: dt} & 0 \\ \sqrt{\tfrac{\gamma}{2}}dt(r_X + i r_P) & 1
\end{array} \right),
\end{split} \ee
which may be used to update the state (using \eqref{state-update} with $\hat{M}_r \rightarrow \hat{M}_\alpha$) 
conditioned on acquiring a measurement record drawn from the probability density $\wp(r_X,r_P|\rho(t)) = \mathcal{N}\text{tr}\left( \hat{M}_\alpha \rho(t) \hat{M}_\alpha^\dag \right)$, where $\mathcal{N}$ is a normalization constant. The measurement operators form a proper POVM \cite{BookNielsen}, in that 
\be 
\frac{dt}{2\pi} \iint_{-\infty}^\infty dr_X \: dr_P \: \hat{M}_\alpha^\dag \hat{M}_\alpha = \openone,
\ee
(i.e.~the readouts we have defined here constitute another complete set of measurement outcomes). 

\par It will be useful to take a closer look at the probability density from which the readouts are drawn. Following the procedure we have typically used in the context of optimal paths (OPs) \cite{Chantasri2013, Chantasri2015, Jordan2015flor, Lewalle2016, Lewalle2018}, we will expand the log of the probability density to $O(dt)$, defining a term 
\be \begin{split} \label{hetG}
\mathcal{G}_{het} =& -\tfrac{1}{2} \left(r_I-\sqrt{\tfrac{\gamma }{2}} x\right)^2-\tfrac{1}{2} \left(r_Q-\sqrt{\tfrac{\gamma }{2}} y\right)^2 \\ &+\tfrac{\gamma}{4} \left(x^2+y^2\right)-\tfrac{\gamma}{2} (z+1)
\end{split} \ee
such that $\wp = e^{C + \mathcal{G} dt + O(dt^2)}$. We see that up to the two last terms in Eq.~\eqref{hetG}, the probability density is Gaussian in both readouts, with variances $1/dt$, and means $ x \sqrt{\gamma/2}$ and $y \sqrt{\gamma/2}$ for $r_X$ and $r_P$, respectively. 
Notice that this corresponds precisely to what we had from the SME, as in \eqref{hetd-ro-sme}; the Gaussian form implicit in \eqref{hetG} is in fact key in demonstrating that the form of the SME \eqref{SME} written in terms of Weiner increments $dW$ is formally suitable for this system.

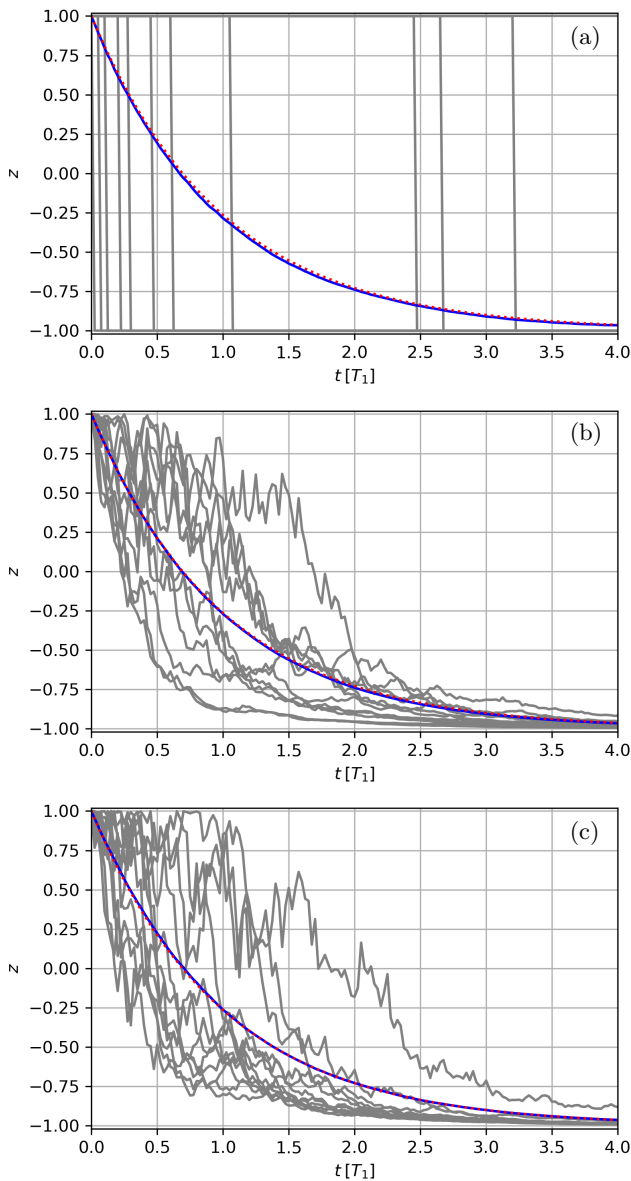
\begin{figure}
    \centering
    \includegraphics[width = .99\columnwidth, trim = {0pt 5pt 10pt 10pt}, clip]{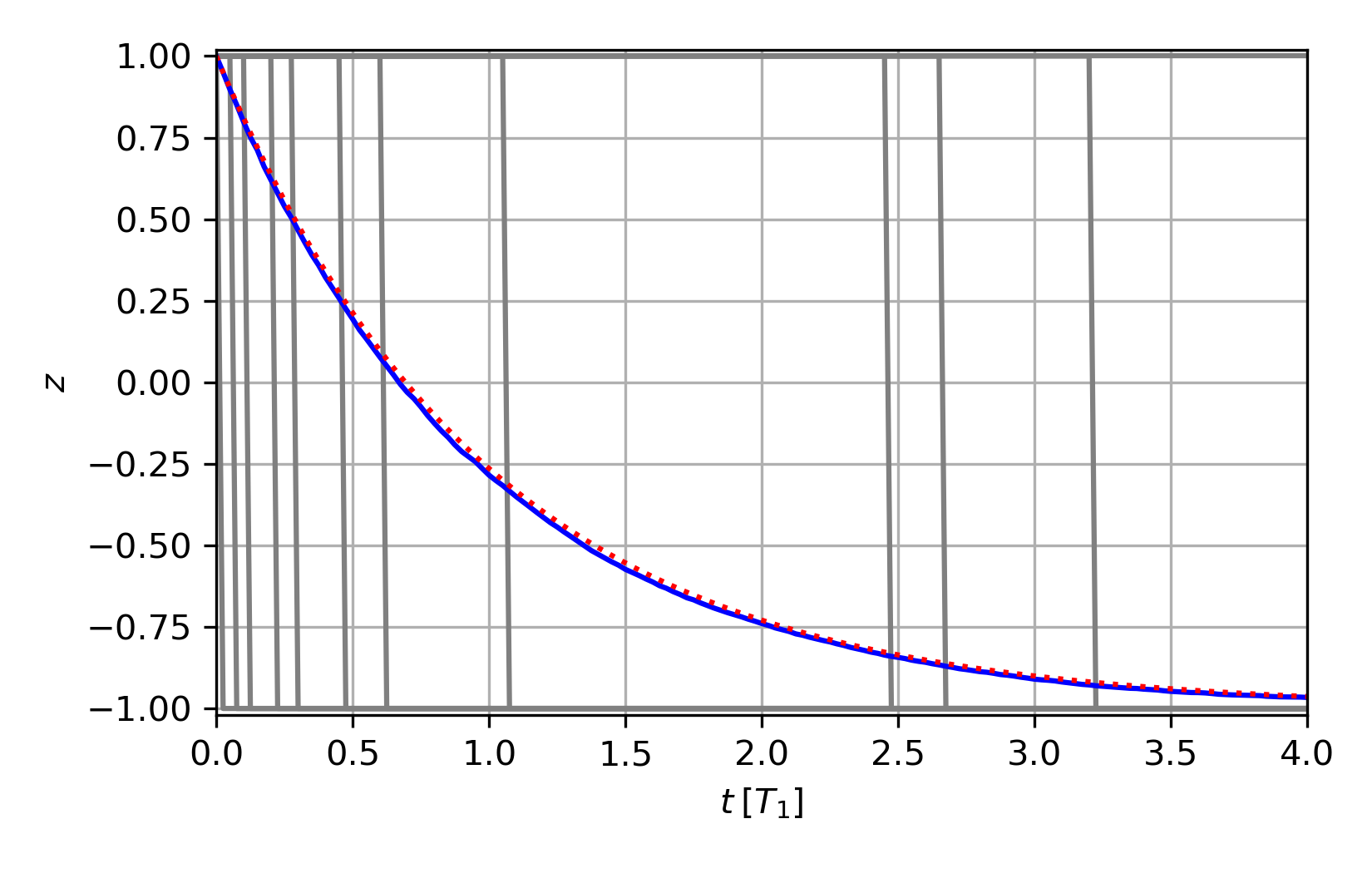} \\
    \includegraphics[width = .99\columnwidth, trim = {0pt 5pt 10pt 10pt}, clip]{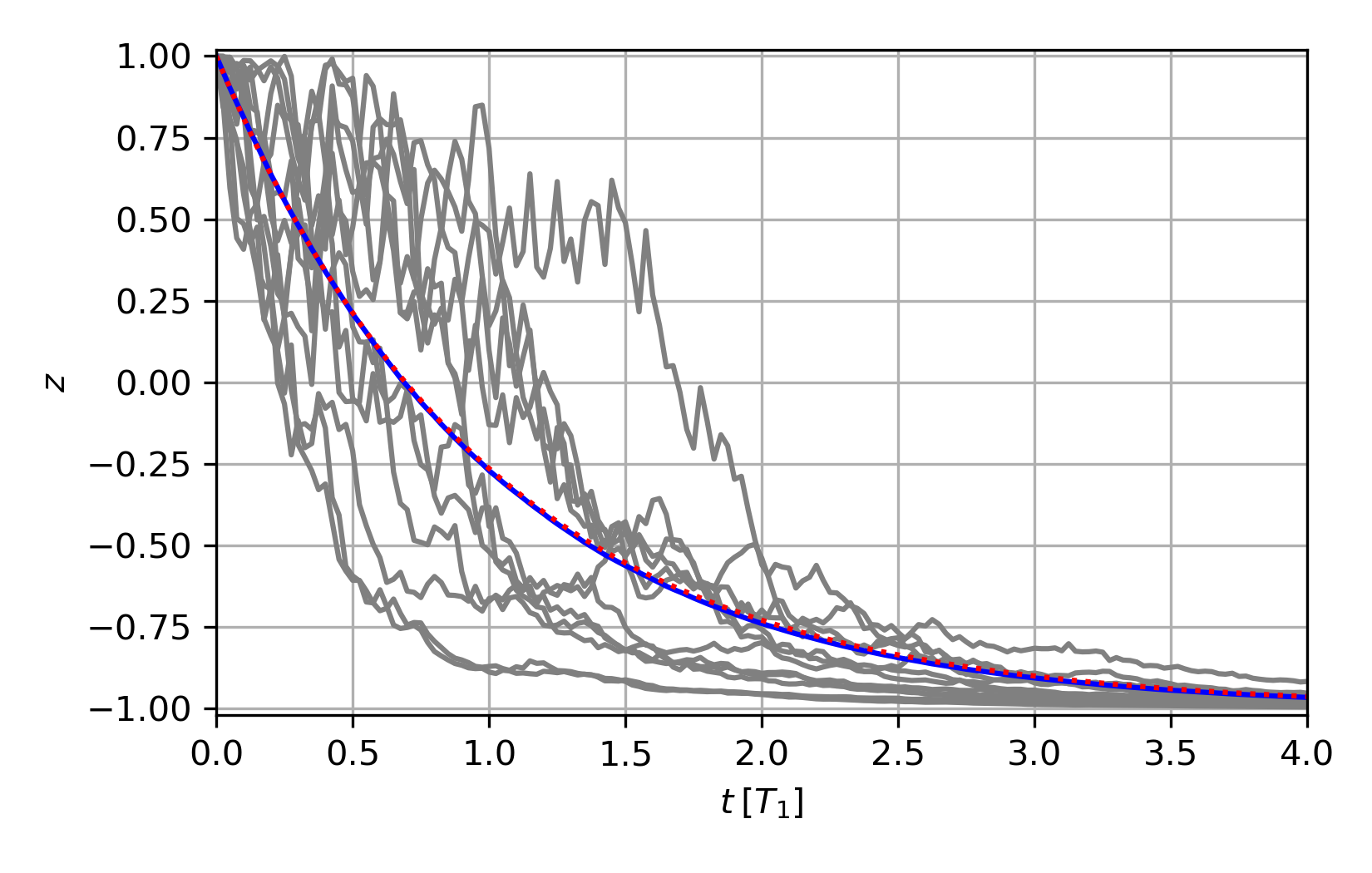} \\
    \includegraphics[width = .99\columnwidth, trim = {0pt 5pt 10pt 10pt}, clip]{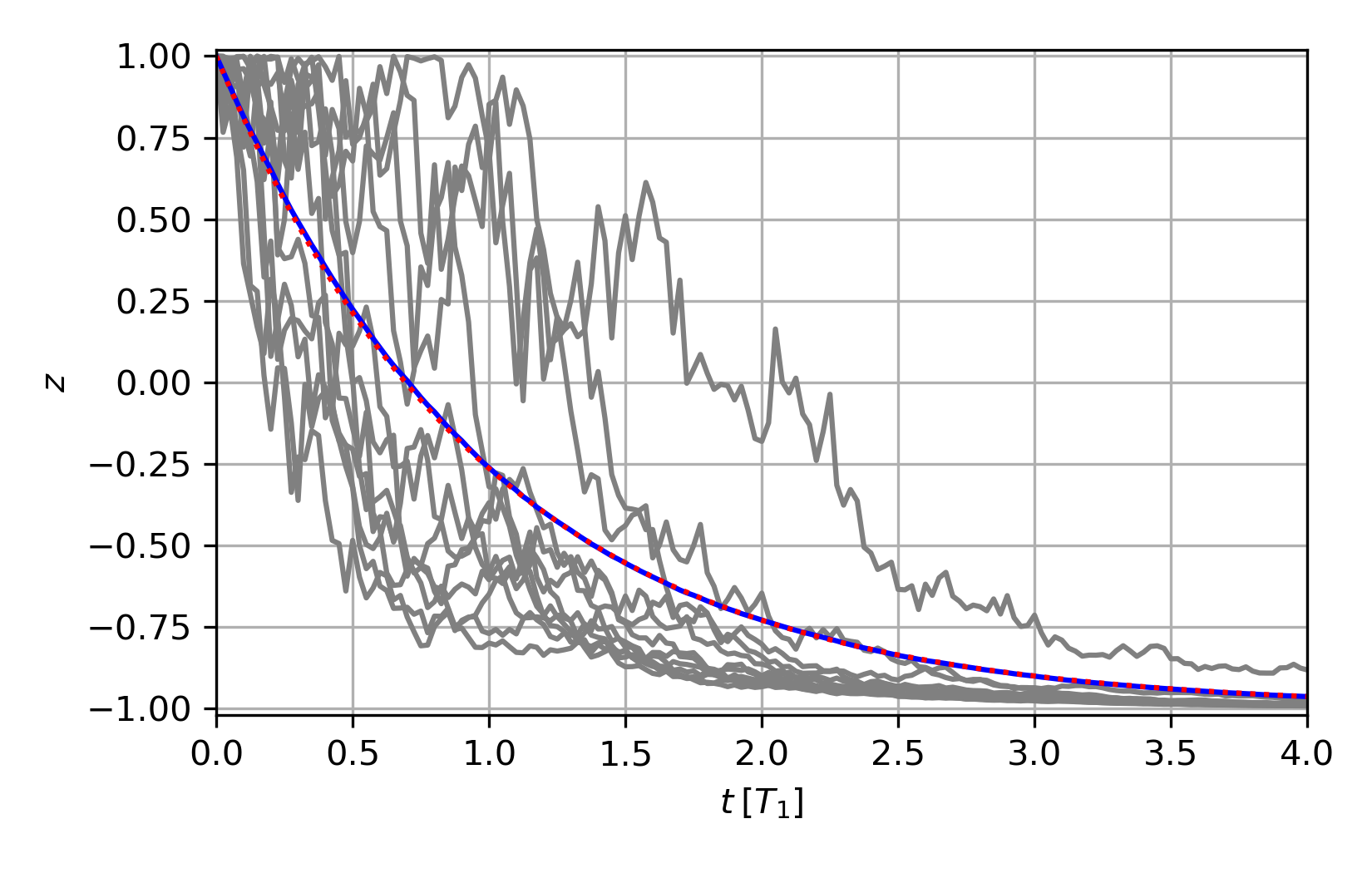} \\
    \begin{picture}(0,0)
    \put(100,447){(a)}
    \put(100,297){(b)}
    \put(100,146){(c)}
    \end{picture} \vspace{-15pt}
    \caption{We show simulations of the decay from $\ket{e}$ to $\ket{g}$ under ideal measurements, including photodetection (a), heterodyne detection for $\theta = 0$ (b), and homodyne detection for $\theta = 0$ (c). In every case, we plot a dozen individual trajectories in grey, the average trajectory over an ensemble of 10,000 simulated trajectories in solid blue, and the unmonitored curve integrated from \eqref{unmonitored-Bloch-eqs} in dotted red. As required by the SME, we see good agreement between the simulated paths averaged over measurement noise realizations (solid blue), and a direct computation of the un--monitored dynamics, which in the present case must simply follow $z(t) = 2e^{-\gamma t} - 1$ (dotted red). We can see the qualitative similarity between the diffusive homodyne and heterodyne trajectories in (b) and (c), respectively, as well as their stark difference with the jump trajectories generated by photodetection (a); these contrasts are clear and important, as is the average dynamics common to all three schemes, which follows from the shared underlying decay process at the heart of all three measurements considered here. All our simulations are performed by applying our Kraus operator methods.
    }
    \label{fig-simul3meas}
\end{figure}

\par We also use the Kraus operator to obtain some equations of motion. Consider the exapansion of the Kraus operator itself to $O(dt)$, which reads
\be \label{approx-mhet} 
\hat{M}_\alpha e^{|\mathbf{r}|^2 dt/4} \approx \openone + dt \underbrace{\left( \begin{array}{cc}
-\tfrac{\gamma}{2} & 0 \\
\sqrt{\tfrac{\gamma}{2}}(r_X+ir_P)  & 0
\end{array}\right)}_{\hat{m}_{\alpha}} + O(dt^2)
\ee 
for $|\mathbf{r}|^2 = r_X^2 + r_P^2$.  
We can strip the Gaussian factor $e^{-|\mathbf{r}|^2 dt/4}$ from the operator for this purpose, since it appears in both the numerator and denominator of the state update expression \eqref{state-update}, and thereby cancels off. Consider the following series of approximations, assuming small $dt$:
\be \label{Kraus-Eqmo-Odt} \begin{split}
\rho(t+dt) &\approx \frac{(\openone + \hat{m}_\alpha dt) \rho(t) (\openone + \hat{m}_\alpha^\dag dt)}{\text{tr}\left((\openone + \hat{m}_\alpha dt) \rho(t) (\openone + \hat{m}_\alpha^\dag dt) \right)} \\
&\approx \rho + dt \left(  \hat{m}_\alpha \rho + \rho \hat{m}_\alpha^\dag - \rho\: \text{tr}\left( \hat{m}_\alpha \rho + \rho \hat{m}_\alpha^\dag\right) \right),
\end{split} \ee
which can then be rearranged according to $\rho(t+dt)-\rho(t) \approx dt \: \dot{\rho}$, such that 
\be \label{Kraus-Eqmo-Odt2}
\dot{\rho} \approx \hat{m}_\alpha \rho + \rho \hat{m}_\alpha^\dag - \rho\: \text{tr}\left( \hat{m}_\alpha \rho + \rho \hat{m}_\alpha^\dag\right).
\ee
This can be expressed in Bloch coordinates by
\begin{subequations} \label{hetd-kraus-eqmo} \be 
\dot{x} = \tfrac{\gamma}{2} x \: z + \sqrt{\tfrac{\gamma}{2}} \left[ r_X (1 + z - x^2) - r_P\: x \: y \right],
\ee \be 
\dot{y} = \tfrac{\gamma}{2} y \: z + \sqrt{\tfrac{\gamma}{2}} \left[ r_P (1 + z - y^2) - r_X \: x \: y \right],
\ee \be 
\dot{z} = \tfrac{\gamma}{2}(z^2-1) - \sqrt{\tfrac{\gamma}{2}} (1+z) \left[ r_X \: x + r_P \: y \right].
\ee \end{subequations}
It is then straightforward to make the substitutions $r_X = x \sqrt{\gamma/2} + \xi_X$ and $r_P = y \sqrt{\gamma/2} + \xi_P$ \eqref{hetd-ro-sme}, and see that these equations from the Kraus operator approach are identical to the Stratonovich equations \eqref{strato-generalform} 
obtained by conversion from the SME approach; this relationship between a Kraus operator based on Bayesian logic, and the SME, is consistent with previous results for this particular measurement \cite{Jordan2015flor}, and other types of continuous qubit measurements leading to diffusive SQTs \cite{Gambetta2008, Chantasri2015, Korotkov2016, Lewalle2016}. 

\par Simulations can be generated by applying the state update rule \eqref{state-update} with $\hat{M}_r \rightarrow \hat{M}_{\alpha}$, with a pair of readouts drawn from Gaussians of means and variances described above, at each timestep. The resulting stochastic trajectories diffuse as expected, and recreate the required decay dynamics on average, as shown in Fig.~\ref{fig-simul3meas}(b).

\subsection{Generalizations}
\par We consider the addition of a Rabi drive to the qubit (i.e.~we now discuss additional tones inducing a unitary rotation in the Bloch sphere) by the addition of a Hamiltonian term $i[\rho,\hat{H}]$ to the SME ($\hbar\rightarrow 1$), or a corresponding operator $\hat{U} = e^{-i\hat{H}dt}$ to the measurement scheme with the Kraus operator (where the resulting equations of motion are insensitive to the order of operations, since they are only to $O(dt)$). 
Without loss of generality, we use $\hat{H} = \delta\hat{\sigma}_z/2 + \Omega \hat{\sigma}_y/2$, where we have denoted the detuning $\delta = \omega_\text{qb}-\omega_\text{dr}$, with $\omega_\text{dr}$ the frequency of the tone\footnote{Note that this description is associated with a frame rotating at frequency $\omega_\text{dr}$, or equivalently the interaction picture with respect to $\hat{H}_\text{frame} = \omega_\text{dr}\hat{\sigma}_z/2$. In the fixed frame, the qubit Hamiltonian in presence of the drive reads $\hat H(t) = \omega_\text{qb}\hat{\sigma}_z/2 + \Omega (i\hat{\sigma}_-e^{i\omega_\text{dr}t}-i\hat{\sigma}^+e^{-i\omega_\text{dr}t})$.}. Such a tone induces a rotation around an axis tilted by an angle $\text{arctan}(\Omega/\delta)$ with respect to the $z$--axis. 
Note that the assumption in our derivation has been that only photons emitted by the qubit enter the transmission line which leads to the measurement apparatus; the simplest way to imagine engineering a system such that this remains valid with the Rabi drive on, is that the drive is being implemented by a tone which is off--resonant with the qubit/cavity/transmission line, such that the qubit photons couple to the output leading the measurement device only, and the drive photons couple to their own output only. 
As discussed earlier, this assumption also requires us to have the cavity resonance far from any of the Mollow triplet peaks, which are centered around $\omega_\text{dr}$ and $\omega_\text{dr}\pm\Omega_\text{eff}$, where $\Omega_\text{eff} = \sqrt{\Omega^2+\delta^2}$ is the generalized Rabi frequency; this regime and assumption is necessary if we want to treat the form of the decay channel as being unaffected by the drive. 
Drives of the type we have discussed apply generically in ``resonance fluoresence'' scenarios \cite{Kimble1976, Mahdi2016, Quijandria2018}, as well as any other situation in which additional tones are present in qubit's cavity (e.g.~to implement additional measurements \cite{Ficheux2018}). 
The situation we have described here is illustrated in Fig.~\ref{fig-exp-diagrams}(c).

\par Note that it is possible to generalize this heterodyne measurement by choosing the phase $\theta$ of the LO. The phase $\theta$ is a relative phase between the signal and LO, so it is equivalent to think of a phase plate having been put in the signal line instead of the LO such that $\hat{a}^\dag \rightarrow e^{-i\theta} \hat{a}^\dag$ in \eqref{kraus-het-gen}, with interference against a fixed pump. Mathematically, we can then assign readouts according to 
\be \label{het-ro-phase}
\alpha = \sqrt{\frac{dt}{2}}e^{i\theta}(r_I - i r_Q),
\ee
which leads to 
\be \begin{split}
r_I &= \sqrt{\tfrac{\gamma}{2}} (x \cos\theta - y \sin\theta) + \xi_I , \text{ and} \\ r_Q &= \sqrt{\tfrac{\gamma}{2}}(y \cos\theta + x \sin\theta) + \xi_Q.
\end{split} \ee
The operators for the SME which match the pair of observables we infer from the means of $\mathcal{G}$ are
\be 
\hat{L}_I = \sqrt{\tfrac{\gamma}{2}} e^{-i\theta} \hat{\sigma}_-, \text{ and } \hat{L}_Q = i\sqrt{\tfrac{\gamma}{2}} e^{-i\theta} \hat{\sigma}_-.
\ee
We see that changing the phase $\theta$ between the signal and LO effectively rotates the quadrature pair we measure.
We have here used notation such that $r_I = r_X$ and $r_Q = r_P$ for the choice $\theta = 0$. The relationship between the Kraus operator equations of motion and SME equations of motion (It\^{o} or Stratonovich) which we found in the $\theta = 0$ case above, hold for arbitrary $\theta$.

\par We have reviewed the most basic features of an idealized heterodyne measurement. For a more advanced treatment of this system, refer to \cite{Jordan2015flor}; we will now turn our attention to applying the framework we have just developed to homodyne measurement.

\section{Single--Qubit Homodyne Fluorescence Trajectories \label{sec-1QHom}}

\par Several experiments \cite{Mahdi2016, Naghiloo2016flor, Tan2017,Mahdi2017Qtherm} and some theory \cite{Bolund2014} have been published about homodyne fluorescence measurement; we will develop our theory examples here far enough to compare them directly with the simplest experimental results.

\subsection{Kraus Operator and Measurement Dynamics}

Homodyne detection again involves interfering our signal with a strong LO. 
Practically, instead of amplifying both quadratures of the resulting signal as in heterodyne detection, homodyne detection involves amplifying one quadrature and de-amplifying the other \cite{Tyc2004} (``phase-sensitive'' amplification). 
This procedure amounts to squeezing out the quadrature that isn't measured, such that 
in the limit of ideal squeezing we project our signal onto a single quadrature's eigenstate, instead of onto a coherent state \cite{Wiseman1996Review, BookWiseman}. 
This yields a single readout signal, rather than the pair which arise in the heterodyne case.
We will follow the same recipe as in the heterodyne case, except that we project onto a final state $\ket{X}$ (the eigenstate of the $\hat{X} = (\hat{a}^\dag + \hat{a})/\sqrt{2}$ operator in the quadrature space), instead of the coherent state $\ket{\alpha}$. 
Again, those unfamiliar with this phase space terminology may wish to consult e.g.~Ref.~\cite{Silberhorn2007}. 
For dimensionless $X$, recall that we have the following solutions to the quantum harmonic oscillator, which models the field mode:
\be 
\ip{X}{0} = \pi^{-\frac{1}{4}} e^{-X^2/2}, \text{ and } \ip{X}{1} =  \pi^{-\frac{1}{4}} \sqrt{2} X e^{-X^2/2}.
\ee
Projecting onto the general fluorescence operator, and suppressing the factors $\pi^{-\tfrac{1}{4}}$ on all terms, we get
\be \label{MXnoconst}
\hat{M}_x = \bra{X} \left( \begin{array}{cc} 
\sqrt{1 - \epsilon} & 0 \\ \sqrt{\epsilon} \hat{a}^\dag & 1
\end{array} \right) \ket{0} = e^{-X^2/2} \left(\begin{array}{cc}  \sqrt{1-\epsilon} & 0 \\ \sqrt{2\epsilon} X & 1 \end{array} \right).
\ee
Then using a readout substituted in according to 
\be \label{hom-rodef}
X \rightarrow \sqrt{\frac{dt}{2}} r,
\ee
we find that the POVM is normalized, i.e.
\be 
\sqrt{\frac{dt}{2\pi}} \int_{-\infty}^\infty dr \: \hat{M}_x^\dag \hat{M}_x = \openone.
\ee
The relationship between $X$ and $r$ is again set based on comparing the readout statistics with LO's shot noise, as discussed in appendix~\ref{app-Noise}.
Those readout statistics can be readily understood from the expression
\be 
\mathcal{G}_{hom} = - \tfrac{1}{2}\left( r - \sqrt{\gamma} x \right)^2 - \tfrac{\gamma}{2}(1+z-x^2),
\ee
which again comes from expanding the logarithm of $\text{tr}( \hat{M}_x \rho \hat{M}_x^\dag )$. 
We infer that projecting onto $\ket{X}$ in the photon space leads to a signal related to $x$ in the qubit space, since $r$ has a mean $\sqrt{\gamma} x$, and variance $1/dt$. 

\par As before, we may take $\hat{a}^\dag \rightarrow \hat{a}^\dag e^{-i\theta}$ to generalize the choice of measured quadrature, yielding an operator
\be \label{MX}
\hat{M}_x = \left(\frac{dt}{2\pi}\right)^\frac{1}{4} e^{-r^2 dt /4} \left(\begin{array}{cc}  \sqrt{1-\gamma dt} & 0 \\ dt \sqrt{\gamma} \: r e^{-i\theta} & 1 \end{array} \right),
\ee 
which still generates a proper POVM. Expanding the log--probability density for the readout gives us
\be \begin{split} \label{homG}
\mathcal{G}_{hom}^\theta =& - \tfrac{1}{2}\left[ r - \sqrt{\gamma} (x\cos\theta - y \sin\theta) \right]^2 \\& - \tfrac{\gamma}{2}(1+z-(x \cos\theta - y \sin\theta)^2);
\end{split} \ee
thus the mean of the Gaussian in $r$ matches the signal given by $\hat{L} + \hat{L}^\dag = \sqrt{\gamma}(\hat{\sigma}_x \cos\theta - \hat{\sigma}_y \sin\theta)$ for the SME operator $\hat{L} = \sqrt{\gamma} e^{-i\theta}\hat{\sigma}_-$. 

\par We proceed to find the equations of motion. Note that we can approximate $\hat{M}_x$ as we did $\hat{M}_\alpha$ \eqref{approx-mhet}, such that
\be 
\left(\begin{array}{cc}  \sqrt{1-\gamma dt} & 0 \\ dt \sqrt{\gamma} \: r e^{-i\theta} & 1 \end{array} \right) \approx \openone + dt \underbrace{\left(
\begin{array}{cc}
-\tfrac{\gamma}{2} & 0 \\ \sqrt{\gamma}\: r e^{-i\theta} & 0
\end{array}
\right)}_{\hat{m}_x} + O(dt^2).
\ee
Then by the logic of \eqref{Kraus-Eqmo-Odt} and \eqref{Kraus-Eqmo-Odt2} 
we may derive equations of motion in terms of Bloch coordinates
\begin{subequations}\label{eqmo-hom} \be 
\dot{x} = \tfrac{\gamma}{2} x z + \Omega z - \delta y + r \sqrt{\gamma} \left[ (1+ z - x^2)\cos\theta + x y \sin\theta \right],
\ee \be 
\dot{y} = \tfrac{\gamma}{2} y z + \delta x + r \sqrt{\gamma} \left[ (y^2-z-1) \sin\theta - x y \cos\theta \right],
\ee \be
\dot{z} = \tfrac{\gamma}{2}(z^2-1) - \Omega x + r \sqrt{\gamma} (z+1)\left[ y \sin\theta - x \cos\theta \right].
\ee \end{subequations}
We have again used a Rabi drive characterized by $\hat{H} = \Omega \hat{\sigma}_y/2 + \delta \hat{\sigma}_z/2$, or $\hat{U} \approx \openone - i \Omega \hat{\sigma}_y dt/2 - i \delta \hat{\sigma}_z dt/2$. 
As above, these equations are consistent with those derived from the SME \eqref{SME}, using $\hat{L} = \sqrt{\gamma} e^{-i\theta}\hat{\sigma}_-$, provided the SME output is correctly interpreted as an It\^{o} equation, whose Stratonovich form then matches the above exactly. 
Simulated trajectories for the case $\theta = 0$ and $\Omega = 0$ are shown in Fig.~\ref{fig-simul3meas}(c), and demonstrate good agreement with expectations, as in the previous cases.

\subsection{Inefficient Measurements \label{sec-inefficient}}
Inefficient measurement is easily included in the SME \eqref{SME}, and is completely described by the dimensionless parameter $\eta \in [0,1]$. In the Kraus operator picture, we must modify the amplitude of the signal going into the measurement apparatus; we will find a case intermediate between perfect measurements \eqref{state-update} and no measurement \eqref{no-measurement}, reflecting that some fraction of the information is lost rather than collected. 
A straightforward way to represent this is with an unbalanced beamsplitter placed in front of our (still otherwise ideal) measurement device, as shown in Fig.~\ref{fig-exp-diagrams}(d). If $\hat{a}^\dag$ creates a photon in the emitted field mode, the beamsplitter transforms it according to 
\be \label{eta-bs}
\hat{a}^\dag \rightarrow \sqrt{\eta} \:\hat{a}_s^\dag+\sqrt{1-\eta} \:\hat{a}_\ell^\dag,
\ee
where the surviving signal $\hat{a}_s^\dag$ goes to the detector with probability $\eta$, but the information in channel $\hat{a}_\ell^\dag$ is lost with probability $1-\eta$, and outcomes (all of which could have occurred) in the latter channel must be traced out. We will do the trace of the lost channel in the Fock basis for simplicity (a sum of two terms is simpler than an integral over a continuous homodyne or heterodyne readout, although averaging over any complete set of hypothetical measurement outcomes is technically correct). 
The scheme we are describing, for homodyne detection with efficiency $\eta$, can be implemented with a pair of operators
\be \label{mxj}
\hat{M}_{xj} = \bra{X_s j_\ell}  \left( \begin{array}{cc} 
\sqrt{1 - \epsilon} & 0 \\ \sqrt{\epsilon \eta} \:\hat{a}_s^\dag+\sqrt{\epsilon(1-\eta)} \:\hat{a}_\ell^\dag  & 1
\end{array} \right) \ket{0 0},
\ee
for Fock states $j = 0,1$ in the lost mode, i.e.
\begin{subequations}
\be 
\hat{M}_{x0} = e^{-X^2/2} \left( \begin{array}{cc}
\sqrt{1-\epsilon} & 0 \\ \sqrt{2 \epsilon \eta} X & 1 
\end{array} \right),
\ee \be 
\hat{M}_{x1} = e^{-X^2/2} \left( \begin{array}{cc}
0 & 0 \\ \sqrt{\epsilon(1-\eta)} & 0 
\end{array} \right),
\ee
\end{subequations}
with a state update rule
\be \label{stateup-eta}
\rho(t+dt) = \frac{\hat{M}_{x0} \rho(t) \hat{M}_{x0}^\dag + \hat{M}_{x1} \rho(t) \hat{M}_{x1}^\dag}{\text{tr}\left(\hat{M}_{x0} \rho(t) \hat{M}_{x0}^\dag + \hat{M}_{x1} \rho(t) \hat{M}_{x1}^\dag \right)}.
\ee
The measured homodyne signal is computed according to projection onto the states $\ket{X}$ exactly as above, and a drive could be added with unitaries in the same manner as above.
The new operators \eqref{mxj} again denote a well--defined measurement, in that they form POVM elements, i.e.
\be 
\int_{-\infty}^{\infty} dX \sum_{j = 0,1} \hat{M}_{xj}^\dag \hat{M}_{xj} \propto \openone.
\ee
We find the same agreement between the expansion of the state update \eqref{stateup-eta} to $O(dt)$, and the SME with finite $\eta$ (converted to its Stratonovich form), as in every case discussed. Thus the description of $\eta$ supposed by Fig.~\ref{fig-exp-diagrams}(d) and \eqref{eta-bs} is entirely equivalent to the description implicit in the SME, and clarifies the meaning of measurement ``inefficiency''. 

\par This picture of inefficiency is also readily connected to scenarios in which several observers simultaneously make measurements, and each gets only partial information \cite{Jacobs2006, Dziarmaga2004, Harrington2019}. One can imagine that an observer lives at each output of the beamsplitter in Fig.~\ref{fig-exp-diagrams}(d), each recieving some proportion of the information about the qubit carried by the decay process as they make measurements. 
If they do not share their results, each will have a different estimate of the qubit's evolution conditioned on their partial information, and tracing out over the other observer's measurement record which they do not have access to. Either of their estimates could be compared to some hypothetical ``true'' evolution which an observer able to access all the relevant measurement records could compute. In practice, it is effectively impossible to have a perfectly efficient measurement in any experiment, and some information is always irretrievably lost to the environment through any channel from which the primary system is not perfectly isolated (generically, this is ``decoherence''). 
The methods we have presented here can readily be adapted to the kind of multiple--observer situation we have just described; this includes situations which involve both jumps and diffusion, due to different observers making different types of measurements (see e.g.~appendix B of \cite{2QFLong}, or \cite{Kuramochi2013}). Such scenarios have recently been fruitfully investigated in the context of quantum state smoothing \cite{Guevara2015, Chantasri2019, Guevara2019}\footnote{Quantum state smoothing is closely related to quantum trajectories; SQTs, as we have presented them in the present text, are a form of ``quantum filtering'' which goes forward in time; in other words, we here only use the measurement record from the system's past to estimate a qubit's state. In the event of an inefficient measurement, quantum state smoothing often allows for a more pure estimate of the system's state to be made at some time, by using the measurement record both before \emph{and after} the time at which the state is estimated.}.

We perform simulations which include measurement inefficiency, which are shown in Fig.~\ref{fig-SimLM}, and discussed further in connection with the ``optimal path'' techniques we develop shortly. 
Measurement inefficiency leads to decay that is qualitatively the same as in the ideal case discussed in Fig.~\ref{fig-simul3meas}, except that instead of trajectories being restricted to pure states on the surface of the Bloch sphere, as in the $\eta = 1$ case, they instead move stochastically on the surface of an ellipsoid which contracts towards $\ket{g}$ over time as information is lost in the $\eta < 1$ case. 
Qualitative agreement between these simulated results shown in Fig.~\ref{fig-SimLM} and those obtained in experiment for either the homodyne \cite{Naghiloo2016flor} or heterodyne \cite{Campagne-Ibarcq2016} detection can be verified at a glance, and a quantitative understanding of this will be developed shortly.

\par As we have now successfully adapted and extended our presentation of basic methods for pertaining to quantum trajectories to the homodyne detection case, and established that they behave correctly, we can proceed by extending our analysis of this system into new examples which can introduce and highlight particular topics in the recent literature.

\section{Special Topics and Further Examples}
We will focus on connections to two areas, using homodyne fluorescence detection as our example of choice; first we describe how this example relates to recent work about the arrow of time in quantum measurement, which connects to work on fluctuation theorems for quantum trajectories, and the growing area of quantum thermodynamics more generally; second we will describe how ``most--likely paths'' can be derived from diffusive quantum trajectory dynamics using a variational principle. 

\subsection{Time reversal symmetry and the arrow of time \label{sec-timereverse}}

How does an arrow of time emerge from microscopically time--reversible physical laws? The issue has been raised in the context of continuous quantum measurements \cite{DresselArrow, SreenathReversal, SreenathFluctuation, Harrington2019}, and applies more broadly across many disciplines within physics~\cite{PhysRevD.32.2489, PhysRevLett.103.080401, lebowitz1993boltzmann}. 
In the quantum measurement case, one could pose this question as a game; a quantum trajectory is shown like a movie, forward and backward, and the goal of the game is to infer the direction in which the movie was originally recorded. 
We will find that the equations of motion are time--symmetric, (e.g.~as in Hamiltonian dynamics), such that both the forward and backward movies both depict legitimate dynamics; this is not the whole story however, as the backward evolution (i.e.~``wavefunction uncollapse'') does not necessarily occur with the same probability \cite{jordan2010uncollapsing,PhysRevLett.97.166805}. 
This leads to a natural discriminator for the arrow of time in terms of the probabilities of occurrence of forward and backward trajectories of the monitored quantum system, as developed in Refs.~\cite{DresselArrow, SreenathReversal}. 
Assuming no prior bias, we could use the measurement record as an additional tool to improve our inference about the direction in which the quantum state dynamics is originally recorded~\cite{DresselArrow} (by analogy, the sound track for a movie could help us understand in which direction it is meant to run). 
Such an approach is fundamentally connected to the time--symmetry of underlying dynamical equations describing the measurement, and connects to the arrow of time analysis pertinent to the thermodynamics of small systems~\cite{chernyak2006path, SreenathFluctuation}. 

Fluorescence appears to exhibit a clear arrow of time, and therefore the time--reversibility of continuously monitored fluorescence dynamics may seem rather surprising. For this reason, here we detail the time symmetry analysis of dynamical equations \eqref{eqmo-hom} which describe homodyne measurement of fluorescence, using the approach presented in Ref.~\cite{SreenathReversal}, wherein similar and detailed analysis was performed for the heterodyne case. 
The time--reversed dynamics can be considered as a legitimate measurement dynamics, starting from the time--reversed final state $\Theta|\psi_{f}\rangle$, evolving through the time--reversed counterpart of the forward sequence of states, back to the time--reversed initial state $\Theta|\psi_{i}\rangle$. 
The measurement operators of the backward dynamics are related to the forward dynamics by a Hermitian conjugate operation, i.e.~$\hat{M}_{B}=\hat{M}_{F}^{\dagger}$; therefore the dynamical equations which describe the backward dynamics are also similar to the retrodicted dynamical equations~\cite{Tan2015}, but starting from the time--reversed final state. 
We may write the retrodicted dynamical equations corresponding to a homodyne measurement, where the quantum state is updated by
\begin{equation}
    \ubar{\rho}(\ubar{t}+\ubar{dt}) = \frac{\hat{M}_{x}^{\dagger}\ubar{\rho}(\ubar{t})\hat{M}_{x}}{\text{tr}[\hat{M}_{x}^{\dagger}\ubar{\rho}(\ubar{t})\hat{M}_{x}]}.
\end{equation}
We have parameterized the single--qubit density matrix $\ubar{\rho}$ with Bloch coordinates according to 
\be 
\ubar{\rho} = \frac{1}{2}\left(\begin{array}{cc} 1+\ubar{z} & \ubar{x}-i\ubar{y} \\ \ubar{x}+i\ubar{y} & 1-\ubar{z} \end{array} \right).
\ee
Using the form of measurement operators given in Eq.~\eqref{MX} The dynamical equations now take the form,
\begin{subequations}\label{eqmo-homback} \be 
\dot{\ubar{x}} = \tfrac{\gamma}{2} \ubar{x} \ubar{z} + r \sqrt{\gamma} \left[ (1- \ubar{z} - \ubar{x}^2)\cos\theta + \ubar{x} \ubar{y} \sin\theta \right],
\ee \be 
\dot{\ubar{y}} = \tfrac{\gamma}{2} \ubar{y} \ubar{z} + r \sqrt{\gamma} \left[ (\ubar{y}^2+\ubar{z}-1) \sin\theta - \ubar{x} \ubar{y} \cos\theta \right],
\ee \be
\dot{\ubar{z}} = \tfrac{\gamma}{2}(\ubar{z}^2-1) + r \sqrt{\gamma} (-\ubar{z}+1)\left[ -\ubar{y} \sin\theta + \ubar{x} \cos\theta \right].
\ee \end{subequations}
Note that the retrodicted equations under the time-reversal operation, $\ubar{x}\rightarrow-x,~\ubar{y}\rightarrow-y,~\ubar{z}\rightarrow-z$, and $\ubar{t}\rightarrow T-t$ (i.e., $\ubar{dt}\rightarrow-dt$) looks exactly like the forward dynamical equations \eqref{eqmo-hom}, demonstrating their time--reversal invariance; we have eliminated the drive characterized by $\Omega$ and $\delta$ in the equations above for brevity, but including it does not affect the result.
Time reversal symmetry of the dynamical equations suggests that the forward dynamics and the reverse dynamics both represent a physical quantum trajectory on the Bloch sphere. 
Given the measurement record, one can associate a probability each to the forward and backward trajectories, which can be used to infer an arrow of time for the measurement dynamics, and subsequently characterize the irreversibility of homodyne measurement of fluorescence using the associated fluctuation theorems \cite{DresselArrow, SreenathReversal, SreenathFluctuation}. 

We can expand on this story somewhat by noting that our diffusive trajectories in Fig.~\ref{fig-simul3meas}(b,c) do not diffuse monotonically downward from $\ket{e}$ towards $\ket{g}$. 
This suggests that the measurement process can actually cause the probability of the qubit being found in the more energetic of its states to rise in some realizations. 
While the average decays monotonically, fluctuations make this question of an arrow of time non--trivial in individual realizations; the probability of sustained re-excitation over a long period is low, but estimates of the arrow of time using only a short window of the evolution cannot necessarily be made with high confidence; rare events can decieve, and something resembling a ``wavefunction uncollapse'' is not merely hypothetical in this system. 
Such behavior has been noted in the literature \cite{Bolund2014}; while it may be initially intuitively challenging, this effect is perfectly correct, and reflects the nature of the information we get about the field when we make a weak ($dt \ll T_1$) quadrature measurement, and its backaction. 
A truly detailed description of the thermodynamics of quantum measurements or trajectories falls beyond our present scope, but is a fascinating area related to the questions we have discussed here, and enjoying increased recent research interest \cite{Cottet2017, Elouard2017_QTherm, Elouard2017_Maxwell, Mahdi2017Qtherm, CyrilEngine2018, Masuyama2018, Monsel2018, Monsel2019, Mohammady2019}. We encourage the curious reader to explore further.

\subsection{A Variational Principle for Quantum Trajectories \label{sec-OPmain}}

Optimal paths (OPs) \cite{Chantasri2013, Chantasri2015, Jordan2015flor, Areeya_Thesis, Lewalle2016, Lewalle2018} have recently been used to elucidate a variety of quantum trajectory phenomena. We will here give a brief overview of their derivation, using the CDJ (Chantasri/Dressel/Jordan) path integral \cite{Chantasri2013}, and then apply the formalism to the homodyne fluorescence examples we have developed above. 

\subsubsection{Derivation of Optimal Paths \label{sec-OP-derivation}}

\par OPs can be understood as the path extremizing the probability to get from one given quantum state $\mathbf{q}_i$ to another $\mathbf{q}_f$ in a particular time interval, under the dynamics due to backaction from the continuous weak quantum measurement. The vector $\mathbf{q}$ parameterizes the quantum state, and here denotes coordinates on the Bloch sphere. Typically OPs will be most--likely paths (MLPs), which maximize the probability of the measurement record connecting the given boundary conditions according to an action--extremization principle. OPs should be confused neither with a globally most--likely path (i.e.~the particular MLP post--selected on the most likely final state after a given time interval), or with an average path. Details about numerical procedures to extract an MLP from data, which corresponds to the theory we are about to develop, can be found in appendix \ref{sec-OPverify}.  

\par To begin, we explain how the path probability can be written in terms of an effective action, which can then be extremized according to a variational principle. 
We may write an expression for the probability of a quantum trajectory, which moves from $\mathbf{q}_i$ to $\mathbf{q}_f$ through a discrete sequence of measurements as
\begin{widetext}
\be \begin{split} \label{jointp}
\mathcal{P} 
(\lbrace \mathbf{q} \rbrace, \lbrace \mathbf{r} \rbrace|\mathbf{q}_i,\mathbf{q}_f) = \delta(\mathbf{q}_i - \mathbf{q}_0) \delta(\mathbf{q}_f - \mathbf{q}_n)
\left\lbrace
\prod_{k=0}^{n-1} \wp(\mathbf{r}_{k}|\mathbf{q}_k) \wp(\mathbf{q}_{k+1}| \mathbf{q}_k,\mathbf{r}_k) \right\rbrace .
\end{split} \ee
The $\delta$--functions at the initial and final points impose the boundary conditions. The indices $k$ run over time, such that if $\rho_k = \rho(t)$, then $\rho(t+dt) = \rho_{k+1}$ and so on. 
The stochastic element of the dynamics arises in drawing the readout from the probability density determined by the denominator of the state update expression \eqref{state-update}, i.e.~$\wp(\mathbf{r}_k|\rho_k) \propto \text{tr}(\hat{M}_{\mathbf{r}_k} \rho_k \hat{M}_{\mathbf{r}_k}^\dag)$, where $\hat{M}_{\mathbf{r}_k}$ could generically be any Kraus operator describing a weak measurement.
We describe the deterministic update of the quantum state \emph{given} the stochastic readout $\mathbf{r}_k$ according to $\wp(\mathbf{q}_{k+1}|\mathbf{q}_k,\mathbf{r}_k) = \delta(\mathbf{q}_{k+1} - \mathbf{q}_k - dt\: \mathcal{F}[\mathbf{q}_k,\mathbf{r}_k])$, where $\dot{\mathbf{q}} = \mathcal{F}[\mathbf{q},\mathbf{r}]$ is an equation of motion, e.g.~like \eqref{eqmo-hom}.
Recall that a $\delta$--function may be written
$
\delta(\mathbf{q}) = (2\pi i)^{-\text{dim}(\mathbf{q})} \int_{-i\infty}^{i\infty} d\mathbf{p} \: \exp\left[ - \mathbf{p}\cdot \mathbf{q} \right]
$, where $\text{dim}(\mathbf{q})$ is the dimension of $\mathbf{q}$, and $d\mathbf{p} = dp_1\; dp_2\; ... \; dp_{\text{dim}(\mathbf{q})}$.
We apply this identity to \emph{all} $\delta$--functions in \eqref{jointp}, such that in the time--continuum limit we have a Feynman--like path integral, in which we effectively sum over all possible quantum trajectories which obey the given boundary conditions, i.e.
\be \begin{split}
\mathcal{P} &= \limit{n}{\infty} \limit{dt}{0} \mathcal{N} \idotsint\limits_{-i\infty}^{i\infty} \left(\prod_{k=0}^{n-1} d\mathbf{p}_k \right) \exp\bigg[ \mathcal{B}
+ \sum_{k=0}^{n-1} \left(-\mathbf{p}_k \cdot (\mathbf{q}_{k+1}  - \mathbf{q}_k - dt \: \mathcal{F}_k) + \ln \wp(\mathbf{r}_k|\mathbf{q}_k) \right) \bigg] \\
&\propto \int \mathcal{D}[\mathbf{p}] \exp \left[\mathcal{B} + \int_0^T dt \left( - \mathbf{p}\cdot \dot{\mathbf{q}} + \mathbf{p}\cdot \mathcal{F}[\mathbf{q},\mathbf{r}] + \mathcal{G}[\mathbf{q},\mathbf{r}] \right) \right] 
\\ &= \int \mathcal{D}[\mathbf{p}] \exp\left[\mathcal{B} + \int_0^T dt( \mathcal{H}(\mathbf{q},\mathbf{p},\mathbf{r}) - \mathbf{p}\cdot \dot{\mathbf{q}}) \right] 
= \int \mathcal{D}[\mathbf{p}] \exp\left(\mathcal{B}+ \mathcal{S}[\mathbf{q},\mathbf{p},\mathbf{r}] \right)
\end{split} \ee 
for $\mathcal{N} = (2\pi i)^{-(n+2) \cdot\text{dim}(\mathbf{q})}$, and where $\mathcal{D}[\mathbf{p}]$ arises from the infinite product of the $d\mathbf{p}_k (2\pi i)^{-\text{dim}(\mathbf{q}_k)}$. We use the shorthand $\mathcal{B} = -\mathbf{p}_{-1}\cdot(\mathbf{q}_0 - \mathbf{q}_i) - \mathbf{p}_n\cdot(\mathbf{q}_n-\mathbf{q}_f)$ for the boundary terms, and the shorthand $\mathcal{G}$ for the expansion to $O(dt)$ of the log--probability for the readouts $\ln \wp(\mathbf{r}|\mathbf{q})$ (see e.g.~\eqref{homG}). This relates a trajectory probability $\mathcal{P}$ to a ``stochastic action'' $\mathcal{S}$. That action is expressed in terms of a Hamiltonian $\mathcal{H} = \mathbf{p}\cdot\mathcal{F} + \mathcal{G}$.

\par We can then see that extremizing the probability corresponds to extremizing $\mathcal{S}$. Action extremization in this case can be expressed much the same way as it is in classical mechanics, such that 
\be \begin{split}
\delta \mathcal{S} &= 0 \\
& = \delta\int_0^T dt \left(-\mathbf{p}\cdot\dot{\mathbf{q}} + \mathcal{H}(\mathbf{q},\mathbf{p},r) \right)  = \int_0^T dt \left( \partl{\mathcal{H}}{\mathbf{q}}{} \cdot \delta\mathbf{q} + \partl{\mathcal{H}}{\mathbf{p}}{} \cdot \delta\mathbf{p} + \partl{\mathcal{H}}{\mathbf{r}}{} \cdot \delta \mathbf{r} - \mathbf{p} \cdot \delta\dot{\mathbf{q}} - \dot{\mathbf{q}} \cdot \delta\mathbf{p} \right) \\
& =\mathbf{p} \cdot \delta\mathbf{q}|_0^T + \int_0^T dt \bigg[ \delta\mathbf{q} \cdot \left(\partl{\mathcal{H}}{\mathbf{q}}{} + \dot{\mathbf{p}} \right) + \delta\mathbf{p} \cdot \left(\partl{\mathcal{H}}{\mathbf{p}}{} -\dot{\mathbf{q}} \right) + \delta \mathbf{r} \cdot \partl{\mathcal{H}}{\mathbf{r}}{} \bigg],
\end{split} \ee
\end{widetext}
which indicates that the OPs obey the equations 
\be \label{OP-general-eqmo}
\dot{\mathbf{q}} = \partl{\mathcal{H}}{\mathbf{p}}{}, \quad \dot{\mathbf{p}} = - \partl{\mathcal{H}}{\mathbf{q}}{}, \quad \partl{\mathcal{H}}{\mathbf{r}}{}\bigg|_{\mathbf{r}^\star} = 0.
\ee
These are Hamilton's usual equations in the Bloch coordinates $\mathbf{q}$ and conjugate variables $\mathbf{p}$, plus an additional equation which stipulates that the stochastic readouts be optimized, leading to some smooth $\mathbf{r}^\star$ instead of the stochastic $\mathbf{r}$. The OPs are thus smooth curves, and are solutions to a Hamiltonian dynamical system of ordinary, rather than stochastic, differential equations. 
The OPs are themselves possible quantum trajectories (since $\dot{\mathbf{q}} = \mathcal{F}$ is preserved, by construction), even though the optimal readouts $\mathbf{r}^\star$ are smooth functions of time, rather than the stochastic readouts which occur in individual runs of an experiment. 
The Hamiltonian structure of the OP dynamics implies that, absent any explicitly time--dependent parameters, there is a conserved ``stochastic energy'' $E = \mathcal{H}$ associated with an OP.

The ``momenta'' $\mathbf{p}$ conjugate to the generalized Bloch coordinates $\mathbf{q}$, which arise in this optimization process, warrant further attention. 
The $\mathbf{p}$ are not directly measurable, but play a substantive mathematical role, in that they effectively generate displacements in the quantum state $\mathbf{q}$ according to the optimal measurement record $\mathbf{r}^\star$. 
They appear as the variables conjugate to $\mathbf{q}$ in the Fourier representation of the $\delta$--functions, and could be understood and Lagrange multipliers in the optimization process.
They are perhaps best understood in analogy with classical optics, however: 
just as we have expressed a diffusive process in terms of an action, the underlying wave process in classical optics can be expressed as an action; extremization in the latter scenario leads to a ray description. 
Our OPs are related to the underlying diffusive process described above in much the same way that a ray description of light is related to the underlying wave optics process\footnote{We are grateful for comments by Prof.~Miguel Alonso which helped us to clarify this point in our own thinking.}.
Diffusive SQTs arrive at final states with a variety of probabilities (and the action representing this is real), whereas optical paths leading to different positions arrive with different phases, exhibiting interference (and the subsequent action in this case is imaginary). 
Continuing with this analogy, we could imagine that rays leave a source in different directions, i.e.~with different wave--vectors (OPs leave their initial state $\mathbf{q}_i$ with a range of initial momenta $\mathbf{p}_i$); 
we may pick out a particular ray or subset of rays by choosing a particular initial wave--vector, or by choosing a final position to which they connect (a particular OP can be selected by choosing a value of $\mathbf{p}_i$, or by choosing a $\mathbf{q}_f$ at a later time). 
Thus, the degree of freedom in choosing an initial momentum $\mathbf{p}_i$ is the same degree of freedom which allows for post--selections to many $\mathbf{q}_f$; the mapping between the two is not necessarily one--to--one (see our work on ``multipaths'' for clarification of this point \cite{Lewalle2016, Mahdi2016, Lewalle2018}). 
In a phase space with $N$ coordinates $\mathbf{q}$ and $N$ momenta $\mathbf{p}$ (a $2N$--dimensional phase space), the set of paths evolving from all $\mathbf{p}_i$ with a fixed $\mathbf{q}_i$ defines an $N$--dimensional Lagrangian manifold (LM) within the phase space; such a manifold can be understood as containing all of the possible dynamics originating from the state $\mathbf{q}_i$ in the OP description, i.e.~such an LM explores the full space of optimal readouts $\mathbf{r}^\star$, just as the underlying diffusive process may explore the full space of stochastic measurement records\footnote{The LM in question has primarily been used in the context of multipath dynamics \cite{Lewalle2016, Mahdi2016, Lewalle2018}; 
there the main concern is whether the projection of the time--evolved LM out of the full OP phase space down into the $\mathbf{q}$--space of final quantum states is one--to--one (a single MLP connects the initial state to the chosen final state) or many--to--one (in which case many OPs may connect the boundary conditions, typically corresponding to different clusters of SQTs in the post--selected distribution).}. 
Our focus now will be on applying this formalism to our homodyne fluorescence measurement; some work in this vein, albeit with a different emphasis, appears in Ref.~\cite{Mahdi2016}.

\subsubsection{Optimal Paths for Homodyne Fluorescence Trajectories}

Notice that the system of equations \eqref{eqmo-hom} can be simplified straightforwardly; $\dot{y} = 0$ if $y = 0$ and $\theta = 0$, and then all the dynamics are in the $xz$--plane of the Bloch sphere. It is then easy to write down a stochastic Hamiltonian  
\be \begin{split}
\mathcal{H}_{hom}^{xz} = & \quad p_x \left( \tfrac{\gamma}{2} x z  + r \sqrt{\gamma} \left[ 1+ z - x^2 \right] \right) \\ &+ p_z \left( \tfrac{\gamma}{2}(z^2-1)  - r \sqrt{\gamma} (z+1)x \right) \\
& - \tfrac{1}{2}\left( r - \sqrt{\gamma} x \right)^2 - \tfrac{\gamma}{2}(1+z-x^2),
\end{split} \ee
for the OP dynamics using formulas we have already derived above (with $\Omega = 0 = \delta$, and $\eta =1$) to describe the ideal measurement, its backaction, and statistics\footnote{Note that we have derived both $\mathcal{F}$ and $\mathcal{G}$ using regular calculus, and are thereby effectively using the Stratonovich form of $\mathcal{F}$, \emph{not} the It\^{o} form which arises directly from the SME \eqref{SME}. 
Using a form of $\mathcal{F}$ which does not transform according to regular calculus would prevent us from performing our OP analysis using typical approaches of classical mechanics (e.g.~canonical transformations), which we find quite undesireable.}.
The optimal readout obeys $\partial_r \mathcal{H}_{hom}^{xz}|_{r^\star} = 0$, which we solve to obtain
\be 
r^\star= \sqrt{\gamma} \left( x + p_x (1+z-x^2)-x p_z (1+z) \right);
\ee
we see that we have the signal $\sqrt{\gamma}x$, plus some additional terms which depend on the conjugate momenta $p_x$ and $p_z$, which implement the optimized effect of the noise (as discussed above).

\begin{figure}
    \includegraphics[width=\columnwidth,trim = {0 0 0 15},clip]{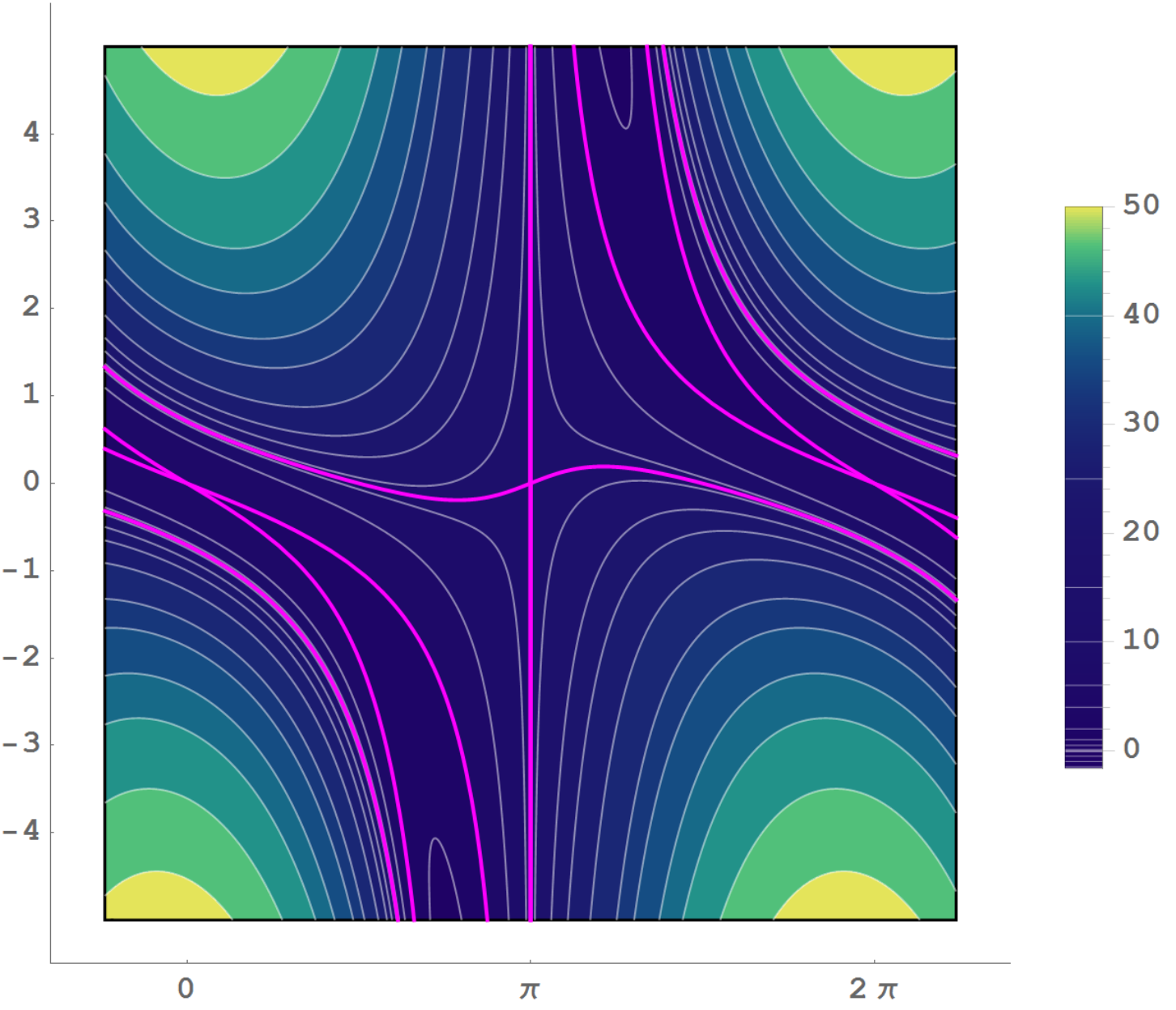}
    \begin{tikzpicture}[overlay]
    \draw[fill = white, draw = white] (-4,6.83) -- (-3.55,6.83)  arc[radius = 0.62, start angle = 180, end angle = 90] -- (-3,8) -- (-4,8) -- cycle;
    \draw[fill = white, draw = white] (2.1,7.83) -- (2.09,7.435)  arc[radius = 0.62, start angle = 90, end angle = 0] -- (2.8,6) -- (3,6) -- (3,8) -- cycle;
    \draw[fill = white, draw = white] (-4,1.683) -- (-3.55,1.675)  arc[radius = 0.62, start angle = 180, end angle = 270] -- (-2.96,0.7) -- (-4,0.7) -- cycle;
    \draw[fill = white,draw = white] (2.11,0.7) -- (2.11,1.067)  arc[radius = 0.609, start angle = 270, end angle = 360] -- (3.2,0.7) -- cycle;
    \draw[line width = 1.2,rounded corners = 20] (-3.55,1.035) rectangle (2.7,7.46);
    \draw[magenta, line width = 0.8, ->] (-2,4.332) -- (-1.9,4.302);
    \draw[magenta, line width = 0.8, ->] (1.1,4.168) -- (1,4.198);
    \draw[magenta, line width = 0.8, ->] (2.2,4.652) -- (2.3,4.612);
    \draw[magenta, line width = 0.8, ->] (-0.425,6) -- (-0.425,6.1);
    \draw[magenta, line width = 0.8, ->] (-0.425,2.4) -- (-0.425,2.3);
    \draw[magenta, line width = 0.8, ->] (-3.0,3.81) -- (-3.1,3.86);
    \draw[magenta, line width = 0.8, ->] (-1.7,3.0) -- (-1.76,3.1);
    \draw[magenta, line width = 0.8, ->] (-1.055,2.7) -- (-1.015,2.6);
    \draw[magenta, line width = 0.8, ->] (0.29,5.6) -- (0.24,5.7);
    \draw[magenta, line width = 0.8, ->] (1.005,5.2) -- (1.06,5.1);
    \draw (-2.945,0.7) -- (-2.945,0.8);
    \draw (2.1,0.7) -- (2.1,0.8);
    \draw (-3.96,1.68) -- (-3.86,1.68);
    \draw (-3.96,6.805) -- (-3.86,6.805);
    \end{tikzpicture}\\
    \begin{picture}(0,0)
    \put(85,200){$E~(\mathrm{MHz})$}
    \put(70,30){$\vartheta$}
    \put(-112,216){$p$}
    \put(-80,160){\color{magenta} \tt X+}
    \put(65,160){\color{magenta} \tt X+}
    \put(-90,96){\color{magenta} \tt X-}
    \put(55,96){\color{magenta} \tt X-}
    \put(-31,114){\color{magenta} \tt Y+}
    \put(1,142){\color{magenta} \tt Y-}
    \put(-96,126){\color{magenta} \tt Y-}
    \put(66,131){\color{magenta} \tt Y+}
    \put(4,195){\color{magenta} \tt Z-}
    \put(-37,63){\color{magenta} \tt Z+}
    \end{picture} \\ \vspace{-10pt}
    \includegraphics[width=\columnwidth,trim = {1 0 7 13},clip]{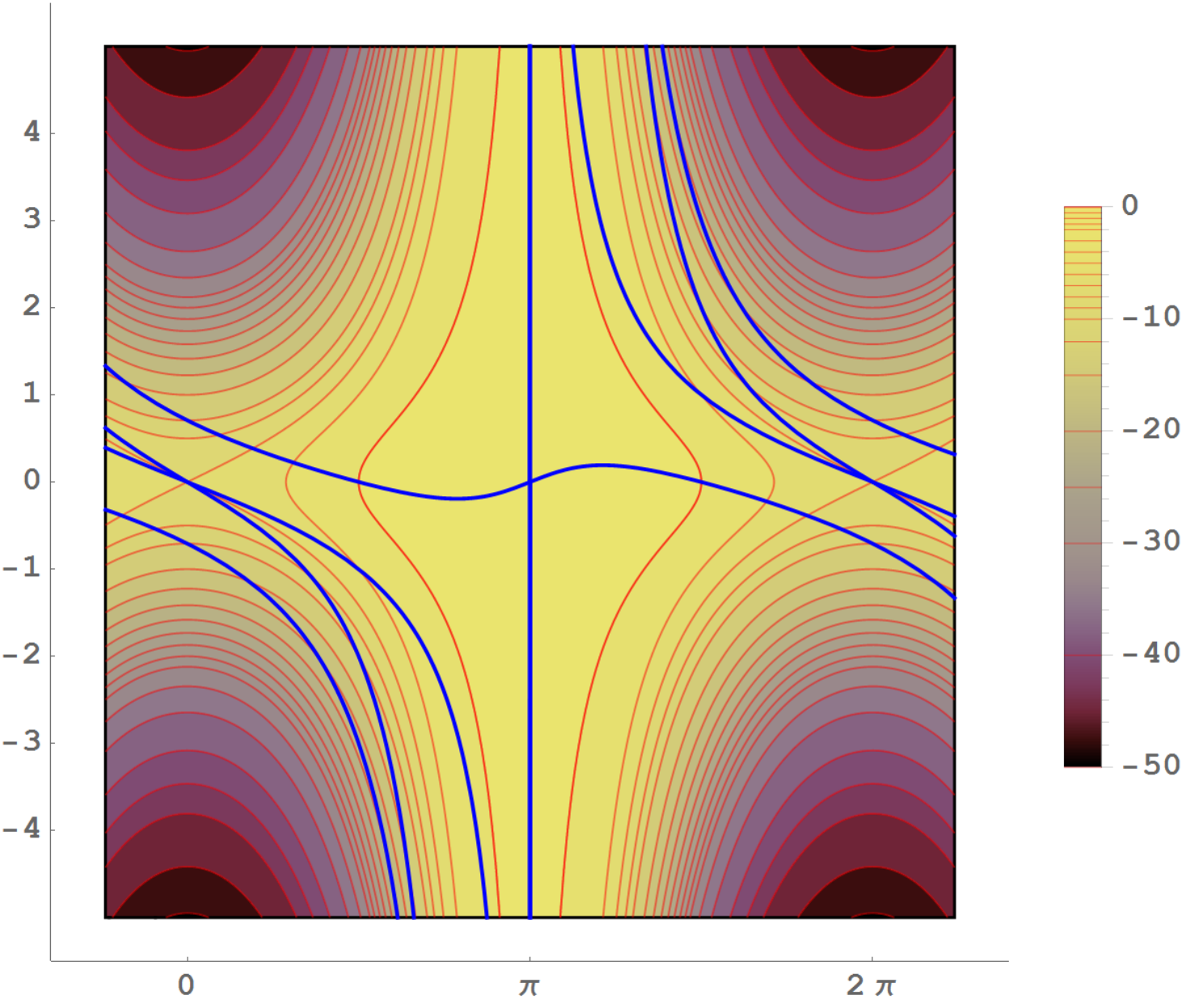}
    \begin{tikzpicture}[overlay]
    \draw[fill = white, draw = white] (-4,6.81) -- (-3.55,6.83)  arc[radius = 0.62, start angle = 180, end angle = 90] -- (-3,7.9) -- (-4,7.9) -- cycle;
    \draw[fill = white, draw = white] (2.1,7.83) -- (2.06,7.44)  arc[radius = 0.62, start angle = 90, end angle = 0] -- (2.8,6) -- (3,6) -- (3,8) -- cycle;
    \draw[fill = white, draw = white] (-4,1.683) -- (-3.55,1.675)  arc[radius = 0.62, start angle = 180, end angle = 270] -- (-2.96,0.7) -- (-4,0.7) -- cycle;
    \draw[fill = white,draw = white] (2.08,0.7) -- (2.07,1.06)  arc[radius = 0.609, start angle = 270, end angle = 360] -- (3.2,0.7) -- cycle;
    \draw[line width = 1.2,rounded corners = 20] (-3.56,1.035) rectangle (2.67,7.46);
    \draw[blue, line width = 0.8, ->] (-2.025,4.325) -- (-1.925,4.295);
    \draw[blue, line width = 0.8, ->] (1.075,4.16) -- (0.975,4.19);
    \draw[blue, line width = 0.8, ->] (2.175,4.635) -- (2.275,4.595);
    \draw[blue, line width = 0.8, ->] (-0.45,6) -- (-0.45,6.1);
    \draw[blue, line width = 0.8, ->] (-0.45,2.4) -- (-0.45,2.3);
    \draw[blue, line width = 0.8, ->] (-3.025,3.81) -- (-3.125,3.86);
    \draw[blue, line width = 0.8, ->] (-1.725,3.0) -- (-1.785,3.1);
    \draw[blue, line width = 0.8, ->] (-1.08,2.7) -- (-1.04,2.6);
    \draw[blue, line width = 0.8, ->] (0.265,5.58) -- (0.215,5.68);
    \draw[blue, line width = 0.8, ->] (0.97,5.2) -- (1.03,5.1);
    \draw (-2.945,0.7) -- (-2.945,0.8);
    \draw (2.07,0.7) -- (2.07,0.8);
    \draw (-3.98,1.69) -- (-3.88,1.69);
    \draw (-3.98,6.795) -- (-3.88,6.795);
    \end{tikzpicture}\\ 
    \begin{picture}(0,0)
    \put(85,200){$\dot{\mathcal{S}}~(\mathrm{MHz})$}
    \put(70,30){$\vartheta$}
    \put(-112,216){$p$}
    \put(-80,160){\color{blue} \tt X+}
    \put(65,160){\color{blue} \tt X+}
    \put(-90,96){\color{blue} \tt X-}
    \put(55,96){\color{blue} \tt X-}
    \put(-31,114){\color{blue} \tt Y+}
    \put(2,142){\color{blue} \tt Y-}
    \put(-96,126){\color{blue} \tt Y-}
    \put(66,131){\color{blue} \tt Y+}
    \put(4,195){\color{blue} \tt Z-}
    \put(-36,63){\color{blue} \tt Z+}
    \end{picture} \vspace{-20pt} \\
    \caption{We show the OP phase space in $\vartheta$ and $p$, for an ideal homodyne measurement of the qubit fluorescence, and with $\gamma = 1~\mathrm{MHz}$ and $\Omega = 0$. We plot the stochastic Hamiltonian in the top panel. Contours are lines of constant ``stochastic energy'' $E = \mathcal{H}^{\vartheta\star}_{hom}$ in $\mathrm{MHz}$, which are solutions to the OP dynamics. Fixed points appear at $\vartheta = 0$ and $\pi$, at $p = 0$; the separatrices which pass through these points are shown in magenta. Those separatrices bound off distinct regions {\tt X}, {\tt Y}, and {\tt Z}; each has some distinct behavior, all of them ultimately lead their paths to the ground state $\theta = \pi$. The regions come in pairs due to the symmetry of the phase space, where {\tt +} denotes that the paths in that region approach the ground state from below (with $\vartheta$ increasing), and their mirror images {\tt -} approach from above (with $\vartheta$ decreasing). We plot $\dot{\mathcal{S}} = \mathcal{H}^{\vartheta\star}_{hom} - p\dot{\vartheta}$ in the bottom panel; this quantity can be regarded as an approximate rate of probability decay, such that paths which spend time in regions of more negative $\dot{\mathcal{S}}$ correspond to sequences of measurement results which are relatively less likely. The OP phase space regions are overlaid on the bottom plot in blue for reference. Comparing the two panels, we see that we can associate paths with larger stochastic $E$ with less--likely dynamics (as a rule of thumb).}
    \label{fig-homPS}
\end{figure}
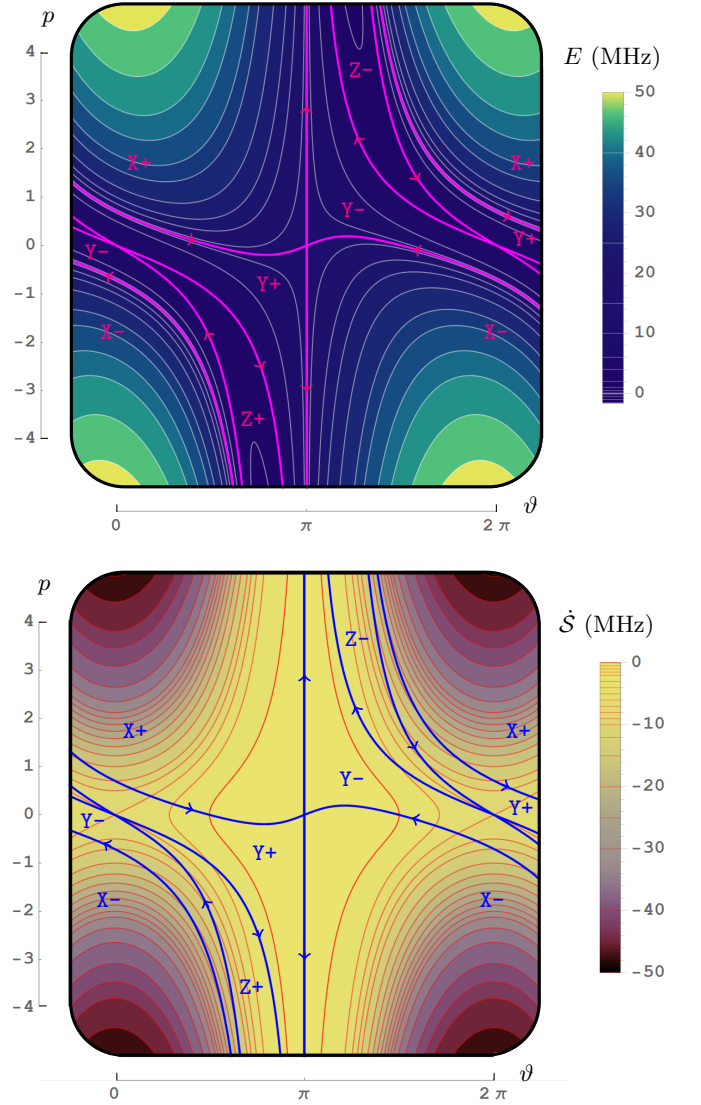

\par We can simplify the equations even more. Consider a change to polar coordinates according to the canonical transformation 
\be 
\begin{array}{lcc}
x & \rightarrow & R \sin \vartheta \\
z & \rightarrow & R \cos \vartheta \\
p_x & \rightarrow & p_R \sin \vartheta + p \cos\vartheta /R \\
p_z & \rightarrow & p_R \cos \vartheta - p \sin\vartheta /R
\end{array}
\ee
which preserves the Poisson brackets between all of the pairs of conjugate variables. Then we see that for the choice $R = 1$ and $p_R = 0$, we have $\dot{R} = \partial_{p_R} \mathcal{H}_{hom}^{\star,R\vartheta} = 0$, meaning that we can look at dynamics purely on the great circle of the Bloch sphere at $R = 1$ (pure states), where states are parameterized entirely by a single coordinate $\vartheta$ (with $\vartheta = 0 \leftrightarrow \ket{e}$ and $\vartheta = \pi \leftrightarrow \ket{g}$). Making this transformation, and subtituting in the optimal readout such that $\mathcal{H}^\star= \mathcal{H}|_{r=r^\star}$, we obtain the Hamiltonian 
\be \begin{split} \label{H_OP_hom}
\mathcal{H}_{hom}^{\vartheta \star} = &p^2 \left(\gamma  \cos \vartheta  +\frac{1}{4} \gamma  \cos (2 \vartheta )+\frac{3 \gamma }{4}\right)\\ & +p \left(\frac{3}{2} \gamma  \sin \vartheta +\frac{1}{2} \gamma  \sin (2 \vartheta )\right)\\ & -\frac{1}{2} \gamma  \cos \vartheta -\frac{1}{4} \gamma  \cos (2 \vartheta )-\frac{\gamma }{4},
\end{split} \ee
which generates the OPs in the simplest case we can consider for this system.

\par The phase space for this Hamiltonian is plotted in Fig.~\ref{fig-homPS}, along with the time--derivative of the stochastic action $\mathcal{S}$
which is extremized by the OP dynamics (effectively, $\dot{\mathcal{S}}$ gives an approximate representation of the probability cost involved with traversing certain regions of the OP phase space). A careful reading of these plots can provide an insightful overview of the system dynamics. 
First, we can immediately infer a rule of thumb: OPs with higher stochastic energy generically correspond to events which occur with lower probabilities (this is true to the extent that regions of large $E$ correspond to regions of more--negative $\dot{\mathcal{S}}$). Secondly, we see that all paths in the OP phase space eventually approach $\ket{g}$ ($\vartheta = \pi$) in the long--time limit, as we expect they must; there are possibilities for this to occur in either direction around the Bloch sphere with some probabilities, but these pure--state OPs never cross through $\vartheta = \pi$. 
The uni--directionality of the flow towards $\ket{g}$ after $t \gg T_1$ reflects our intuition that there should be a statistical arrow of time in the measurement--induced dynamics, as is discussed above and in detail elsewhere \cite{SreenathReversal, SreenathFluctuation}. 
A particular point in the phase space is worthy of further attention; the unstable fixed point at $\vartheta = 0$ and $p = 0$ describes an OP which is stationary at $\ket{e}$ for all time; this does not violate our intuition however, since the probability cost involved with sitting at that point is greater than for sitting very close to $\ket{g}$, such that it is still virtually impossible to post--select on a state still at $\ket{e}$ after $t \gg T_1$. 
As noted before, it is possible for paths to start near the ground state, and re--excite, passing through $\ket{e}$ before asymptotically approaching $\ket{g}$ again around the other side of the Bloch sphere; while such behavior corresponds to relatively rare events (a relatively low--probability post--selection is required), the possibility of such events is readily visible in the OP phase portrait\footnote{For example, starting at $\vartheta_0 = \pi-1$, a globally more--likely path which we would regard as typical might arise from post--selecting on $\vartheta_f = \pi -0.01$ ($\approx$ ground) at some later time, picking out a solution from {\tt X+}. 
However, it is possible to post--select on some state like $\vartheta_f = -\pi+0.01$ (also $\approx$ ground) instead, selecting a path from {\tt X-} or {\tt Y-}; this reveals the possibility of a much rarer set of events, corresponding to an OP which circles back through the excited state before decaying towards ground, from the opposite direction as compared with the more typical set of paths. Post--selections drawing out these dynamics can generically select OPs from regions {\tt X} or {\tt Y} of the phase portrait, as detailed in Fig.~\ref{fig-homPS}; those in region {\tt Z} can only partially re-excite, before turning around to decay.}. 
Further details about OPs this system, for the case of $\Omega \neq 0$, can be found in \cite{Mahdi2016}, and more detailed investigations of the corresponding heterodyne cases can be found in \cite{Jordan2015flor}.

\subsection{Optimal Paths for Inefficient Measurements, and Connections to Experiments \label{sec-OPineff}}

\begin{figure*}
\begin{picture}(1,235)
\put(-258,110){
\begin{tabular}{cc}
\includegraphics[width = 0.235\textwidth, trim = {5 5 15 10},clip]{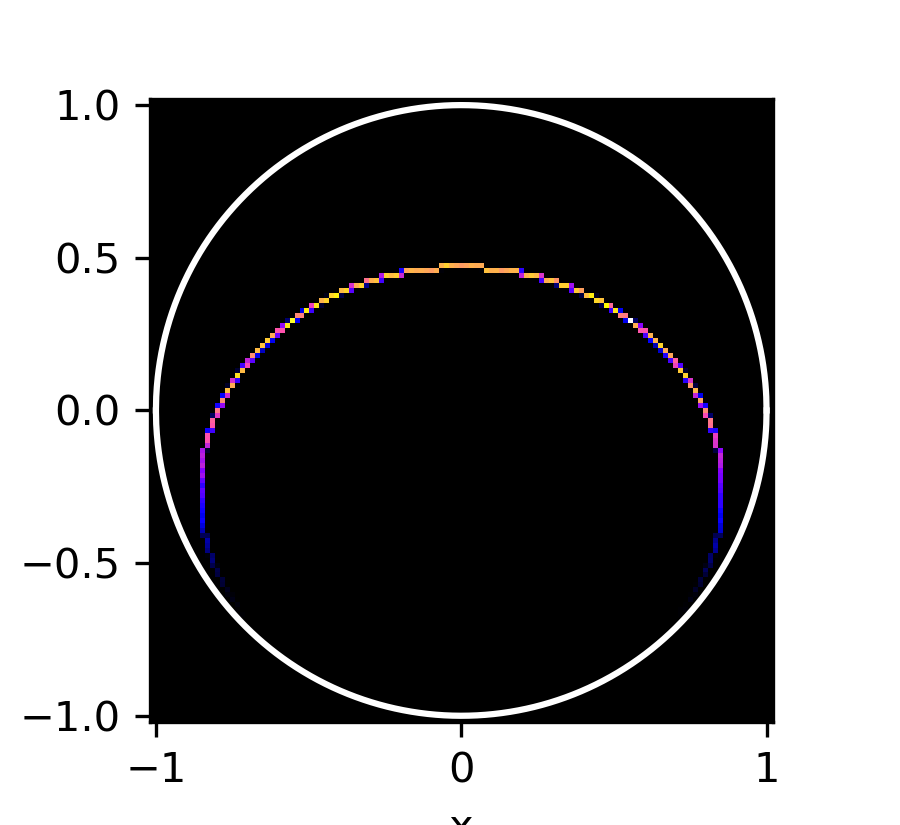} &
\includegraphics[width = 0.235\textwidth, trim = {5 5 15 10},clip]{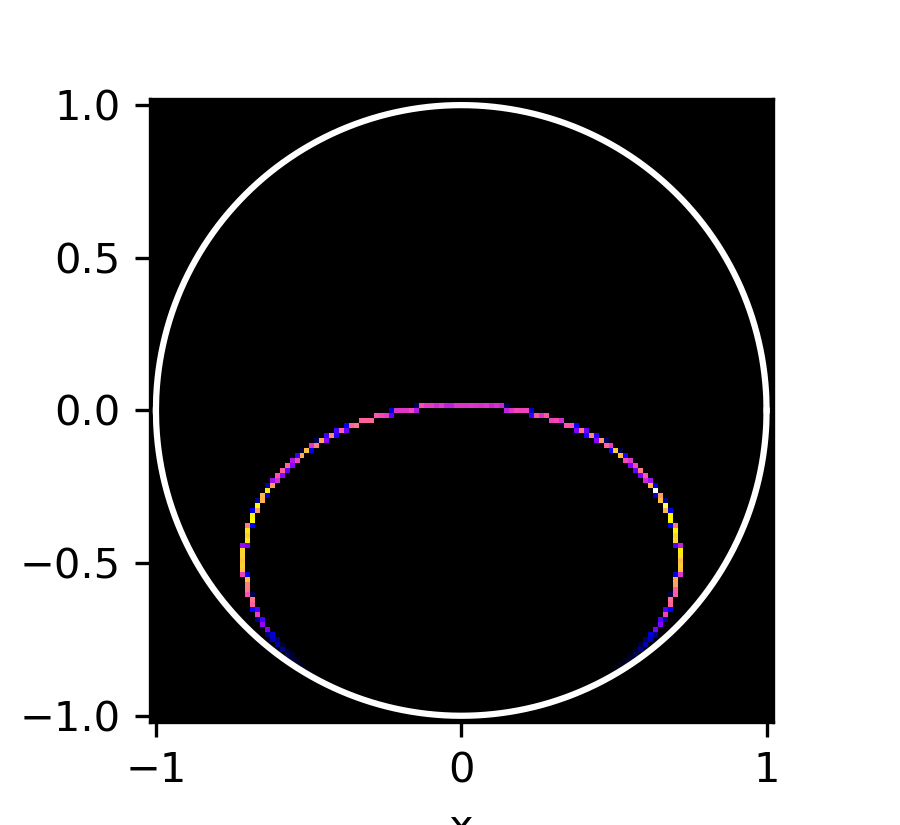} \\
\includegraphics[width = 0.235\textwidth, trim = {5 5 15 10},clip]{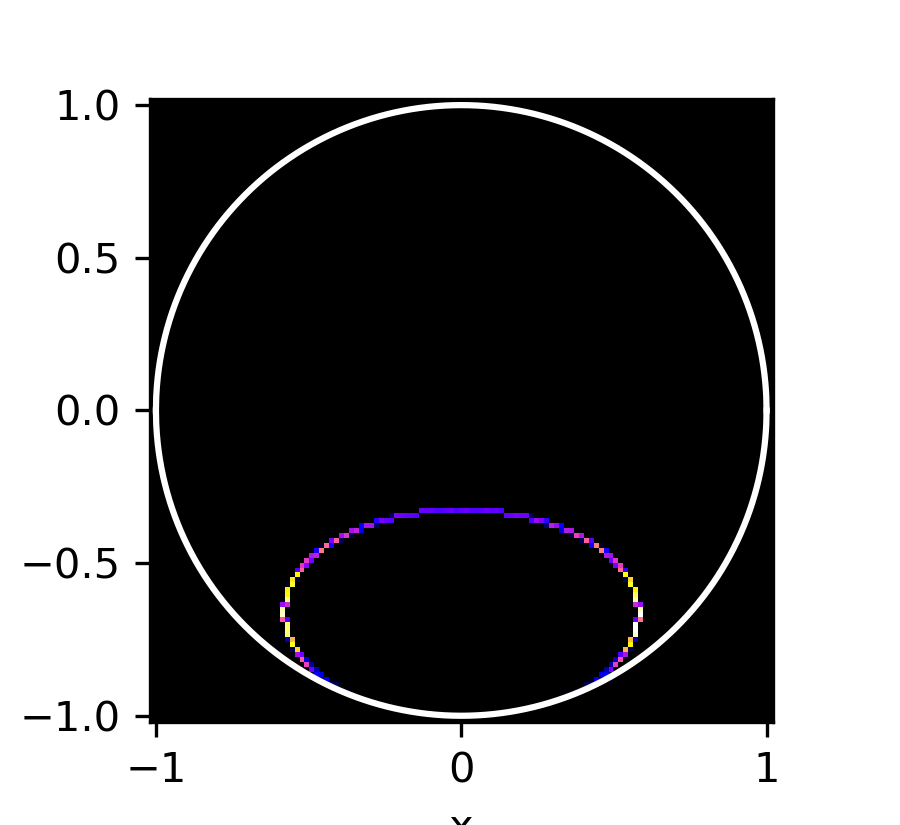} &
\includegraphics[width = 0.235\textwidth, trim = {5 5 15 10},clip]{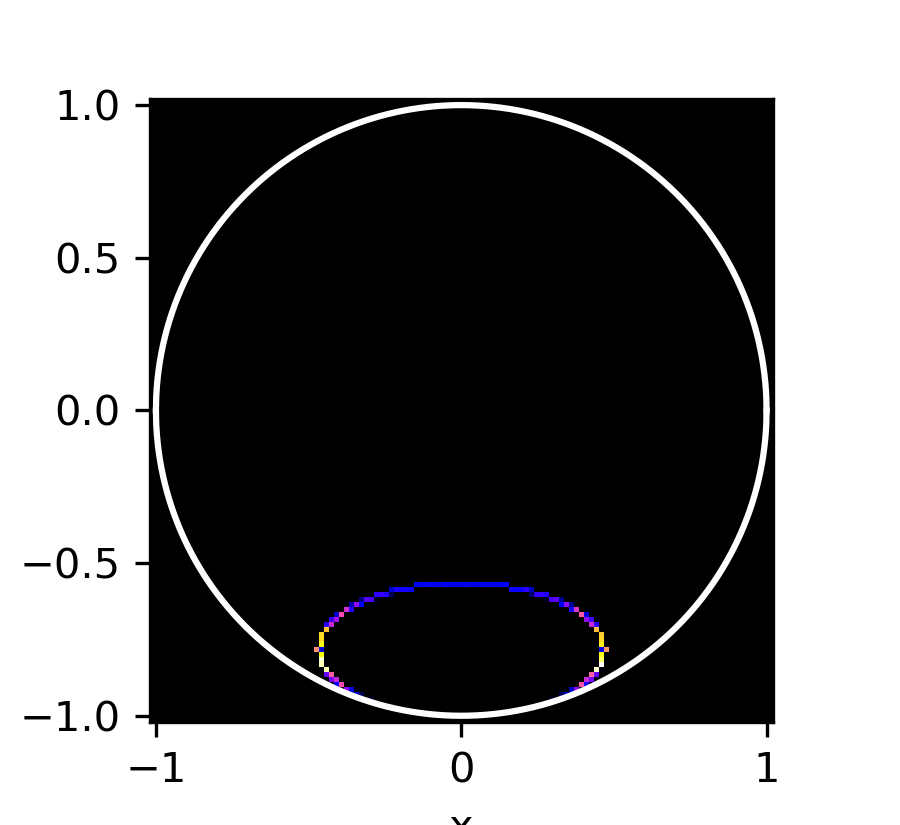}
\end{tabular}}
\put(212,25){\includegraphics[height = .25\textheight]{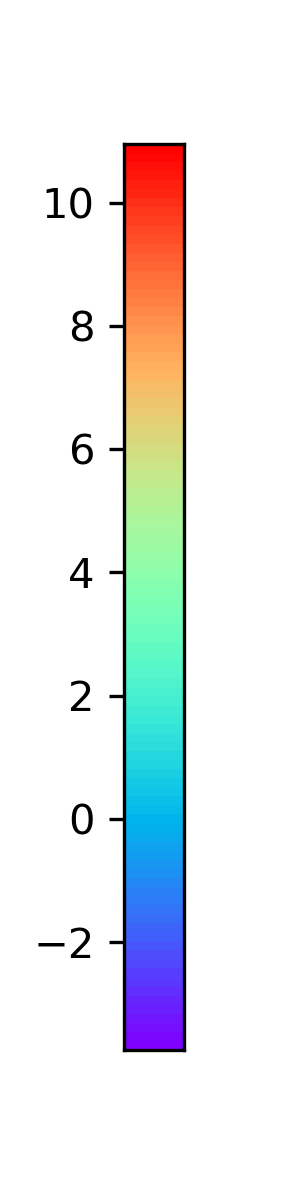}}
\put(-12,-7){\includegraphics[width = 0.45\textwidth, trim={10 10 11 15},clip]{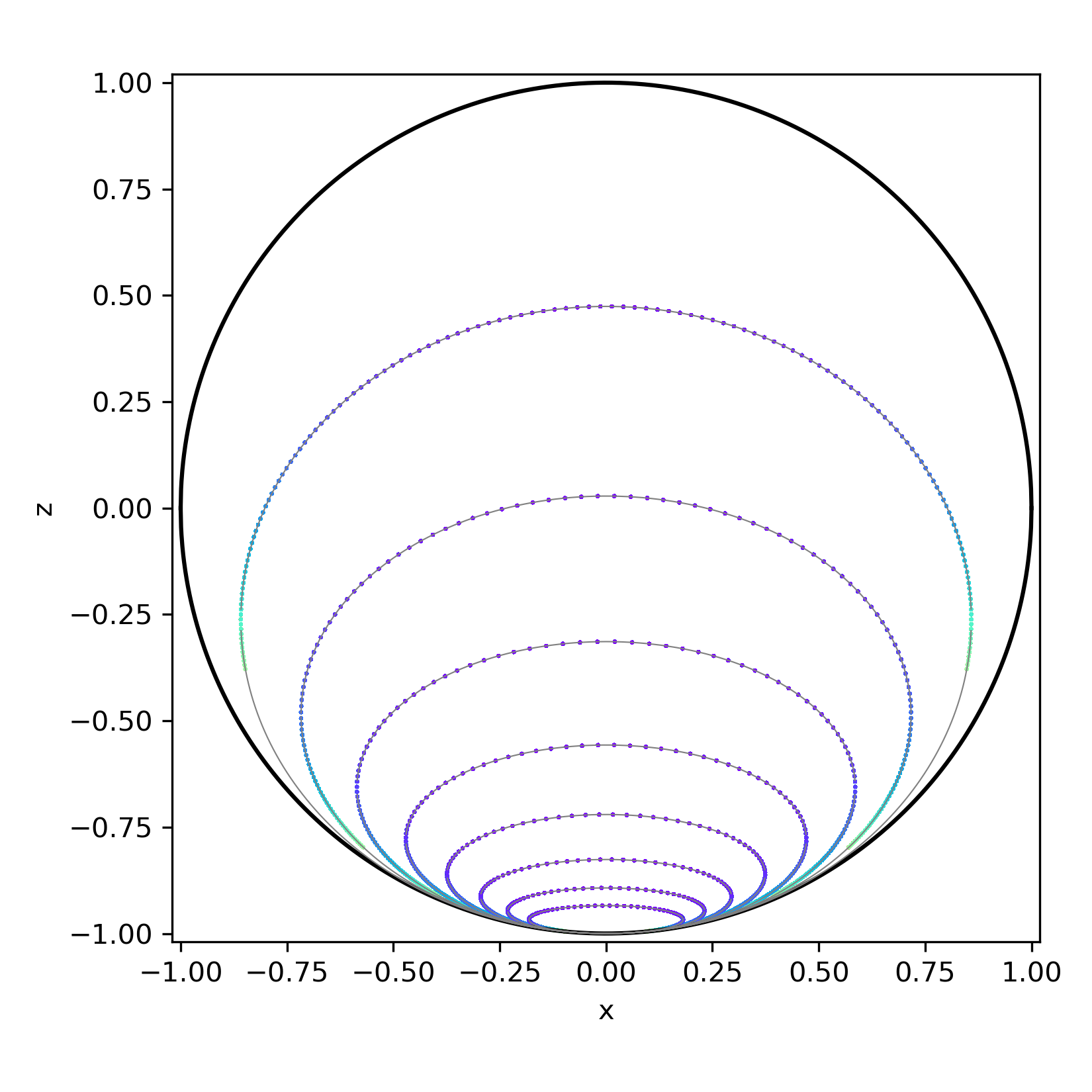}}
\put(198,202){(e)}
\put(217.5,178){$E$ [$\gamma$]}
\put(-155,210){\color{white} (a)}
\put(-150,131){\color{white} $x$}
\put(-229,211){\color{white} $z$}
\put(-203,203){\color{white} $t = T_1/2$}
\put(60,163){\color{cadetgrey} $t = T_1/2$}
\put(-32,210){\color{white} (b)}
\put(-27,131){\color{white} $x$}
\put(-106,211){\color{white} $z$}
\put(-73,203){\color{white} $t = T_1$}
\put(72,121){\color{cadetgrey} $t = T_1$}
\put(-155,94){\color{white} (c)}
\put(-150,17){\color{white} $x$}
\put(-229,96){\color{white} $z$}
\put(-205,87){\color{white} $t = 3T_1/2$}
\put(72,91){\color{cadetgrey} $t = 3T_1/2$}
\put(-32,94){\color{white} (d)}
\put(-27,17){\color{white} $x$}
\put(-106,96){\color{white} $z$}
\put(-75,87){\color{white} $t = 2 T_1$}
\put(81,68){\color{cadetgrey} $t = 2 T_1$}
\put(92,50){\color{cadetgrey} etc.}
\end{picture}
\caption{We show the evolution of the qubit state from $\ket{e}$, under the dynamics described by \eqref{eqmo-hometa} and \eqref{stoH-hometa}, for a realistic measurement efficiency $\eta = 0.45$. We show the density of simulated quantum trajectories at different times in (a--d); our methods reproduce the ellipses \eqref{z-ellipse}, demonstrating quantitative agreement with the relevant experimental literature \cite{Naghiloo2016flor, Tan2017}. In (e), we plot the Lagrangian manifold projected down into the $xz$ Bloch plane from the OP phase space at different times, starting with $t = T_1/2$, up to $t = 4T_1$ in increments of $T_1/2$. The known analytic solutions \eqref{z-ellipse} are plotted in grey, such that we can see the exact correspondence between the sampled LM, literature, and our simulations. 
The ellipsoids here and in Fig.~\ref{fig-unmon} bear obvious similarities; we stress however, that there are several qualitative differences between these two cases. In the case of Fig.~\ref{fig-unmon} we lose all the information, and many initial states, which originate deterministic solutions to \eqref{unmonitored-Bloch-eqs}, are required to see the decaying ellipses; in the present case, the ellipsoids arise from a \emph{single} initial state, due to different measurement records which contain partial information from the qubit's optical environment.
}\label{fig-SimLM}
\end{figure*}
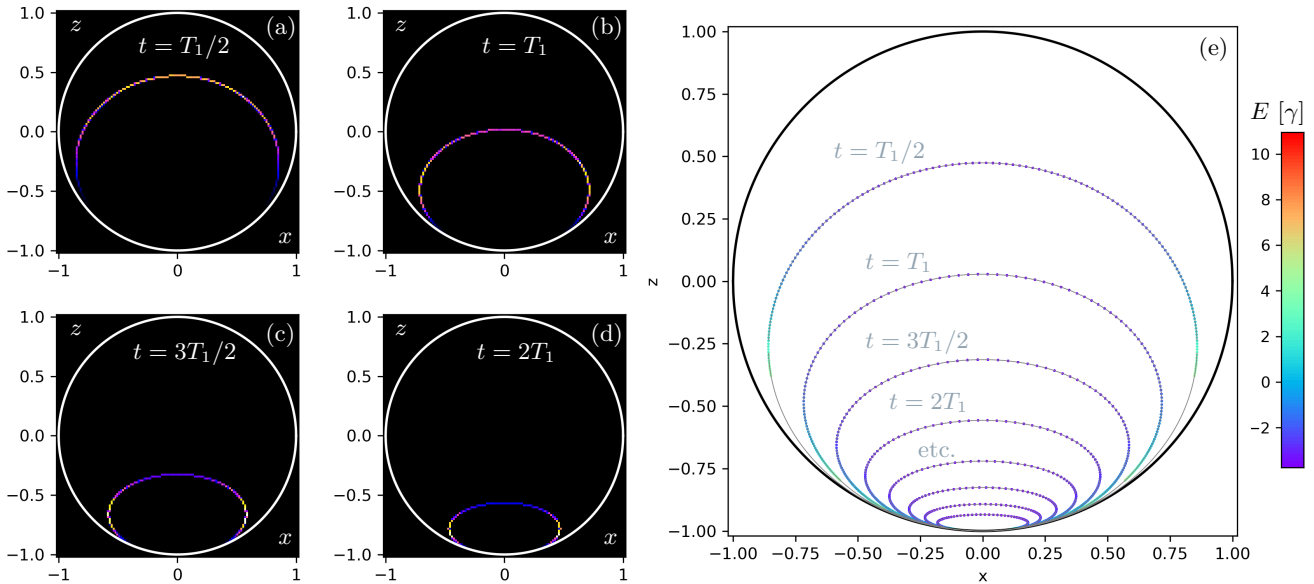

\par We develop a final example; in order to compare OPs directly to experimental results, we need to introduce measurement inefficiency. In the homodyne case with $\theta = 0$ and $\Omega= 0 = \delta$, corresponding to the SME readout $r = \sqrt{\eta \gamma}\: x + \xi$, and the equations of motion from either expanding \eqref{stateup-eta} to $O(dt)$, or converting the requisite SME from It\^{o} to Stratonovich, are
\begin{subequations} \label{eqmo-hometa}
\be \label{dotxeta}
\dot{x} = - r x^2 \sqrt{\gamma  \eta }+ r (z+1) \sqrt{\gamma  \eta }+\tfrac{\gamma}{2}  x (\eta +\eta  z-1) = f_x,
\ee \be 
\dot{y} = y \left( \tfrac{\gamma}{2} (\eta +\eta  z-1)-r x \sqrt{\gamma  \eta }\right) = f_y,
\ee \be \label{dotzeta}
\dot{z} =(z+1) \left(\tfrac{\gamma}{2}  (\eta +\eta  z-2)-r x \sqrt{\gamma  \eta }\right) = f_z .
\ee
\end{subequations}
We immediately see that for this choice of measured quadrature, the $y$ component of the dynamics can be eliminated with the choice $y = 0$, leaving only dynamics in the $xz$--plane of the Bloch sphere; we will assume $y=0$ for the remainder of this section. These assumptions, with imperfect $\eta$, give us the simplest version of this system that can be compared directly with existing experiments. The last piece we need is an understanding of the probability density function from which the readouts are drawn; using the same methods as above, we find
\be 
\mathcal{G}_{hom}^\eta = -\tfrac{1}{2}\left(r - x \sqrt{\eta \gamma} \right)^2 + \tfrac{1}{2}\eta\gamma \left( x^2 - z -1\right).
\ee
Thus we see that simulations involve repeated state updates as per \eqref{stateup-eta}, with readouts drawn at each step from a Gaussian of mean $x \sqrt{\eta \gamma}$ and with variance $dt^{-1}$. Optimal paths are derived from a stochastic Hamiltonian
\be \label{stoH-hometa}
\mathcal{H}_{hom}^\eta = p_x f_x + p_z f_z + \mathcal{G}_{hom}^\eta,
\ee
where $f_x$ and $f_z$ are the RHS of \eqref{dotxeta} and \eqref{dotzeta}, respectively. Our aim below will be to elucidate the basic aspects system dynamics, and show that our simulations and OPs match relevant results in the experimental literature.

A particularly important feature of the dynamics under homodyne fluorescence detection (absent a Rabi drive or other dynamics) is that all trajectories are constrained to an ellipse in the Bloch sphere at any given time \cite{Naghiloo2016flor, Tan2017} (and a similar ellipsoid is apparent in the heterodyne case \cite{PCI-2016-2}). The functional form of these ellipses has been derived in the literature \cite{Tan2017}, and follows
\be \label{z-ellipse}
z(t) = \frac{1}{u(t)}\left[ 1 \pm \sqrt{1- u(t) x^2} \right] -1,
\ee
for the time--dependent function
\be \label{u-ellipse-decay}
u(t) = \eta + (u_0 - \eta)e^{\gamma t},
\ee
where $u_0$ is set by the initial state according to
\be 
u_0 = \frac{2}{1+z_0} - \frac{x_0^2}{(1+z_0)^2}.
\ee
For example, with the initial state $\ket{e}$, we have $u_0 = 1$ at $t=0$, and at any time $t>0$ all possible trajectories evolving from $\ket{e}$ under dynamics from the inefficient homodyne measurement can be found on the ellipse \eqref{z-ellipse} (as a function of $x$); 
this ellipse is initially the great circle bounding the $xz$--plane of the Bloch sphere, and decays towards the ground state according to the time dependence \eqref{u-ellipse-decay}. 
We develop this example in Fig.~\ref{fig-SimLM}. 
In the left four panels (a--d) we show the density of simulated trajectories originating at $\ket{e}$ after different evolution times; 
we find essentially perfect agreement between the histograms of these simulated trajectory densities, and the analytic curves \eqref{z-ellipse} known from the literature. 
We stress that in departure from many other quantum measurement scenarios, there are final states on which it is impossible to post--select in the present system; 
typically some states appear very rarely in the dynamics, but here large regions of the Bloch sphere are forbidden entirely. 

We conclude by demonstrating the connection between the Lagrangian manifold from the OP phase space we described above, and the ellipses we have just described. The relative simplicity of the dynamics under inefficient homodyne fluorescence measurement make this an ideal example with which to illustrate the concepts discussed above.
In Fig.~\ref{fig-SimLM}(e) we show the projection of the LM originating at $\ket{e}$ into the $xz$--plane of the Bloch sphere (i.e.~we evolve the OP equations sampled across the initial LM, and then flatten the two--dimensional manifold, which lives in the four--dimensional phase space, into a plot that appears in the coordinates $x$ and $z$ only, at selected times).
We then see that we have exact agreement between the LM and the analytic curves \eqref{z-ellipse}, consistent with the fact that the OPs are themselves possible quantum trajectories.
This reinforces our statements about the consistency between the methods reviewed here, and the broader literature on continuous monitoring of fluorescence, but also serves to illustrate the role of the initial momenta $\mathbf{p}_i$ in the OP formalism. 
Choosing a particular $\mathbf{p}_i$ selects particular boundary conditions from the possible multitude, and the complete set of $\mathbf{p}_i$ contained in the LM index a complete set of possibilities for the OPs originating at a particular state.
While the LM in question may at first seem a somewhat abstract mathematical object, we are here able to highlight its physical character. 

\section{Closing Remarks \label{sec-conclude}}

We have given an overview of many useful methods and insights that arise from considering continuous quantum measurement, emphasizing examples in which we track a quantum emitter's state by gathering and measuring spontaneously--emitted photons. 
We have focused on a Kraus operator approach to this problem, most similar to that developed in Ref.~\cite{Jordan2015flor}, and made connections to a corresponding stochastic master equation description throughout. 
Many of the issues which arise in treating this particular type of system are common to stochastic quantum trajectories in general, and we have consequently addressed many of the important principles and typical problems one needs to become aware of when entering this research area. 
We have also been able to use the fluorescence examples above to offer accessible illustrations and introductions to selected advanced topics of contemporary interest; for example, we have been able to make comments to help the interested reader engage with work on the arrow of time in quantum trajectories, or understand how to generate and interpret trajectories which follow an optimal measurement record.

\par We can take a larger view of the processes we have described. We are accustomed to talking about the fluorescence process in terms of the emission of individual photons at particular times, from a sudden jump in the qubit state, i.e.~we typically discuss fluorescence in language which lends itself naturally to the photodetection case shown in Fig.~\ref{fig-simul3meas}(a). 
This notion of a jump and photon emerging at a particular time were mathematically enforced in the framework presented above by choosing outcomes in the Fock basis of the field, and with perfect measurement efficiency, we can say that we have collected complete information about the output mode. 
The other measurements we have discussed, are however, just as ``complete'' (in the sense that we have a POVM, and can ascribe a pure state to the qubit at any time in the $\eta = 1$ case; even in the $\eta < 1$ case, we are able to assign a state to the qubit which is consistent with a subsequent tomographic verification \cite{PCI-2016-2, Naghiloo2016flor}). The trajectories generated by the heterodyne and homodyne measurement schemes do not readily admit interpretation in terms of a photon emerging at any particular time; there is no single point along a trajectory in Fig.~\ref{fig-simul3meas}(b or c) to which we can point and say ``this is when the photon was emitted''. 
However, we can still ascribe a stochastic evolution to the qubit state, which agrees with any sensible check we know how to perform, and that evolution evidently depends on \emph{how} or \emph{what type of} environmental information was collected; the dynamics \emph{on average} reflect the same decay statistics we are used to regardless of the measurements (which is always the case in such continuous quantum measurement problems, and more or less the only assumption we made to formulate the model we have used). 
In other words, we the observer can dynamically assign our qubit a state\footnote{One could say that we, the observer, are updating our probabalistic prediction about outcomes of a future measurement; this ``best guess'' about possible future outcomes is made based on some preparation procedure, past measurement record, and the rules we typically refer to as ``quantum mechanics''.} which may e.g.~begin at $\ket{e}$ and eventually wind up in $\ket{g}$ according to the typical statistics of spontaneous emission; none of the other details of our photodetection story carry over in a simple way, however, to the case of generalized measurements. 
All of this is related to the fact that the emission process entangles the qubit with the field mode (see e.g.~the transition from \eqref{pure-initstate} to \eqref{pure-firststepstate}); this correlation allows us, as observers, to infer an emitter state based on the field, but also means that the type of information we infer about the emitter depends on what kind of state we project the field into (we must choose what kind of question to ask the field, and this will affect the kind of answer we can subsequently expect). 
Such issues get to the heart of quantum mechanics, and quantum trajectories generally serve as an excellent point of departure for such discussions about the foundations and interpretations of the theory, and the role of an observer probing a system which is otherwise sufficiently isolated so as to behave ``quantumly''.

\par There are, of course, many related topics which we have not been able to cover at all, but which we hope may become substantially easier for a new reader to digest with the foundation we have developed above. We highlight three in particular, in addition to those of a thermodynamic character \cite{Elouard2017_QTherm, Elouard2017_Maxwell, Mahdi2017Qtherm, Masuyama2018}, pertaining to quantum state smoothing \cite{Guevara2015, Gough2019-2, Chantasri2019, Guevara2019}, or the optimal path methods \cite{Chantasri2013, Chantasri2015, Jordan2015flor, Areeya_Thesis, Lewalle2016, Lewalle2018} we have already mentioned. 
First, some of the main interest in quantum trajectories is geared towards its applications to quantum control; we point out some literature which adapts the types of measurements we have described above to this purpose \cite{WangWiseman2001, PCI-2016-2}.
Second, dispersive measurements \cite{Murch2015teach, Gambetta2008, Murch2013, Weber2014}, allowing for measurements of e.g.~$\hat{L} \propto \sigma_z$, are extremely common in the literature; unlike the measurements we have described here, a qubit in a cavity (or coupled to a resonator) is directly probed with a pulse which reveals information about its state; virtually everything in terms of general approach we have developed above can be carried over to this case however. 
Such measurements have also been utilized in contemporary feedback control problems \cite{Taylor2017, Gourgy2018, Minev2018}, or in simultaneously weakly measuring multiple non--commuting observables \cite{Leigh2016, Lewalle2016, Lewalle2018, Atalaya2017}. 
Combinations of dispersive and fluorescence measurements have been realized experimentally \cite{Ficheux2018}. 
An introduction to a simple model of dispersive measurements, similar in spirit to that above, can be found in the appendix of Ref.~\cite{Lewalle2016} and references therein. 
Third, direct extensions of the measurements we have discussed here to the two--qubit case can serve as a springboard to study measurement--induced entanglement generation between a pair of emitters \cite{Mascarenhas2011, Santos2012, 2QFShort, 2QFLong}, and the decay of entangled states open to the influence of decay channels and/or measurements \cite{YuEberly_2004, Mintert2005, Viviescas2010, Mascarenhas2010, Carvalho2011}. 
``Bell state measurements'' are essential in recent tests of quantum mechanics and local realism \cite{WisemanLoopholeFreeReview}.
As such, we hope that our present work may help new readers to better digest a wide variety of literature concerning quantum measurements, open quantum systems, and beyond.

\begin{acknowledgements}
We acknowledge funding from NSF grant no.~DMR-1809343, and US Army Research Office grant no.~W911NF-18-10178. PL acknowledges additional support from the US Department of Education grant No.~GR506598 as a GAANN fellow, and thanks the Quantum Information Machines school at \'{E}cole de Physique des Houches for their hospitality during part of this manuscript's preparation. 
We are grateful to Joe Murphree for helpful comments. 
Analyses of the type described in appendix~\ref{sec-OPverify} and Fig.~\ref{fig-MLP_verify} have benefited from numerous conversations with Areeya Chantasri, and some versions of the underlying numerical methods have benefited from work by John Steinmetz and Kurt C.~Cylke.
Numerical methods underlying Figs.~\ref{fig-unmon}, \ref{fig-simul3meas}, \ref{fig-SimLM}, and \ref{fig-MLP_verify} have been implemented in Python 2.7.
\end{acknowledgements}

\appendix
\section*{APPENDICES}

\section{LO Power and Photocurrent Fluctuations \label{app-Noise}}

We consider the logic behind the scaling of the readouts in time in the dyne measurements, i.e.~we justify the expressions \eqref{het-rodef} and \eqref{hom-rodef}. The argument we offer follows directly from Ref.~\cite{Jordan2015flor}. Dyne measurements involve interfering the signal beam (which here contains only zero or one photon) with a strong coherent state LO. Recall that a coherent state
\be \label{coherent-state}
\ket{\alpha}  = e^{-|\alpha|^2/2} \sum_{n = 0}^\infty \frac{\alpha^n}{\sqrt{n!}} \ket{n}
\ee
has a mean photon number 
\be 
\dxval{N} = \exval{\alpha}{a^\dag a}{\alpha} = |\alpha|^2
\ee
with fluctuations
\be 
\Delta N = \sqrt{\dxval{a^\dag a a^\dag a}-\dxval{a^\dag a}^2} = |\alpha| = \sqrt{\dxval{N}}.
\ee
This is a direct result of the fact that a coherent state generates the Poisson statistics for photon arrival times; this is typical for arrival times pertaining to any random process characterized by a constant average rate. A constant underlying rate is, of course, consistent with an assumption that the LO has constant power (on average, up to the quantum fluctuations). If $h \nu$ is the energy per photon at frequency $\nu$ in the LO beam, then the LO power corresponds to the average photon number arriving in a time interval $dt$ according to $\dxval{P} = \dxval{N} h \nu / dt$, and we have fluctuations $\dxval{N} \pm \sqrt{\dxval{N}}$ or
\be 
\dxval{P} \pm \Delta P = \frac{h \nu}{dt}\left( \dxval{N} \pm \sqrt{\dxval{N}} \right).
\ee
We can talk about the fluctuations either in photon number, or LO power, which are
\begin{subequations}
\be  
\Delta N = \sqrt{\dxval{N}} = \sqrt{\frac{dt \dxval{P}}{h\nu}},\text{ or} 
\ee \be \Delta P = \frac{h \nu}{dt} \sqrt{\dxval{N}} = \sqrt{\frac{h \nu \dxval{P}}{dt}} 
\ee \end{subequations}
Thus we stress that for fixed $\nu$ and $\dxval{P}$, the fluctuations in the LO photon number go like $\sqrt{dt}$, and the fluctuations in the LO power go like $1/\sqrt{dt}$. In other words, for fixed average LO power, there are fluctuations in the subsequent photocurrent with variance $1/dt$, and it is precisely these physics which motivate the assignments \eqref{het-rodef} and \eqref{hom-rodef} in modeling the heterodyne and homodyne measurements, respectively. Those fluctuations are well approximated as Gaussian for large $\dxval{N}$ (the Poisson distribution converges to a Gaussian for large numbers). 
That our formalism leads to agreement with standard tools like the SME, and with experimental data (e.g.~as in Fig.~\ref{fig-SimLM}), further justifies the use of the intuition we provide here, connecting our measurement noise to the shot noise of the measurement LO / amplification pump.

\section{Connection Between Optimal Paths and Simulation / Data \label{sec-OPverify}}

\begin{figure}
    \centering
    \includegraphics[width = .7\columnwidth]{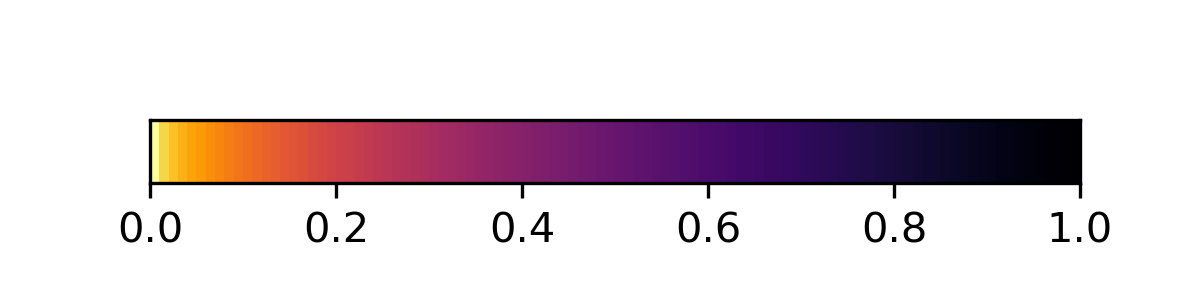} \\
    \includegraphics[width= .9\columnwidth, trim = {24 23 22 36},clip]{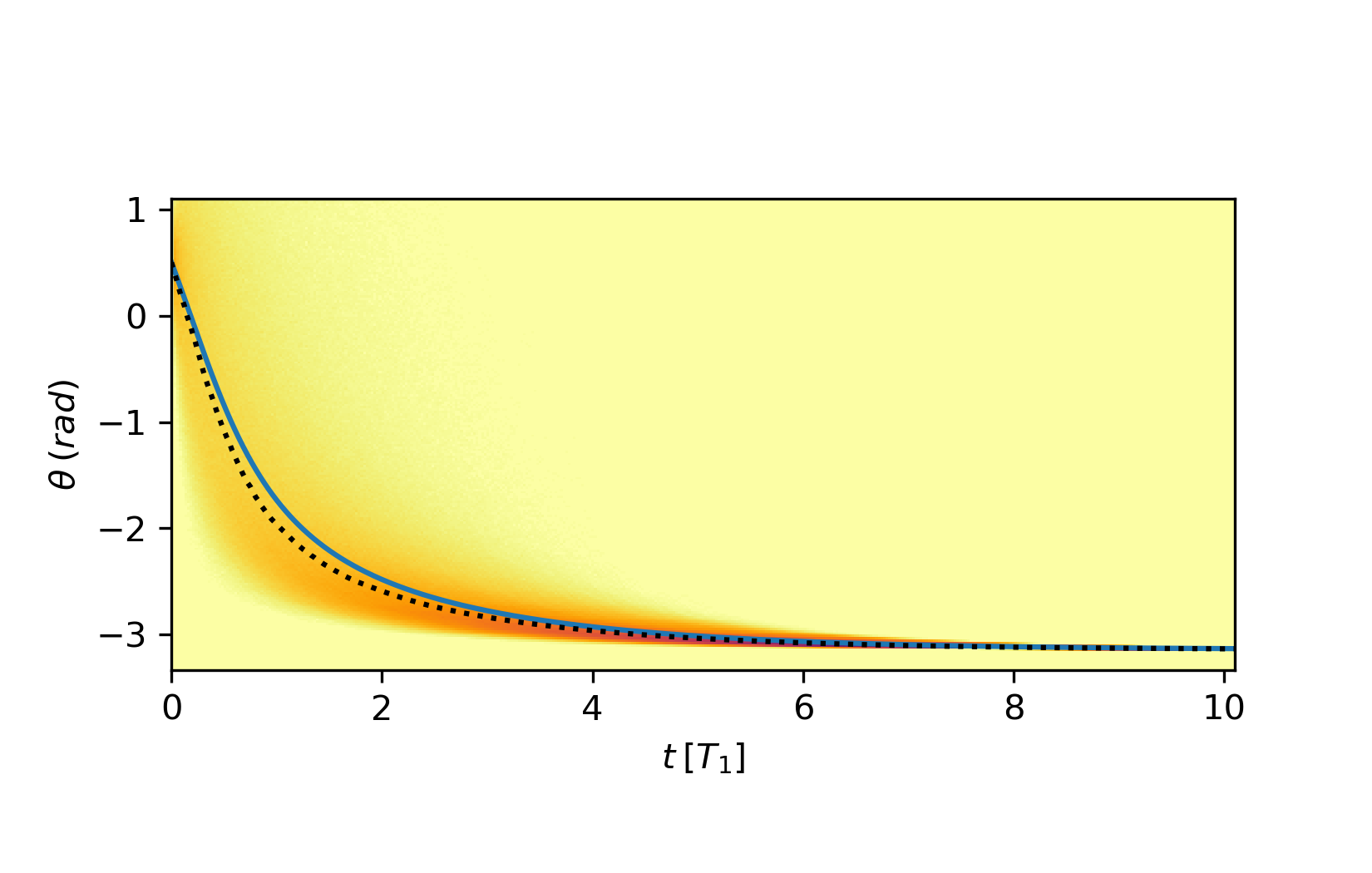} \\
    \begin{picture}(0,0)
    \put(87,108){(a)}
    \put(-113,95.5){$\vartheta$}
    \put(99,95.5){\textbf{--}$\ket{e}$}
    \put(99,34){\textbf{--}$\ket{g}$}
    \end{picture} \vspace{-15pt} \\
    \includegraphics[width = .9\columnwidth, trim = {57 0 51 20}, clip]{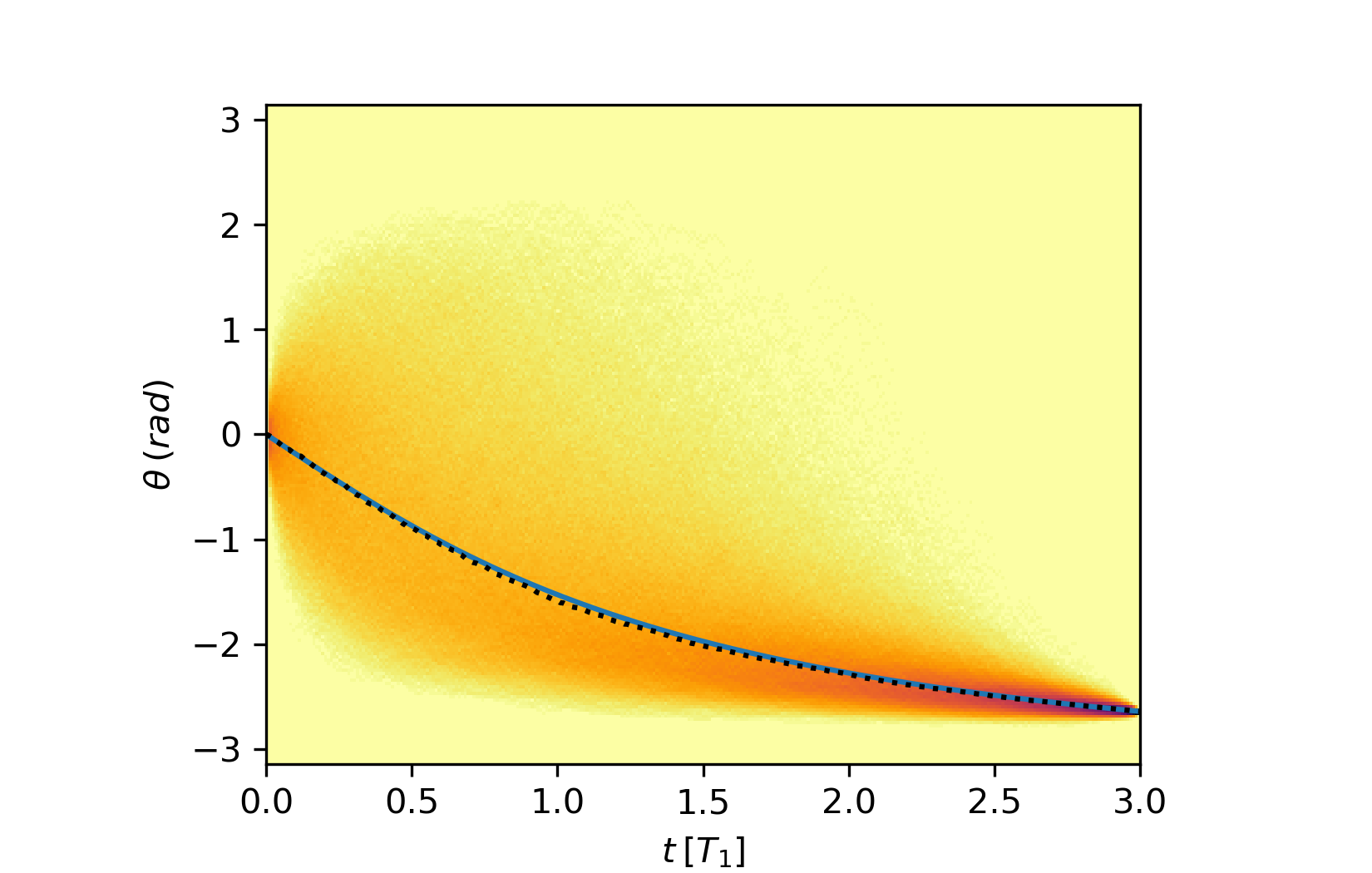} \\
    \begin{picture}(0,0)
    \put(85,171){(b)}
    \put(-115,105){$\vartheta$}
    \put(98,105){\textbf{--}$\ket{e}$}
    \put(98,32){\textbf{--}}
    \put(103,38){$\ket{g}$}
    \put(98,178.5){\textbf{--}}
    \put(103,172.5){$\ket{g}$}
    \end{picture} \vspace{-10pt}
    \caption{We compare ideal (pure--state) simulations against OPs generated by \eqref{H_OP_hom}, between a few boundary conditions. All times ($x$--axes) are expressed in units of $T_1$, and all cases shown use $\Omega =0$. We plot the density of trajectories initialized at a given state $\vartheta_i$, and post--selected to within $\vartheta_f \pm 0.01$ at the final time on each plot; the density scale, which is normalized between the initial and final timesteps, is shown on the accompanying colorbar. The ``experimental MLP'' extracted from the post--selected SQTs, as described in sec.~\ref{sec-OPverify}, is shown over the density in dotted black, while the corresponding curve derived from the stochastic Hamiltonian \eqref{H_OP_hom} is shown in solid blue. In (a), we use $\vartheta_i = \tfrac{1}{2}$, and $\vartheta_f = -\pi+0.01$ at $T = 10.1 T_1$ (the post--selection window keeps SQTs which satisfy $\vartheta(10.1) \in [-\pi,-\pi+0.02]$). These boundary conditions are chosen such that the MLP shown approximately follows the separatrix in Fig.~\ref{fig-homPS} from a state near $\ket{e}$, \emph{over} the excited state, before decaying towards $\ket{g}$. 
    In (b), we look at a somewhat less likely case, where trajectories are initialized at $\vartheta_i = 0$ (the excited state $\ket{e}$), and the post--selection is applied at $\vartheta_f = -\pi+\tfrac{1}{2}$ at $T = 3 T_1$; so we are looking at anomalously slow decay, which is still hovering a bit above the ground state $\ket{g}$ after several decay times. We note that we have excellent agreement between the theoretical and numerical MLPs. It is also apparent that paths which dip ``below'' our chosen $\vartheta_f$ and then rise back to it are exceedingly rare; there are far more events in this PSSE which decayed partially to the ``wrong'' side of the Bloch sphere at early times and then came back over $\ket{e}$ to meet the final boundary condition.
    }\label{fig-MLP_verify}
\end{figure}
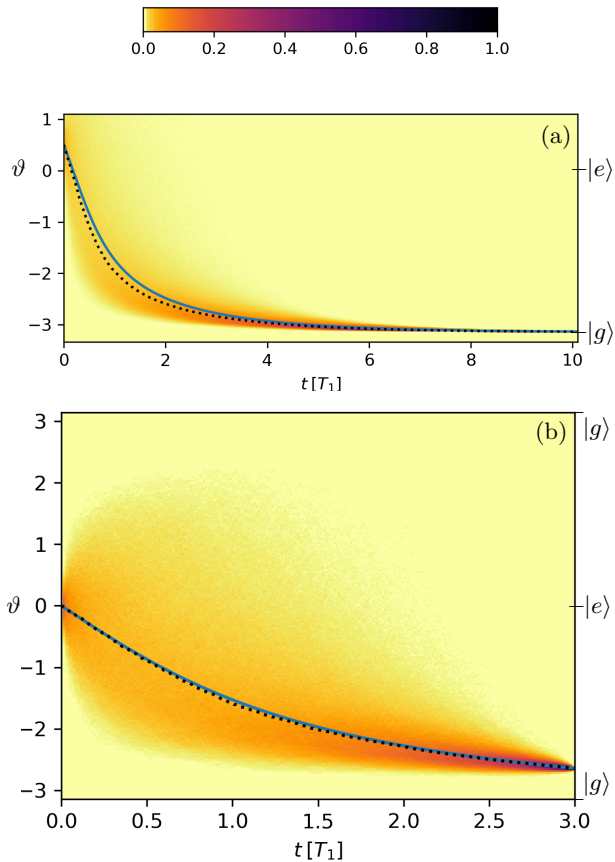

Our expectation in deriving most--likely paths (MLPs) is that, in typical cases, they should correspond approximately to a highest--probability peak in the post--selected sub--ensemble of trajectories connecting an initial and final state (see comments below about some of the subtleties, however).
Our aim here is to describe the procedure by which we approximate this concept in extracting a MLP from simulated SQTs or data, and show by example that such results are in good correspondence with those given by the CDJ optimization (see Sec.~\ref{sec-OP-derivation}). This has been performed in a variety of cases elsewhere \cite{Weber2014, Mahdi2016, Lewalle2018}, and we here describe and perform the requisite analysis for solutions of \eqref{H_OP_hom} and the corresponding simulation.

We generically suppose that we are given a simulated ensemble of SQTs $\lbrace{\rho(t)}\rbrace$, initialized at a particular $\rho_0$; such trajectories are necessarily sampled over some small but discreet timestep $dt$, consistent with the simulation procedures described in the main body of the text above. We describe the numerical manipulations we perform on such a set of $\lbrace{\rho(t)}\rbrace$ to extract an ``experimental MLP'' which may be compared directly with theory. These go as follows:
\begin{enumerate}
    \item We begin by imposing the final boundary condition, i.e.~we post--select on the desired $\rho_T$, at a later time $T$. This means that we must pick a distance measure $D(\rho_1,\rho_2)$ between quantum states (e.g.~fidelity, Bures distance, or similar), and keep the sub--ensemble for which $D(\rho(t=T),\rho_T) \leq W$, where $W$ is some small widow or allowed tolerance about the chosen final state. We'll call the post--selected sub--ensemble (PSSE) $\lbrace \rho(t) \rbrace_{ps}$; this is, by construction, the set of trajectories which connect $\rho_0$ to $\rho_T$.
    \item A simple intuition about the meaning of the MLP is that it should follow a densest cluster of trajectories in $\lbrace \rho(t) \rbrace_{ps}$; in order to approximate this concept of a ``densest cluster'' numerically, we must rank each SQT in $\lbrace \rho(t) \rbrace_{ps}$ according to its distance to all other SQTs in $\lbrace \rho(t) \rbrace_{ps}$. It is useful to construct a matrix of elements $\mathfrak{D}_{nm}$, where $n$ and $m$ are indices which run over the SQTs in $\lbrace \rho(t) \rbrace_{ps}$; we write
    \be 
    \mathfrak{D}_{nm} = \sum_k D(\rho_{n,k},\rho_{m,k}),
    \ee
    where the sum over $k$ runs over all the timesteps between $t = 0$ and $t = T$. The matrix $\mathfrak{D}$ will be symmetric as long as the distance measure $D$ is symmetric (we strongly discourage the use of any asymmetric distance measures for our present purposes). Then each SQT can be assigned a distance score relative to all other elements of $\lbrace \rho(t) \rbrace_{ps}$ according to $\bar{\mathfrak{D}}_n = \sum_m \mathfrak{D}_{nm}$, where a relatively smaller value of $\bar{\mathfrak{D}}_n$ indicates that trajectory $n$ of is closer to other trajectories in the post--selected set. These distances scores thus allow us to rank all of the trajectories in the PSSE.
    \item The final step in the procedure is a simple average; we take the the closest--clustered 5\%-10\% of trajectories in $\lbrace \rho(t) \rbrace_{ps}$, (those with the smallest 5\%-10\% of $\bar{\mathfrak{D}}_n$), and average them. The idea is that this approximates the smooth curve following the densest cluster of SQTs in $\lbrace \rho(t) \rbrace_{ps}$.  
\end{enumerate}
We apply this procedure, and compare with the analytic solutions to \eqref{H_OP_hom}, for a few selected boundary conditions, in Fig.~\ref{fig-MLP_verify}.

\par We close with a few remarks about the procedure we have just described. We lack a formal proof that the numerical procedure just outlined necessarily always converges to the optimization we perform by the CDJ path integral method \cite{Chantasri2013}. 
We nonetheless see in Fig.~\ref{fig-MLP_verify} that the present case continues to support the agreement between SQTs (from either simulation, or real data) and OPs, which have been successfully compared in many other scenarios as well \cite{Weber2014, Mahdi2016, Lewalle2018}. 
An important feature of the numerical procedure we have described is that it ranks trajectories by considering their entire evolution connecting $\mathbf{q}_i$ and $\mathbf{q}_f \tilde{\pm} W$, rather than looking for some piecewise optimization, or explicitly following a peak trajectory density at particular times. This reflects the character of the variational approach we have used in the theory; invoking $\delta \mathcal{S} = 0$ leads to an optimization over an entire trajectory duration. While in many simple cases some piecewise numerical optimization could get us similar results, we would expect this simplified picture to fail in other cases, as it makes a substantial conceptual departure from the theory we want these numerics to match.
In terms of practical considerations, we point out that the ``experimental MLP'' procedure above leads to an attrition in the number of SQTs used at each step, first due to post--selection, and then due to the ranking procedure. In order to obtain smooth results, it is necessary that at least several hundred SQTs make it into the final, most--closely clustered group which are averaged in step 3 above. The number of trajectories before post--selection, the size of the post--selection window, and the overall probability for trajectories to reach the desired final state, all play a role in these final numbers. The finer the post--selection window, the more initial trajectories are required at the beginning in order to obtain smooth results. Post--selections on very rare events may be prohibitively difficult to verify in practice, simply due to the overwhelming amount of data that would need to be collected / generated in order to have an appropriate number of SQTs in the PSSE.

\bibliography{refs}
\end{document}